\documentclass[a4paper,11pt]{article}

\usepackage[utf8x]{inputenc}
\usepackage[centertags]{amsmath}
\usepackage{amssymb,amsthm,epsfig,psfrag}
\usepackage{graphicx}
\usepackage{dsfont}
\usepackage{color}
\usepackage{pstricks-add}
\usepackage{fullpage}
\usepackage{etex}

\usepackage{bm}
\usepackage{epic}
\usepackage{hyperref}
\usepackage{young}
\usepackage{multirow}
\usepackage{fancybox}

\usepackage{fix-cm}

\usepackage[nosort]{cite}
\usepackage[bulletsep]{collref}
\usepackage{tensor}

\def \be  {\begin{equation}}
\def \ee  {\end{equation}}
\def \ba  {\begin{eqnarray}}
\def \ea  {\end{eqnarray}}
\def \baa {\begin{eqnarray*}}
\def \eaa {\end{eqnarray*}}
\def \bb  {\begin {thebibliography} }
\def \eb  {\end{thebibliography}}
\def \lab #1 {\label{#1}}

\newcommand{\nn}{\nonumber}

\newcommand{\beq}{\begin{equation}}
\newcommand{\eeq}{\end{equation}}
\newcommand{\beqa}{\begin{eqnarray}}
\newcommand{\eeqa}{\end{eqnarray}}

\newcommand{\RR}{\mathcal{R}}
\newcommand{\MM}{\mathcal{M}}

\newcommand{\NN}{\mathcal{N}}

\newcommand{\sfrac}[2]{{\textstyle\frac{#1}{#2}}}
\newcommand{\half}{\sfrac{1}{2}}

\newcommand{\quarter}{\sfrac{1}{4}}
\newcommand{\alg}[1]{\mathfrak{#1}}

\renewcommand{\a}{\alpha}
\newcommand{\da}{{\dot{\alpha}}}
\newcommand{\db}{{\dot{\beta}}}
\renewcommand{\b}{\beta}
\newcommand{\la}{\lambda}
\newcommand{\tla}{\tilde\lambda}
\newcommand{\g}{\gamma}

\newcommand \widebar [1] {\overline{#1}}

\newcommand \vev [1] {\left\langle{#1}\right\rangle}

\newcommand{\eqn}[1]{(\ref{#1})}

\newcommand{\osca}{\mathbf{a}}
\newcommand{\oscb}{\mathbf{b}}

\begin{document}

\thispagestyle{empty}

\begingroup\raggedleft\footnotesize\ttfamily
HU-Mathematik-2013-12\\ 
HU-EP-13/33\\ 
AEI-2013-235\\
DESY 13-488\\
ZMP-HH/13-15\\

\vspace{15mm}
\endgroup

\begin{center}
{\Large\bfseries Spectral Parameters for Scattering Amplitudes in $\mathcal{N}=4$ 
Super Yang-Mills Theory\par}%
\vspace{15mm}

\begingroup\scshape\large 
Livia Ferro${}^{1}$, Tomasz \L ukowski${}^{2}$,
Carlo Meneghelli${}^{3}$,\\[0.2cm]
Jan Plefka${}^1$,
Matthias Staudacher${}^{2,4}$
\endgroup
\vspace{7mm}

\textit{${}^1${
Institut f\"ur Physik,
Humboldt-Universit\"at zu Berlin,
Newtonstra{\ss}e 15, 12489 Berlin, Germany}\\[0.2cm]
${}^2${
Institut f\"ur Mathematik und Institut f\"ur Physik, Humboldt-Universit\"at zu Berlin,\\
IRIS Adlershof, 
Zum Gro\ss en Windkanal 6, 12489 Berlin, Germany}\\[0.2cm]
${}^3${
Fachbereich Mathematik, Universit\"at Hamburg, Bundesstra\ss e 55, 20146 Hamburg,\\
\& Theory Group, DESY, Notkestra\ss e 85, 22603 Hamburg, Germany}\\[0.2cm]
${}^4${
Max-Planck Institut f\"ur Gravitationsphysik, Albert-Einstein-Institut,\\
Am M\"uhlenberg 1, 14476 Potsdam, Germany
}}\\[0.4cm]
{\tt ferro,plefka$\bullet$physik.hu-berlin.de, carlo.meneghelli$\bullet$desy.de, lukowski,staudacher$\bullet$mathematik.hu-berlin.de}

\vspace{8mm}

\textbf{Abstract}\vspace{5mm}\par
\begin{minipage}{14.7cm}
Planar $\mathcal{N}=4$ Super Yang-Mills theory appears to be a quantum integrable four-dimensional conformal theory. This has been used to find equations believed to describe its exact spectrum of anomalous dimensions. Integrability seemingly also extends to the planar space-time scattering amplitudes of the $\mathcal{N}=4$ model, which show strong signs of Yangian invariance. However, in contradistinction to the spectral problem, this has not yet led to equations determining the exact amplitudes. We propose that the missing element is the spectral parameter, ubiquitous in integrable models. We show that it may indeed be included into recent on-shell approaches to scattering amplitude integrands, providing a natural deformation of the latter. Under some constraints, Yangian symmetry is preserved. Finally we speculate that the spectral parameter might also be the regulator of choice for controlling the infrared divergences appearing when integrating the integrands in exactly four dimensions.
\end{minipage}\par
\end{center}
\newpage

\section{Introduction and Overview}

Despite its seeming complexity on the Lagrangian level the maximally supersymmetric Yang-Mills theory ($\mathcal{N}=4$ SYM) \cite{Brink:1976bc,Gliozzi:1976qd} 
might be the simplest interacting
four-dimensional quantum field theory known. Its global
 Poincar\'e symmetry is maximally enhanced
to the $\mathcal{N}=4$ superconformal symmetry group generated by the super-algebra
$\alg{psu}(2,2|4)$. 
Excitingly, this Yang-Mills model with local gauge symmetry
group $SU(\mathrm{N})$ appears
to be integrable in 't~Hooft's planar $\mathrm{N}\to\infty$  limit.
This property has been instrumental for finding closed equations for the exact spectrum of
planar anomalous dimensions of local gauge invariant composite operators. These equations then also determine the exact planar two-point correlation functions of the theory. 
The key to this solution lies in a reformulation as an integrable one-dimensional system exhibiting features of both quantum spin chains as well as two-dimensional sigma models. 
The refined formalism of the Quantum Inverse Scattering 
Method centered around the Yang-Baxter equation could then be applied to 
find the spectrum exactly in the 't~Hooft coupling constant $\lambda=g^{2}_{\text{YM}}\, \mathrm{N}$. See \cite{Beisert:2010jr} for a recent, fairly up-to-date review. Indeed integrability here accounts for an extension of the superconformal symmetry to an infinite-dimensional algebra of Yangian
type. 

Integrability, being a property of the large $\mathrm{N}$ planar theory, is not visible at the Lagrangian 
level, where only the classical superconformal symmetry $\alg{psu}(2,2|4)$ is implemented. Instead it manifests itself at the level
of ``observables''. By these we mean any gauge invariant expectation
value of the quantum field theory such as an $n$-point function of local operators, a vacuum
expectation value of a Wilson loop operator or, in further abuse of the word, a scattering amplitude. So far there is no universal
integrability theory for all such observables of  $\mathcal{N}=4$ SYM available. In this work we focus
on the sector of scattering amplitudes, explaining and extending the results of a recent letter of ours \cite{Ferro:2012xw}. 

There have been spectacular advances in the structural understanding and computation of  planar scattering amplitudes in
$\mathcal{N}=4$ SYM in recent years. 
In particular the development of novel techniques in the form of
on-shell recursion relations for the determination of tree-level amplitudes 
\cite{Britto:2004ap,Britto:2005fq} and on-shell super-amplitudes \cite{ArkaniHamed:2008gz,Brandhuber:2008pf,Elvang:2008na} along with the method of 
generalized unitarity applied to one and higher-loop amplitudes \cite{Bern:1994zx,Bern:1994cg} 
have led to the analytic construction of all tree-amplitudes in
closed form \cite{Drummond:2008cr} and to many high-multiplicity and high-loop results 
(for
recent reviews see \cite{Dixon:2011xs,Roiban:2010kk,Drummond:2010km,Alday:2010kn}). 
Scattering amplitudes in $\mathcal{N}=4$ SYM enjoy a dual superconformal symmetry \cite{Drummond:2008vq} reflecting the duality
of amplitudes to light-like polygonal Wilson loops 
\cite{Alday:2007hr,Drummond:2007aua,Brandhuber:2007yx,Drummond:2007cf}.
The dual $AdS_{5}\times S^{5}$ string theory explanation at strong coupling of this
symmetry was provided in terms of a 
combination of bosonic and fermionic T-dualities \cite{Berkovits:2008ic,Beisert:2008iq}. 
In fact these developments have led to the, in principle, exact construction of the four and five-gluon scattering amplitudes at all orders in  $\lambda$, where the conjectured exact cusp anomalous dimension \cite{Beisert:2006ez} enters as an input, however. 
The closure of the
ordinary and dual superconformal symmetry algebras leads to the Yangian symmetry algebra
$Y[\alg{psu}(2,2|4)]$ under which tree-level super-amplitudes are invariant \cite{Drummond:2009fd} for non-collinear external momenta \cite{Bargheer:2009qu}. Yangian algebras 
are infinite-dimensional Hopf algebras with a level structure (algebra filtration) built upon
a semi-simple Lie algebra or super-algebra at level zero. They are a manifestation of integrability.
At the one-loop level a complicated deformation of the Yangian generators arises 
\cite{Sever:2009aa,Beisert:2010gn}.
It appears that the Yangian algebraic structure is deeply connected to the Gra{\ss}mannian formulation of tree-level scattering amplitudes which was 
pioneered in \cite{ArkaniHamed:2009dn,Mason:2009qx}. This formulation solves the super BCFW
recursion relation, it is Yangian invariant and even unique under
certain assumptions \cite{Drummond:2010qh,Drummond:2010uq,Korchemsky:2010ut}. Further developments led to the construction of the all-loop \emph{integrand} for any super-amplitude upon employing a 
conjectured generalized all-loop BCFW recursion relation going beyond the tree-level case
\cite{ArkaniHamed:2010kv,ArkaniHamed:2010gh}.
In the remarkable recent work \cite{ArkaniHamed:2012nw} a novel way of constructing
amplitudes in terms of on-shell diagrams was put forward. These on-shell diagrams
may be built up as planar graphs starting from two types of trivalent (=cubic) vertices representing the
maximally helicity violating (MHV) and anti-MHV three-particle super-amplitudes. 
The graph edges correspond to cut propagators, i.e.~they involve one-dimensional delta functions putting the carried momenta on-shell. Finally all internal bosonic and fermionic
degrees of freedom are integrated
over. 
At first sight this construction leads to an increase in the complexity of calculating amplitudes, for example
the four-point MHV 
tree-level amplitude is given by a one-loop on-shell box diagram, whereas the one off-shell loop
correction is represented via a five-loop on-shell diagram. However, if true
it yields an interesting constructive way of obtaining the entire $S$-matrix
 of $\mathcal{N}=4$ SYM from on-shell data alone. 
Furthermore the work of \cite{ArkaniHamed:2012nw} yields unexpected connections to mathematical structures
 such as the  positive Gra{\ss}mannian and cluster algebras, ubiquitous in modern mathematical physics.
 
In the present paper we aim at unifying these developments with the observation of Zwiebel
\cite{Zwiebel:2011bx} who connected the tree-level four-point MHV scattering amplitude to the one-loop dilatation operator of $\mathcal{N}=4$ SYM. The latter being the Hamiltonian of an integrable spin-chain is generated by an R-matrix satisfying the Yang-Baxter equation, 
the cornerstone of the Quantum Inverse Scattering Method, see e.g.~\cite{Faddeev:1996iy}. Indeed from this viewpoint a Yangian
symmetry algebra is not the starting point but rather a consequence of the existence
of an R-matrix with the monodromy matrix encapsulating the Yangian algebra. 
Motivated by this we consider
the Yang-Baxter equation for the scattering amplitude problem which operates on
 the triple tensor product of one fundamental
and two $\mathcal{N}=4$ super-oscillator realizations of $\mathfrak{gl}(4|4)$
corresponding to the on-shell degrees of freedom of external legs. In the scattering problem, these oscillators correspond to either the
on-shell chiral superspace spinors parametrizing the momentum and helicity degrees of freedom, or alternatively
their ``quarter'' Fourier transforms to super-twistor space variables.
The matrix elements of the well-known R-matrix intertwining two oscillator $\mathfrak{gl}(N|M)$ representations
are found explicitly. They may also be encoded in a kernel, which
turns out to take the form of a $2\to 2$ scattering amplitude in the
Gra{\ss}manian formulation deformed by three complex parameters. We call
this object the harmonic R-matrix.
The parameters in turn are identified with deformed complex helicities
of the on-shell legs
subject to certain conservation conditions. Even more we show
that there is yet a more fundamental level of these findings building upon a deformation of the 
``atoms'' of the on-shell diagram construction of \cite{ArkaniHamed:2012nw}.  They correspond
to matrix kernels spectrally deforming the $2\to 1$ (MHV${}_{3}$) and $1\to 2$ ($\widebar{\text{MHV}}_{3}$) three-particle scattering amplitudes. 
These ``three-point harmonic R-matrices'' may then
be inserted into the suitably generalized on-shell diagram constructions of 
\cite{ArkaniHamed:2012nw} via a box diagram, and one indeed recovers the $2\to 2$ harmonic 
R-matrix. Finally, unifying the on-shell diagram constructions of \cite{ArkaniHamed:2012nw} with our spectral
parameter deformation leads us to the formulation of generalized Yang-Baxter equations for a ``generalized'' R-matrix with $k$ incoming and $n-k$ outgoing particles.

While the spectral parameter deformation of amplitudes we report on surely represents an interesting
mathematical structure in itself, two pressing questions arise. What is the physical interpretation of the deformation, and what is its practical use, if any? Here we can offer an insight already reported on in our letter \cite{Ferro:2012xw}. As discussed above the methods of \cite{ArkaniHamed:2010kv,ArkaniHamed:2010gh,ArkaniHamed:2012nw} allow for
the in principle general construction of \emph{unregulated} off-shell loop integrals using 
the generalized BCFW recursion and the on-shell diagrammatics.
Due to the IR divergencies present in these integrals these, however, 
are generically ill-defined and from that perspective useless unless a regulating
prescription is provided.
Traditionally dimensional reduction is used. For $\mathcal{N}=4$ SYM, a natural alternative
is the Higgs regulator of \cite{Alday:2009zm}, which has recently been introduced and successfully employed in \cite{Henn:2010bk,Henn:2010ir,Lipstein:2013xra}. See also the further, novel dual-superconformally invariant regulator proposed in \cite{Bourjaily:2013mma}. 
While dimensional regularization breaks all conformal symmetries
of the theory, the Higgs regulator of \cite{Alday:2009zm} and the recently proposed IR regulator of \cite{Bourjaily:2013mma} preserve 
(extended) dual conformal symmetry. All three prescriptions, however, obviously break 
standard conformal symmetry through the introduction of a novel scale. Improving on this, we provide
initial evidence that the amplitudes suitably deformed by a spectral parameter regulate the theory while {\it preserving} the full superconformal symmetry. In particular, the regulating scale is not externally introduced, but rather automatically tuned by the external kinematical data.
We demonstrate this mechanism for the concrete
example of the one-loop correction to the MHV${}_{4}$ amplitude.  Here the spectral parameter
deformation applied to the on-shell diagram formalism leads  to an analytic regularization
of the one-loop box integral induced by the deformed helicities of external and internal
particles. This is exciting and certainly merits further studies. 

Somewhat ironically, our specific choice of spectral regulator does break dual conformal invariance. However, this is not necessarily a problem, and there are two potential, distinct ways out. Firstly, dual conformal invariance was an ``unexpected'' symmetry in the first place, and there is no physical reason for it to hold exactly.
In this context one should note that while integrable spin chains are certainly based on an underlying Yangian algebraic structure,  the spectrum of spin chains is not Yangian invariant, i.e.~the integrable Hamiltonian does not commute with the Yangian generators.
Secondly, as we will show, there is a way to choose the spectral parameters to obtain Yangian invariant deformations of on-shell diagrams for integrands. Interestingly, the corresponding choice then renders the integrals infrared divergent. However, it is possible that the divergence is a global one sitting in front of a properly defined deformed all-loop amplitude, which might well render it manageable.

The paper is organized as follows. In section \ref{Sec:ch1_FourPoint} we explain how to solve the Yang-Baxter equation for the cases of present interest. Section \ref{Sec:ch1_intro} explains the general method, and introduces suitable Schwinger-type oscillators useful for both the spectral problem as well as for scattering amplitudes. Section \ref{Sec:ch1_HarmonicAction} is somewhat outside the main line of development, and illustrates the method of section \ref{Sec:ch1_intro} with a novel derivation of the known R-matrix of the $\mathcal{N}=4$ one-loop spectral problem in the oscillator basis. In section \ref{Sec:ch1_RmatrixGrass} we solve the Yang-Baxter equation in the form of a Gra{\ss}mannian integral, providing, as shown in section \ref{Sec:Deformationsoffourpoint}, the sought deformation of the tree-level MHV${}_{4}$ amplitude. In section \ref{Sec:ch2_ThreePoints} we find analogous deformations of the Hodges-type on-shell three-point vertices \cite{Hodges}, show that they satisfy certain bootstrap equations familiar from the theory of two-dimensional integrable quantum field theories and relate the 
spectral parameters to representation labels. We also demonstrate in section \ref{Sec:gluing.fourpt} that four suitably deformed three-point vertices may be combined to recover the deformed four-point amplitude of section \ref{Sec:Deformationsoffourpoint}, in natural generalization of the undeformed case. In section \ref{Sec:ch3_ManyPoint} we study generic on-shell diagrams. After a succinct discussion of the rather involved undeformed formalism in section \ref{Sec:ch3_Undeformed}, we explain in section \ref{Sec:ch3_Deformations} that the construction of  \cite{ArkaniHamed:2012nw} again very naturally lifts to our deformation. We also find in section \ref{Sec:ch3_Moves} that certain graphic ``moves'' inherent to the on-shell formalism remain valid under the deformation, after implementing certain restrictions on the allowed set of spectral parameters. In section \ref{Sec:ch4_Symmetries} we study the all-important issue of the symmetries of our deformation. We find that {\it locally}, i.e.~on the level of the Lie-algebra, superconformal symmetry survives the deformation. Excitingly, the full Yangian symmetry may also be maintained, if the same restrictions on the set of spectral parameters found in \ref{Sec:ch3_Moves} hold. Finally, in section \ref{Sec:ch5_OneloopReg} we explain in some detail our findings on the spectral regularization of the one-loop four-point amplitude already reported in \cite{Ferro:2012xw}. As already mentioned, the regularization is superconformally invariant, but ``mildly'' breaks dual conformal invariance. Some further technical issues are delegated to three appendices.


\section{R-Matrices, Spin Chains, and Amplitudes} 
\label{Sec:ch1_FourPoint}

\subsection{Yang-Baxter Equation, R-Matrices, and Oscillator Realizations}
\label{Sec:ch1_intro}
In this section we construct $\mathfrak{gl}(N|M)$ invariant solutions to the Yang-Baxter equation corresponding to so-called oscillator realizations reviewed  below.
These solutions, called R-matrices, are obtained as intertwiners of two such realizations.
Expressions for R-matrices of this type have been known for some time and were derived as part of the Quantum Inverse Scattering Method program, see \cite{Kulish:1981gi}, 
and are  well understood in the framework of quantum groups \cite{Drinfeld:1985rx,Drinfeld:1986in,Drinfeld:1987sy}.
They take the form
\begin{equation}\label{LeningradR}
 \mathbf{R}_{12}(z)\,=\,(-1)^{\mathbb{J}}\,\frac{\Gamma(\mathbb{J}+1+z)}{\Gamma(\mathbb{J}+1-z)}\,, 
\qquad \qquad
\mathcal{C}_2\,=\,\mathbb{J}\,(\mathbb{J}+1)\,,
\end{equation}
where $\mathcal{C}_2$ is the quadratic Casimir operator.
This expression first appeared in \cite{Kulish:1981gi} as an interwiner of $\mathfrak{sl}(2)$
highest weight representations and is the relevant R-matrix for the
1-loop integrable spin-chain in planar $\mathcal{N}=4$ SYM \cite{Beisert:2003yb}.
Here we will write these R-matrices in two alternative forms more suitable for our purposes, namely the so-called ``harmonic action'' form as well as the Gra\ss mannian integral form. 
The former is particularly convenient to act on spin-chain states and 
allows for a first principles derivation of the harmonic action form of the Hamiltonian presented
in \cite{Beisert:2003jj}. The latter can be interpreted as  a deformation of the four-point tree-level amplitude in  $\mathcal{N}=4$ SYM.

\begin{figure}[ht]
\begin{center}
\scalebox{0.35}{\input{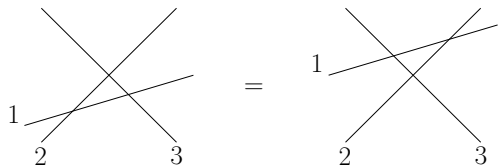}}
\end{center}
\caption{Yang-Baxter equation.}
\label{Fig:YBEclassic}
\end{figure}

Let us start by recalling some basic facts about the Yang-Baxter equation
\begin{equation}
\label{YBe}
\mathbb{R}_{23}(z_1) \mathbb{R}_{13}(z_2) \mathbb{R}_{12}(z_2-z_1)=\mathbb{R}_{12}(z_2-z_1) \mathbb{R}_{13}(z_2) \mathbb{R}_{23}(z_1).
\end{equation}
In \eqref{YBe} $\mathbb{R}_{ij}(z)$ are linear operators acting on the tensor product of three vector spaces 
$\mathbb{V}_1\otimes \mathbb{V}_2\otimes \mathbb{V}_3$ and each $\mathbb{R}_{ij}(z)$ acts non-trivially only on  
$\mathbb{V}_i\otimes \mathbb{V}_j$. The parameter $z$ takes complex values  and is called spectral parameter.
This equation plays a central role in the theory of quantum integrable models.
It can be interpreted as a factorization condition for multiparticle scattering in 1+1 dimensions,
where the  R-matrix is regarded as a two-particle scattering matrix. 
In this case the spectral parameter is related to the rapidity of the two dimensional on-shell particle.
In an integrable system the order in which particles scatter does not matter, the three-particle scattering can be written in two different orders and the result does not depend on which of the two we choose. 
The graphical representation of \eqref{YBe} is presented in figure \ref{Fig:YBEclassic}.

The Yang-Baxter equation \eqref{YBe} is an over-determined system of equations and there is no known complete classification of its solutions.
Luckily, there are recipes to systematically construct solutions of \eqref{YBe}.
The basic idea is to characterize R-operators as solutions to some \emph{linear} equation.  
Let us see how this works in the case of our interest.
Let  $\mathbb{V}_3=\mathbb{C}^{N|M}$ correspond to the fundamental representation of  $\mathfrak{gl}(N|M)$ and take
\begin{equation}
\label{Ri3}
\mathbb{R}_{i3}(z)\,\mapsto\,\mathbf{L}_i(z)\,:=\,
 z\,\mathbf{1}\,+\sum_{\mathcal{A},\mathcal{B}}\,
(-1)^{\mathcal{B}}\,J_{i\,
\mathcal{B}}^{\mathcal{A}}\,\,e^{\mathcal{B}}_{\mathcal{A}}\,,
\qquad  \qquad 
\mathbb{R}_{12}(z)\,\mapsto\, \mathbf{R}_{12}(z)\,,
\end{equation}
where $\mathcal{A}$ and $\mathcal{B}$ are $\mathfrak{gl}(N|M)$ indices, $\mathcal{A},\mathcal{B}=1,\ldots, N|N+1,\ldots,N+M$ and $(-1)^{\mathcal{B}}$ encodes grading. 
Here $J_{i\,\mathcal{B}}^{\mathcal{A}}$ are the generators of the $\mathfrak{gl}(N|M)$ algebra written in the representation $i$ and $e^{\mathcal{B}}_{\mathcal{A}}$ are the generators in the fundamental representation. The choice of $\mathbb{R}_{i3}$ in \eqref{Ri3}  corresponds to a specific representation of the Yangian algebra $Y(\mathfrak{gl}(N|M))$, see e.~g.~\cite{Bazhanov:2010jq} for more details. 

After the substitution \eqref{Ri3}, the Yang-Baxter equation \eqref{YBe} becomes a linear equation for the R-matrix $\mathbf{R}_{12}$. 
According to standard procedure, after expanding this equation in e.~g.~$z_1$ and asking for its validity for every spectral parameter, we are left with two equations. One of them encodes the $\mathfrak{gl}(N|M)$ invariance of the R-matrix
\begin{equation}
\label{glNM}
\left[\mathbf{R}_{12}(z), J_1 + J_2\right]=0\,,
\end{equation}
while the other reads
\begin{equation}
\label{seceq}
\sum_\mathcal{B}\Big( 
(-1)^{\mathcal{B}} \mathbf{R}_{12}(z) \, J_{1 \mathcal{B}}^{\mathcal{A}} \, J_{2 \mathcal{C}}^{\mathcal{B}} - (-1)^{\mathcal{B}} \, J_{2 \mathcal{B}}^{\mathcal{A}} \, J_{1 \mathcal{C}}^{\mathcal{B}} \mathbf{R}_{12}(z) \Big)
= 
z \big(J_{2 \mathcal{C}}^{\mathcal{A}}\, \mathbf{R}_{12}(z)- \mathbf{R}_{12}(z)J_{2 \mathcal{C}}^{\mathcal{A}}\big) .
\end{equation}
If $\mathbf{R}_{12}(z)$ satisfies  \eqref{glNM} and \eqref{seceq} also $\rho(z)\,\mathbf{R}_{12}(z)$ does,
where $\rho(z)$ is any function of $z$. In order to further restrict this freedom one can impose the  
unitarity condition 
\begin{equation}\label{unitarity}
\mathbf{R}_{12}(z)\mathbf{R}_{21}(-z)=1\,.
\end{equation}
However, note that this still does not completely fix the normalization $\rho(z)$.

The remaining part of this section is devoted to finding the solution of equations \eqref{glNM} and \eqref{seceq} in the case when spaces labelled $\mathbf{1}$ and $\mathbf{2}$ in \eqref{YBe} 
correspond to oscillator realizations that we are now going to describe.
The idea of oscillator realizations apparently goes back to J.~Schwinger. For the ${\mathcal N}=4$ SYM case it was first introduced by \cite{Gunaydin:1984fk}. The bilinear combinations
\begin{equation}\label{oscreal}
J^{\mathcal{A}}_{\mathcal{B}}\,=\,\bar{\osca}^\mathcal{A}\, \osca_\mathcal{B}\,,
\end{equation} 
satisfy the  $\mathfrak{gl}(N|M)$  commutation relations
\begin{equation}\label{glNMcommrel}
[J^{\mathcal{A}}_{\mathcal{B}},J^{\mathcal{C}}_{\mathcal{D}}\}=\delta_{\mathcal{B}}^{\mathcal{C}} \,J^{\mathcal{A}}_{\mathcal{D}}-(-1)^{\mathcal{(A+B)(C+D)}}\,\delta^{\mathcal{A}}_{\mathcal{D}}\,J^{\mathcal{C}}_{\mathcal{B}}\,,
\end{equation}
provided that $\bar{\osca}^\mathcal{A}\,, \osca_\mathcal{B}$ satisfy the graded Heisenberg algebra
\begin{equation}
[\osca_\mathcal{A},\bar{\osca}^\mathcal{B}\}\,=\,\delta^{\mathcal{B}}_{\mathcal{A}}\,,
\qquad \qquad
[\osca_\mathcal{A},\osca_\mathcal{B}\}\,=\,0\,,
\qquad \qquad
[\bar{\osca}^\mathcal{A},\bar{\osca}^\mathcal{B}\}\,=\,0\,.
\end{equation} 
Notice that the element $\mathbf{C}\,:=\,\sfrac{1}{2}\,\sum_{\mathcal{C}}\,\bar{\osca}^\mathcal{C}\, \osca_\mathcal{C}$ is central, i.~e.~it commutes with all $\mathfrak{gl}(N|M)$ generators 
\eqref{oscreal}.
The next step is to choose a representation of the Heisenberg algebra.
Upon using Fock space representations, oscillator realizations are particularly convenient to describe 
unitary highest weight representations of  $\mathfrak{u}(n,\dot n|m+\dot m)$.
Let us rename the oscillators as
\begin{equation}\label{particlehole}
\left(\bar{\osca}^{\mathsf{A}}\,,\,\osca_{\mathsf{A}}\right)\,:=\,
\left(\bar{\osca}^{\mathsf{A}}\,,\,\osca_{\mathsf{A}}\right)\,,
\qquad \qquad
\left(\bar{\oscb}_{\dot{\mathsf{A}}}\,,\,\oscb^{\dot{\mathsf{A}}}\right)\,:=\,
\left(\osca_{\dot{\mathsf{A}}}\,,\,-(-1)^{\dot{\mathsf{A}}}\,\bar{\osca}^{\dot{\mathsf{A}}}\right)\,,
\end{equation}
where we split indices into two sets
\begin{equation}
 \mathsf{A},\,\mathsf{B},\dots\,\in\,\{1,\dots,n\,\big{|}\,N+1,\dots,\,N+m\}\,,
\qquad
 \dot{\mathsf{A}},\,\dot{\mathsf{B}},\dots\,\in\,\{n+1,\dots,N\,\big{|}\,N+m+1,\dots,\,N+M\}\,.
\end{equation}
The new oscillators  satisfy again the Heisenberg algebra.
The redefinition \eqref{particlehole} is known in the condensed matter literature as a particle-hole transformation. 
It also plays an important role in the context of scattering amplitudes.
After this transformation, the  $\mathfrak{gl}(N|M)$  generators \eqref{oscreal}  read
\renewcommand{\arraystretch}{1.5}
\begin{equation}\label{algebragen.noncom}
\begin{pmatrix}
 J^{\mathsf{A}}_\mathsf{B} & J^{\mathsf{A}}_{\dot{\mathsf{B}}}\\ \vspace{0.2cm}
J^{\dot{\mathsf{A}}}_\mathsf{B} & J^{\dot{\mathsf{A}}}_{\dot{\mathsf{B}}}
\end{pmatrix}\,=\,
\begin{pmatrix}
 \bar{\osca}^{\mathsf{A}}\,\osca_\mathsf{B} & \bar{\osca}^{\mathsf{A}}\,\bar{\oscb}_{\dot{\mathsf{B}}}\\
-(-1)^{\dot{\mathsf{A}}}\,\oscb^{\dot{\mathsf{A}}}\,\osca_\mathsf{B} & -(-1)^{\dot{\mathsf{A}}(1+\dot{\mathsf{B}})}\,\bar{\oscb}_{\dot{\mathsf{B}}}\,\oscb^{\dot{\mathsf{A}}}
-(-1)^{\dot{\mathsf{A}}}\,\delta^{\dot{\mathsf{A}}}_{\dot{\mathsf{B}}}
\end{pmatrix}\,,
\end{equation}
\renewcommand{\arraystretch}{1}
and 
\begin{equation}
 \label{centrC}
\mathbf{C}\,=\,\frac{1}{2}\,\left(
\bar{\osca}^{\mathsf{A}}\osca_{\mathsf{A}}-\bar{\oscb}_{\dot{\mathsf{A}}}\oscb^{\dot{\mathsf{A}}}-\dot{n}+\dot{m}\right)\,.
\end{equation}
Next we introduce a Fock vacuum 
\begin{equation}
\osca_{\mathsf{A}}\,|0\rangle\,=\,0\,,
\qquad
\oscb^{\dot{\mathsf{A}}}\,|0\rangle\,=\,0\,.
\end{equation}
The Fock space $\mathcal{F}$ is generated by the repeated action of the creation operators $\bar{\osca}^{\mathsf{A}}, \bar{\oscb}_{\dot{\mathsf{A}}}$ on the Fock vacuum.
An important feature of such representations is that they are not irreducible.
Irreducible representations are labelled by eigenvalues of $\mathbf{C}$ denoted by $\frac{s}{2}$ in the following.
Accordingly, the  Fock space decomposes as 
\begin{equation}\label{decomposesas}
 \mathcal{F}\,=\,\bigoplus_s\,\mathcal{V}_s\,,
\qquad \qquad 
 \mathcal{V}_s\,:=\,\Big\{\prod_{i=1}^{n_\osca}\prod_{j=1}^{n_{\oscb}}\bar{\osca}^{\mathsf{A}_i}\bar{\oscb}_{\dot{\mathsf{A}}_i}|0\rangle:n_\osca-n_\oscb-\dot n+\dot m=s\Big\}\,.
\end{equation}
Notice that each $\mathcal{V}_s$ is infinite dimensional as long as both $n$ and $\dot n$ are different from zero.
In this case $\mathcal{V}_s$ coincides with  the totally symmetric\footnote{Here symmetrization of indices is understood as symmetrization over bosonic indices and antisymmetrization over fermionic ones.} representations of $\mathfrak{u}(N|M)$.
A second important feature of these representations is that the tensor product $\mathcal{V}_{s_1}\otimes \mathcal{V}_{s_2}$ is multiplicity free.

We conclude this part with some remarks on the specific case of $\mathfrak{psu}(2,2|4)$, which is 
relevant for $\mathcal{N}=4$ SYM. It is obtained from $\mathfrak{u}(2,2|4)$ by removing two generators.
The first, called central charge, commutes with all the generators of $\mathfrak{u}(2,2|4)$ and appears
on the right hand side of the commutation relations. 
In the case of oscillator realizations this generator coincides with the central element $\mathbf{C}$
defined in \eqref{centrC}. 
The second, referred to as hypercharge, acts as an automorphism of the algebra.
It never appears on the right hand side of the commutation relation. For this reason there exist $\mathfrak{psu}(2,2|4)$
invariants that are not invariant under the action of the hypercharge.

In the context of scattering amplitudes the generators of the Heisenberg algebra can be realized as differential 
operators as follows
\begin{equation}\label{carlosconvention}
 \bar{\osca}^{\mathsf{A}}\,\sim\,\lambda^{\alpha}\qquad
 \bar{\oscb}_{\dot{\mathsf{A}}}\,\sim\,\left(\,\tilde{\lambda}^{\dot\alpha}\,,\,\,\eta^A\,\right)\qquad
 \osca_{\mathsf{A}}\,\sim\,\frac{\partial}{\partial\,\lambda^{\alpha}}\qquad
 \oscb^{\dot{\mathsf{A}}}\,\sim\,\left(\,\frac{\partial}{\partial\,\tilde{\lambda}^{\dot\alpha}}\,,\,\frac{\partial}{\partial\,\eta^A}\,\right)\,,
\end{equation}
acting on the super-helicity spinor space. 
See section \ref{Sec:ch4_Symmetries} for more details.
This superspace is chiral as the fermionic coordinates $\eta$ pair up with $\tilde{\lambda}$ 
rather than $\lambda$.  
Functions, or more generally distributions, on $p$ copies of this space can be
 characterized by eigenvalues of $\mathbf{C}_i$, see \eqref{centrC}, as
\begin{equation}
 f(\{u_i\,\lambda_i,\,u_i^{-1}\,\tilde{\lambda}_i,\,u_i^{-1}\,\eta_i\})\,=\,
 u_1^{-2\,h_1}\dots\,u_n^{-2\,h_p}\,f(\{\lambda_i,\,\tilde{\lambda}_i,\,\eta_i\})\,,
\qquad \qquad
u_i \in \mathbb{C}^{*}\,.
\end{equation}
One calls  the numbers $h_i$ super-helicities. They are connected to the eigenvalues of  $\mathbf{C}_i$ as
$c_i=\frac{s_i}{2}=1-h_i$.
For scattering amplitudes in $\mathcal{N}=4$ SYM the central charge vanishes for each particle, in other words $h_i=1$. In the present work this condition will be relaxed.
The hypercharge $\mathbf{B}:=\sfrac{1}{2}\sum_i\eta_i\sfrac{\partial}{\partial\eta_i}$
measures   the deviation of the super-amplitudes from being maximally helicity violating (MHV).
More precisely $B=4+2k$ for N$^k$MHV super-amplitudes.


\subsection{Harmonic Action Form of the R-Matrix and the  \texorpdfstring{$\mathcal{N}=4$ Spin Chain}{}}
\label{Sec:ch1_HarmonicAction}
To illustrate the use of the Yang-Baxter equation, we present a small intermezzo somewhat outside the main scope of this paper, and directly derive the ``harmonic action'' form of the $\mathcal{N}=4$ SYM one-loop Hamiltonian introduced in \cite{Beisert:2003jj}. It shows its great usefulness for finding the spectrum of the spin chain that emerges in the spectral problem of $\mathcal{N}=4$ SYM. Corresponding to an integrable nearest neighbor Hamiltonian density, one should be able to obtain its form by taking a logarithmic derivative of a suitable R-matrix, see \cite{Faddeev:1996iy}. In this section we then rederive this well-known R-matrix by solving the Yang-Baxter equation \eqref{YBe}, finding again \eqref{LeningradR}, but expressed directly in the oscillator basis.

In order to solve the Yang-Baxter equation we introduce a family of operators which acts on the tensor product of two Fock spaces $\mathcal{F}$. Such operators are manifestly invariant under the
 $\mathfrak{gl}(n|m)\oplus \mathfrak{gl}(\dot{n}|\dot{m})$ subalgebra of $\mathfrak{gl}(N|M)$
and are given by\footnote{This formalism has been developed in joint discussions with Rouven Frassek. See \cite{NRYMtoappear}, where these hopping operators are also employed.} 
\begin{align}\label{Hopdef}
&\mathbf{Hop}_{k,l,m,n}\,=\,
:\,
\frac{(\bar{\osca}_2\,\osca^1)^{k}}{k!}\,
\frac{(\bar{\oscb}^2\,\oscb_1)^l}{l!}\,
\frac{(\bar{\osca}_1\,\osca^2)^m}{m!}\,
\frac{(\bar{\oscb}^1\,\oscb_2)^n}{n!}
\,:\,\,=
\\
=
\frac{1}{k!\,l!\,m!\,n!}\,
&\bar{\osca}_2^{\mathsf{A}_1}\ldots \bar{\osca}_2^{\mathsf{A}_k}\,
\bar{\oscb}^2_{\dot{\mathsf{A}}_1}\ldots \bar{\oscb}^2_{\dot{\mathsf{A}}_l}\,
\bar{\osca}_1^{\mathsf{B}_1}\ldots\bar{\osca}_1^{\mathsf{B}_m}\,
\bar{\oscb}^1_{\dot{\mathsf{B}}_1}\dots \bar{\oscb}^1_{\dot{\mathsf{B}}_n}\,
\oscb_2^{\dot{\mathsf{B}}_n}\ldots \oscb_2^{\dot{\mathsf{B}}_1}\,
\osca^2_{\mathsf{B}_m}\ldots \osca^2_{\mathsf{B}_1}\,
\oscb_1^{\dot{\mathsf{A}}_l}\ldots \oscb_1^{\dot{\mathsf{A}}_1}\,
\osca^1_{\mathsf{A}_k}\ldots \osca^1_{\mathsf{A}_1}\nonumber
\end{align}
where $:(\dots):$ denotes normal ordering
and the indices $1,2$ refer to the two copies of the Fock space.
It is easy to check that the action of these operators boils down to moving $k+l$ oscillators from the first side to the second, and $m+n$ oscillators in opposite direction\footnote{To be more precise a chiral half of $\mathbf{Hop}$ acts as follows:
\begin{equation}
\mathbf{hop}_{k,m}\,\bar{\osca}_{p_1}^{\mathsf{A}_1}\dots \bar{\osca}_{p_{n_a}}^{\mathsf{A}_{n_a}}\,|0,0\rangle\,=\,
\sum_{\{p'_1,\dots, p'_{n_a}\}_{\star}}\,
\,\bar{\osca}_{p'_1}^{\mathsf{A}_1}\dots \bar{\osca}_{p'_{n_a}}^{\mathsf{A}_{n_a}}\,|0,0\rangle\,,
\qquad 
\mathbf{hop}_{k,m}\,=\,:\,
\frac{(\bar{\osca}_2\,\osca^1)^{k}}{k!}\,
\frac{(\bar{\osca}_1\,\osca^2)^m}{m!}\,:\,
\end{equation}
where $p\in\{1,2\}$ and the sum runs over all $\{p'_1,\dots, p'_{n_a}\}$ that differ from 
$\{p_1,\dots, p_{n_a}\}$ by changing exactly $k$ indices from $p_i=1$ to $p'_i=2$ and $m$ indices from $p_i=2$ to $p_i'=1$.}.
For this reason we refer to such operators as ``hopping operators''.

Any $\mathfrak{gl}(N|M)$ invariant operator on $\mathcal{F}\otimes \mathcal{F}$
can be expressed as a linear combination of hopping operators.
The coefficients  of this expansion are  functions of $\mathbf{C}_1$, $\mathbf{C}_2$, see \eqref{centrC}, and of $\mathbf{N}$, which is the total number operator. Invariance under $\mathfrak{gl}(N|M)$,
in particular under the off diagonal generators in \eqref{algebragen.noncom},
further restricts these coefficients. 
In general, we do not have to assume that the representations are identical in both spaces, i.~e.~that
the eigenvalues of $\mathbf{C}_1$ and  $\mathbf{C}_2$, denoted by  $\sfrac{s_1}{2}$ and $\sfrac{s_2}{2}$, are the same.
Nevertheless, in this section we focus our attention  only on the case $s_1=s_2$. In particular, only in that case the R-matrix 
we construct possesses the so-called regularity properties, namely $\mathbf{R}_{12}(0)=\mathbf{P}_{12}$ with $\mathbf{P}_{12}$ being the graded permutation of two spaces. This allows to extract the  Hamiltonian density $\mathbf{H}_{12}$ as 
$\mathbf{R}_{12}(z)=\mathbf{P}_{12}(1+z\,\mathbf{H}_{12}+\mathcal{O}(z^2))$. 

As pointed out in the previous section, the Yang-Baxter equation encodes two equations.
The first, expressing $\mathfrak{gl}(M|N)$ invariance of the R-matrix \eqref{glNM}, is solved by expanding
the R-matrix  in the hopping basis as
\begin{equation}\label{Rhop}
\mathbf{R}_{12}(z)=\sum_{k,l,m,n}\,\alpha_{k,l,m,n}^{(\mathbf{N})}\,\mathbf{Hop}_{k,l,m,n}\,
\end{equation}
and imposing, with $I:=\sfrac{k+l+n+m}{2}$, 
\begin{equation}\label{alphafromsl}
 \alpha_{k,l,m,n}^{(\mathbf{N})}\,=\,\delta_{k+n,l+m}\,(-1)^{(k+l)(m+n)}\,\alpha_{\sfrac{k+l+n+m}{2}}^{(\mathbf{N})}\,,\qquad \qquad
\alpha_{I}^{(\mathbf{N+2})}\,+\,\alpha_{I+1}^{(\mathbf{N+2})}\,=\,\alpha_{I}^{(\mathbf{N})}\,.
\end{equation}
Notice that $\mathbf{N}$ is always even in the present situation.
The fact that \eqref{alphafromsl} ensures invariance of the R-matrix can be checked by direct calculations similar to the one in the  appendix of \cite{Beisert:2003jj}.

The second equation takes the form \eqref{seceq}, where the $\mathfrak{gl}(N|M)$ generators are taken as in 
\eqref{algebragen.noncom}. As \eqref{seceq} is $\mathfrak{gl}(N|M)$ covariant and the R-matrix is 
$\mathfrak{gl}(N|M)$ invariant, it is sufficient to solve for a single component of \eqref{seceq}.
Without loss of generality we restrict the open indices in \eqref{seceq} to correspond to the diagonal blocks in 
\eqref{algebragen.noncom}.
After normal ordering the oscillators appearing in the resulting expression one finds that 
\eqref{seceq} is satisfied if and only if 
\begin{equation}\label{eq2foralpha}
\frac{\alpha^{(\mathbf{N})}_I}{\alpha^{(\mathbf{N})}_{I-1}}\,=\,-\,\frac{I}{I-\sfrac{1}{2}\,\mathbf{N}-z}\,.
\end{equation}
The general solution to this recursion relation, in conjunction with \eqref{alphafromsl}, is given by
\begin{equation}
\alpha_{I}^{(\mathbf{N})}=\rho(z)
\frac{(-1)^{I+\sfrac{1}{2}\,\mathbf{N}}\,\Gamma(I+1)}
{\Gamma(I+1-z-\sfrac{1}{2}\,\mathbf{N})\Gamma(z+\sfrac{1}{2}\,\mathbf{N}+1)}\,,
\end{equation}
where $\rho(z)$ is any function of the spectral parameter.
We further  require that the R-matrix satisfies the unitarity relation \eqref{unitarity}.
A solution is given by
\begin{equation}
\rho(z)\,=\,\Gamma(1+z)\Gamma(1-z).
\end{equation}
The final form of the R-matrix is 
\begin{equation}\label{RmatrixHopfinalform}
 \mathbf{R}_{12}(z)\,=\,
\sum_{I=0}^{\infty}\,
\frac{(-1)^{I+\sfrac{1}{2}\,\mathbf{N}}\,\Gamma(1+z)\Gamma(1-z)\,\Gamma(I+1)}
{\Gamma(I+1-z-\sfrac{1}{2}\,\mathbf{N})\Gamma(z+\sfrac{1}{2}\,\mathbf{N}+1)}\,
\mathbf{Hop}_I\,,
\end{equation}
where
\begin{equation}
 \mathbf{Hop}_I\,:=\,
\sum_{k,\,l,\,m,\,n}\,(-1)^{(k+l)(m+n)}\,
\delta_{k+n,I}\,\delta_{m+l,I}\,
        \mathbf{Hop}_{k,l,m,n}\,.
\end{equation}
One can easily check that the constructed R-matrix possesses the regularity property 
\begin{equation}
\mathbf{R}_{12}(0)=\mathbf{Hop}_{n_{\osca_1},n_{\oscb_1},n_{\osca_2},n_{\oscb_2}}=\mathbf{P}_{12}\,,
\end{equation} 
where $\mathbf{P}_{12}$ is the graded permutation acting on the tensor product of the two spaces. 

The integrable Hamiltonian density is then obtained as usually by evaluating the logarithmic derivative of 
the R-matrix at $z=0$. Applying this recipe to \eqref{RmatrixHopfinalform} we find 
\begin{equation}
\label{HopH}
\mathbf{H}_{12}\,=\,-h(\sfrac{1}{2}\mathbf{N})\,+\,
\sum_{I=1}^{\infty}\,\,\frac{\Gamma(I)\,\Gamma(1+\sfrac{1}{2}\mathbf{N}-I)}{\Gamma(1+\sfrac{1}{2}\mathbf{N})}\,\,\mathbf{Hop}_I\,,
\end{equation}
where $h(j)$ are harmonic numbers defined as  $h(j)=\sum_{k=1}^j\,\sfrac{1}{k}$.
We may now compare \eqref{HopH} to the result in \cite{Beisert:2003jj}. 
Let us notice that the action of any $\mathbf{Hop}_{k,l,m,n}$ changes the oscillator numbers as
\begin{equation}
(n_{\osca_1},n_{\oscb_1},n_{\osca_2},n_{\oscb_2})
\stackrel{\mathbf{Hop}_{k,l,m,n}}{\longrightarrow}
(n_{\osca_1}-k+m,n_{\oscb_1}-l+n,n_{\osca_2}-m+k,n_{\oscb_2}-n+l).
\end{equation} 
Then the labels of representations are equal to $s_1=n_{\osca_1}-n_{\oscb_1}-\dot n+\dot m$ and
 $s_2=n_{\osca_2}-n_{\oscb_2}-\dot n+\dot m$ before and after the action of $\mathbf{Hop}$, provided that $k+n=m+l$. This is analogous to the central charge condition in the harmonic action formula. Thus our result reproduces the harmonic action form of the Hamiltonian, up to an overall minus sign.


\subsection{Gra\ss mannian Form of the R-Matrix}
\label{Sec:ch1_RmatrixGrass}

After this successful warm-up, let us return to the main focus of our paper. As announced at the beginning of this section, the Yang-Baxter equation can also be solved using yet another formalism, namely the Gra\ss mannian approach. 
It conveniently establishes a direct relation between integrability and the scattering amplitudes of $\mathcal{N}=4$ SYM.
Indeed, the kernel of the resulting R-matrix turns out to be a deformation of the tree-level four-point MHV amplitude.
A first hint at such a connection was discovered in  \cite{Zwiebel:2011bx}, where the complete one-loop dilatation operator of the $\mathcal{N}=4$ model was related to its tree-level scattering amplitudes.
In particular, the one-loop Hamiltonian of the nearest-neighbor spin chain encoding the spectral problem was shown to be technically related to the tree-level four-point amplitude. 
In order to deepen and precise this relation, we will use a formalism for the tree-level scattering amplitudes first proposed in \cite{ArkaniHamed:2009dn}, where
the leading singularities of the $\NN=4$ SYM $\mathrm{N^{k-2}MHV}$ $n$-point amplitudes were described by a Gra\ss mannian integral in twistor space.
Since then, the formalism has been considerably refined, resulting in a reformulation of tree- and loop-level amplitudes in terms of Gra\ss mannian integrals and rather powerful
on-shell diagrammatic techniques \cite{ArkaniHamed:2012nw}.
Let us briefly review this approach. The general, formal Gra\ss mannian integral relevant to scattering amplitudes, after fixing a certain GL$(k)$ symmetry \cite{ArkaniHamed:2009dn}, reads
\be
\mathcal{G}_{n,k} = \int \frac{\prod_{a=1}^{k} \prod_{i=k+1}^{n} dc_{ai}}{(1\ldots k)(2\ldots k+1)\ldots(n \ldots n+k-1)} \,
\prod_{a=1}^k \delta^{4|4}\left( \mathcal{Z}^{\mathcal{A}}_a - \sum_{i=k+1}^n c_{ai} \mathcal{Z}^{\mathcal{A}}_{i}\right),
\label{ACCK}
\ee
where $\mathcal{Z}_i^{\mathcal{A}}$ are the super-twistor variables
$\mathcal{Z}_i^{\mathcal{A}}=(\tilde \mu^{\alpha}_{i},{\tilde \lambda}^{\dot\alpha}_{i}, \eta^{A}_{i})$, with $\tilde \mu^{\alpha}_{i}$
 the Fourier conjugate to $\lambda_{i}^{\alpha}$. The integration is over an (unspecified) set of contours for the, in general complex, integration variables $c_{ai}$.
The integrals $\mathcal{G}_{n,k}$ associated to a given amplitude are labeled by the two numbers $n$ and $k$. Here $n$ is the total number of external particles, and $n-2 k$ the total helicity of the amplitude. The parameters $c_{ai}$ are the non-trivial entries of a $k \times n$ matrix 
\beq\label{C-matrix}
C=\left(
\begin{array}{ccc|cccc}
{} & {} & {}  & -c_{1,k+1} & -c_{1,k+2} & \cdots  & -c_{1,n} \\
{} & \mathds{1}_{k \times k}  & {} & \vdots  & \vdots & \ddots & \vdots \\
{} & {} & {}  &  -c_{k,k+1} & -c_{k,k+2} & \cdots & -c_{k,n} 
\end{array}
\right), 
\ee
where the mentioned GL$(k)$ symmetry allows to fix $k^2$ of the $k n$ parameters of $C$, corresponding to the trivial submatrix $\mathds{1}_{k \times k}$. 
The denominator consists of the cyclic product of the minors  $\MM_i=(i\, i+1...i+k-1)$ (= $k \times k$ sub determinants) of the rectangular matrix $C$. 
The demonstration that this object is Yangian invariant \cite{Drummond:2010qh} was an important first hint at its connection to integrability, a link to be significantly strengthened in the following. 

 This requires considering a generalization of the Gra\ss mannian integral \eqref{ACCK}, as we will again be looking for 
appropriate solutions to the Yang-Baxter equation. 
To make contact with the above description, we simply relate the coordinates $\mathcal{Z}^{\mathcal{A}}$, which for the special case of $\mathcal{N}=4$ SYM are just the super-twistors defined below \eqref{ACCK}, as well as their derivatives $\frac{\partial}{\partial \mathcal{Z}^{\mathcal{A}}}$ to the Schwinger oscillators via\footnote{With respect to the convention in \eqref{carlosconvention} there is an additional particle-hole transformation on all 4+4 oscillators, which may be implemented by an 8-fold Fourier transform.}
\begin{equation}\label{twistorosc}
\bar \osca^{\mathcal{A}}\leftrightarrow \mathcal{Z}^{\mathcal{A}}\,,\qquad \osca_{\mathcal{A}}\leftrightarrow \frac{\partial}{\partial \mathcal{Z}^{\mathcal{A}}} \,.
\end{equation}
The  $\mathfrak{gl}(N|M)$ generators then become, as e.g.~in \cite{Witten:2003nn},
\begin{equation}
\label{generators}
J_{i\,\mathcal{B}}^{\mathcal{A}}=\mathcal{Z}_{i}^{\mathcal{A}}\frac{\partial}{\partial\mathcal{Z}_i^{\mathcal{B}}} \,.
\end{equation}
For the moment, the approach is purely algebraic and formal, and we avoid the issues of the appropriate reality conditions for \eqref{twistorosc} and of the inner product of the states built from these variables.
We would now like to show that a modified version of the Gra\ss mannian integral \eqref{ACCK} is the formal integral kernel
of the R-matrix solution to the Yang-Baxter equation \eqref{YBe}.
This  integral kernel $\mathcal{R}$ of the R-matrix $\mathbf{R}_{12}(z)$ is defined by its action $\circ$ on an arbitrary function $g$ depending on two  variables
\beq\label{rkernel}
(\mathbf{R}_{12}(z) \circ g)(\mathcal{Z}_3,\mathcal{Z}_4) := \int \, d^{N|M}\mathcal{Z}_1 d^{N|M}\mathcal{Z}_2 \, 
\RR(z;\mathcal{Z}_3,\mathcal{Z}_4|\mathcal{Z}_1,\mathcal{Z}_2)\, g(\mathcal{Z}_1,\mathcal{Z}_2) \,,
\eeq
where the integration domain, clearly related to the reality conditions and the inner product, is left unspecified for the moment. Note that the kernel $\mathcal{R}$, in contradistinction to the operator $\mathbf{R}_{12}$, now depends on five variables, namely in addition to the spectral parameter $z$ also on $\mathcal{Z}_1,\mathcal{Z}_2$ and $\mathcal{Z}_3,\mathcal{Z}_4$, which we associate, respectively, to the two incoming and two outgoing particles. 

As there are many different solutions to the Yang-Baxter equation, depending on which representations of $\mathfrak{gl}(N|M)$ we study, we need to specify the labels of representations for all four particles in order to get a particular (and ideally unique in its matrix structure) solution. For the oscillator realization discussed earlier this means that we have to specify one number $s_i$, which is the total number of oscillators corresponding to the $i$-th particle. This translates, via \eqref{twistorosc}, to a constant degree of homogeneity $s_i$ in the  variables $\mathcal{Z}_i$. 
These representation labels $s_i$ then serve as parameters for the R-matrix kernel. 
Taking into account the correct degrees of homogeneity, \eqref{generators} allows to turn the two equations (\ref{glNM}) and (\ref{seceq}) derived from the expansion of the Yang-Baxter equation into easily solvable differential equations for the kernel of the R-matrix.

Let us start from the $\mathfrak{gl}(N|M)$ invariance of $\mathbf{R}_{12}(z)$. Translating \eqref{glNM} via \eqref{generators},\eqref{rkernel} to $\RR(z)$, one finds, after abbreviating $\RR(z):= \RR(z;\mathcal{Z}_3,\mathcal{Z}_4|\mathcal{Z}_1,\mathcal{Z}_2)$,   
\beq\label{invariance0}
\int \, d^{N|M}\mathcal{Z}_1 d^{N|M}\mathcal{Z}_2 \, 
\left\{ 
\RR(z) \left(\mathcal{Z}^{\mathcal{A}}_1 \frac{\partial}{\partial \mathcal{Z}_1^{\mathcal{C}}}+ 
\mathcal{Z}^{\mathcal{A}}_2 \frac{\partial}{\partial \mathcal{Z}_2^{\mathcal{C}}}\right) -
\left(\mathcal{Z}^{\mathcal{A}}_3 \frac{\partial}{\partial \mathcal{Z}_3^{\mathcal{C}}}+
\mathcal{Z}^{\mathcal{A}}_4 \frac{\partial}{\partial \mathcal{Z}_4^{\mathcal{C}}}\right) \RR(z)
\right\}\, 
g(\mathcal{Z}_1,\mathcal{Z}_2)=0 \,.
\eeq
In the interest of a more compact notation, we write this in the succinct symbolic form
\begin{equation}\label{invariance1}
\RR(z) \left(\mathcal{Z}^{\mathcal{A}}_1 \frac{\partial}{\partial \mathcal{Z}_1^{\mathcal{C}}}+ 
\mathcal{Z}^{\mathcal{A}}_2 \frac{\partial}{\partial \mathcal{Z}_2^{\mathcal{C}}}\right) -
\left(\mathcal{Z}^{\mathcal{A}}_3 \frac{\partial}{\partial \mathcal{Z}_3^{\mathcal{C}}}+
\mathcal{Z}^{\mathcal{A}}_4 \frac{\partial}{\partial \mathcal{Z}_4^{\mathcal{C}}}\right) \RR(z) = 0  \,,
\end{equation}
i.e.~the integration over the kernel as well as its action on an arbitrary function $g$ is understood. A similar and hopefully obvious short-hand notation will be used in many of the following equations in this section.

As we will do in all our calculations, we want to rewrite \eqref{invariance0} (and thus its short form \eqref{invariance1}) as a differential equation for the kernel $\RR(z)$. We accomplish it by integrating the first two terms in \eqref{invariance0} by parts, and find
\begin{equation}
\label{gl_inv}
\sum_{i=1}^4 \mathcal{Z}^{\mathcal{A}}_i \frac{\partial}{\partial \mathcal{Z}_i^{\mathcal{C}}} \, \RR(z) = -2 (-1)^{\mathcal{A}}\, \delta^{\mathcal{A}}_{\mathcal{C}} \, \RR(z) \,,
\end{equation}
where $(-1)^{\mathcal{A}}$ encodes grading as in section \ref{Sec:ch1_intro}.
Any time we perform an integration by parts, we require that the boundary terms vanish.
Inspired by the Gra\ss mannian formulation and by the results of \cite{Drummond:2010uq,Korchemsky:2010ut}, 
we find that the most general formal solution to equation \eqref{gl_inv}  can be given as a formal Gra\ss mannian integral on a multi-contour of the form
\begin{equation}
\label{kernel1}
\RR(z) = \int \prod_{a=1,2}\prod_{i=3,4} dc_{ai} \, F(C) \,
\delta^{N|M}(\mathcal{Z}_1 - \sum_{j=3}^4 c_{1j} \mathcal{Z}_{j}) \, \delta^{N|M}(\mathcal{Z}_2 -  \sum_{k=3}^4 c_{2k} \mathcal{Z}_{k}) \,,
\end{equation}
where $F(C)\equiv F(c_{13}, c_{14}, c_{23}, c_{24})$ is a generic function depending on the four complex parameters $c_{13}, c_{14}, c_{23}, c_{24}$. 
For sake of simplicity we denote
\begin{equation}\label{simplicity}
\delta_1 := \delta^{N|M}(\mathcal{Z}_1 - c_{13} \mathcal{Z}_3 - c_{14} \mathcal{Z}_4), \qquad \delta_2 :=
\delta^{N|M}(\mathcal{Z}_2 - c_{23} \mathcal{Z}_3 - c_{24} \mathcal{Z}_4) \,.
\end{equation}
We now show that the function $F(C)$ is almost uniquely determined by imposing the known homogeneity properties as well as the Yang-Baxter equation. Homogeneity in the $\mathcal{Z}_i$ immediately fixes $F(C)$ to the form
\beq
\label{ansatz_F}
F(C) = c_{13}^{\alpha} c_{14}^{\beta} c_{23}^{\gamma} c_{24}^{\delta}\,  \tilde f\left(\frac{c_{13} c_{24}}{c_{14} c_{23}}\right),
\eeq
where now $\tilde f$ is a function of just one cross-ratio instead of four variables. Clearly one of the four exponents, say $\delta$, is arbitrary, since we can rewrite this as
\beq
\label{ansatz_Fmod}
F(C) = c_{13}^{\alpha-\delta} c_{14}^{\beta+\delta} c_{23}^{\gamma+\delta} \left(\frac{c_{13} c_{24}}{c_{14} c_{23}}\right)^{\delta}  \tilde f\left(\frac{c_{13} c_{24}}{c_{14} c_{23}}\right)
=:c_{13}^{\alpha-\delta} c_{14}^{\beta+\delta} c_{23}^{\gamma+\delta}  f\left(\frac{c_{13} c_{24}}{c_{14} c_{23}}\right),
\eeq
where $f$ is again just a function of the cross-ratio. The exponents $\alpha, \beta, \gamma$ and $\delta$ are related to the precise degrees of homogeneity $s_i$.  
These can be determined by the following four equations, differing in form for incoming and outgoing legs, respectively:
\begin{equation}\label{scaling}
\mathcal{R}(z) \, \mathcal{Z}^{\mathcal{A}}_i \frac{\partial}{\partial \mathcal{Z}_i^{\mathcal{A}}} = s_i \, \mathcal{R}(z),\, i=1,2 \,, \qquad
\mathcal{Z}^{\mathcal{A}}_i \frac{\partial}{\partial \mathcal{Z}_i^{\mathcal{A}}} \, \mathcal{R}(z)  = s_i \, \mathcal{R}(z), \, i=3,4 \,.
\end{equation}
Let us present  the explicit calculation for the case $i=1$. Expressing $\mathcal{R}(z)$ in \eqref{scaling} by \eqref{kernel1},\eqref{simplicity} we have
\beq\label{firststep}
 \int \prod_{a=1,2}\prod_{i=3,4} dc_{ai} \, F(C) \, \delta_1 \, \delta_2 \, \mathcal{Z}^{\mathcal{A}}_1 \frac{\partial}{\partial \mathcal{Z}_1^{\mathcal{A}}} 
= s_1   \int \prod_{a=1,2}\prod_{i=3,4} dc_{ai} \, F(C) \, \delta_1 \delta_2 \,.
\eeq
Integrating the l.h.s.~by parts w.r.t.~$\mathcal{Z}_1^{\mathcal{A}}$, the differential operator 
$\mathcal{Z}^{\mathcal{A}}_1 \frac{\partial}{\partial \mathcal{Z}_1^{\mathcal{A}}}$
acts on the delta function $\delta_1$. 
The action on the  variables $\mathcal{Z}^{\mathcal{A}}_1$ can now be exchanged for an action on the complex parameters $c_{ai}$ \cite{Drummond:2010qh}. After a short calculation we obtain from \eqref{firststep}
\beq
 \int \prod_{a=1,2}\prod_{i=3,4} dc_{ai} \, F(C) \left(c_{13} \frac{\partial}{\partial c_{13}}  +  c_{14} \frac{\partial}{\partial c_{14}} \right) \delta_1 \delta_2 
= s_1  \int \prod_{a=1,2}\prod_{i=3,4} dc_{ai} \, F(C) \, \delta_1 \delta_2\,.
\eeq
After integrating the l.h.s.~by parts once more, this time w.r.t.~$c_{13}$ and $c_{14}$, such that the differential operator 
$\left(c_{13} \frac{\partial}{\partial c_{13}}  +  c_{14} \frac{\partial}{\partial c_{14}} \right)$
acts on the function $F(C)$, and using \eqref{ansatz_F}, we finally find
\beq\label{lin1}
\alpha + \beta +2 = -s_1 \,.
\eeq
We can proceed analogously for the other scaling operators, i.e.~the cases $i=2,3,4$, which provide us with the three additional equations 
\beq\label{lin234}
\gamma + \delta +2 = -s_2 \,, \qquad \alpha + \gamma +2 = -s_3 \,, \qquad \beta + \delta +2 = -s_4 \,.
\eeq
Interestingly, the four equations in \eqref{lin1},\eqref{lin234} do not have a solution for $\alpha, \beta, \gamma,\delta$ for generic $s_1,s_2,s_3,s_4$. However, in view of \eqref{ansatz_Fmod}, the three equations in \eqref{lin234} are easily solved as $\alpha-\delta=s_2-s_3$, $\beta+\delta=-s_4-2$, and $\gamma+\delta=-s_2-2$. This yields
\beq
\label{Ftilde}
F(C) = c_{13}^{s_2-s_3} \, c_{14}^{-s_4 - 2} \, c_{23}^{-s_2 - 2}  f\left(\frac{c_{13} c_{24}}{c_{14} c_{23}}\right),
\eeq
while substitution of these values into \eqref{lin1} leads to the following constraint on the representation labels
\beq\label{in=out}
s_1+s_2=s_3+s_4\,.
\eeq 

We are left with finding the function $f$ of the cross-ratio in \eqref{Ftilde}. It may be fixed, up to an undetermined multiplicative function of the spectral parameter $z$, by using \eqref{seceq}, which, as we may recall, was a direct consequence of the Yang-Baxter equation. Using the same techniques we just employed, we may derive a further equation for the kernel.
After repeated integration by parts, and use of the commutation relations to rewrite the second-order operators in such a way that operators with contracted indices act on $\RR(z)$ first,
one arrives at
\begin{equation}
\left( \mathcal{Z}_1^{\mathcal{A}} \frac{\partial}{\partial \mathcal{Z}_2^{\mathcal{C}}} \mathcal{Z}^{\mathcal{B}}_2   \frac{\partial}{\partial \mathcal{Z}_1^{\mathcal{B}}} 
-\mathcal{Z}^{\mathcal{A}}_4 \frac{\partial}{\partial \mathcal{Z}_3^{\mathcal{C}}} \mathcal{Z}^{\mathcal{B}}_3 \frac{\partial}{\partial \mathcal{Z}_4^{\mathcal{B}}}
\right) \RR(z) + (1-z) \left(\mathcal{Z}^{\mathcal{A}}_4 \frac{\partial}{\partial \mathcal{Z}_4^{\mathcal{C}}}  + \mathcal{Z}^{\mathcal{A}}_2 
\frac{\partial}{\partial \mathcal{Z}_2^{\mathcal{C}}} + (-1)^{\mathcal{C}} \delta^{\mathcal{A}}_{\mathcal{C}} \, \right)\, \RR(z) = 0 \, .
\end{equation}
Proceeding as before, after inserting the expression \eqref{kernel1} for the kernel, the contracted differential operators in the variables $\mathcal{Z}^\mathcal{A}$ can again be exchanged 
for differential operators in the variables $c_{ai}$ when acting on the delta functions:
\begin{equation}
\label{trade}
\mathcal{Z}^{\mathcal{B}}_2   \frac{\partial}{\partial \mathcal{Z}_1^{\mathcal{B}}}\delta_1\delta_2 \rightarrow - \sum_{i=3}^4 c_{2i}  \frac{\partial}{\partial c_{1i}} \delta_1\delta_2\, ; 
\qquad \mathcal{Z}^{\mathcal{B}}_3   \frac{\partial}{\partial \mathcal{Z}_4^{\mathcal{B}}}\delta_1\delta_2 \rightarrow \sum_{a=1}^2 c_{a4}  \frac{\partial}{\partial c_{a3}} \delta_1\delta_2\, .
\end{equation}
After integration by parts in the variables $c_{ai}$ one easily arrives at an equation of the type 
\begin{equation}
\label{independence}
\left(\ldots\right)^\mathcal{A} \, (\partial_{\mathcal{C}}\delta_1) \, \delta_2 + \left(\ldots\right)^\mathcal{A} \, \delta_1 (\partial_{\mathcal{C}}\delta_2) = 0 \,,
\end{equation}
where we should recall the condensed notation \eqref{simplicity}.
Since the two terms in \eqref{independence} are linearly independent, the two summands have to vanish separately, 
implying two equations. Let us write out the first one in more detail. It reads
\begin{equation}
\int \, \prod_{a=1,2}\prod_{i=3,4} dc_{ai}  \, \left(c_{13} \, \sum_{b=1}^2 c_{b4}  \frac{\partial}{\partial c_{b3}} F(C)  + (1-z) \, c_{14} \, F(C) \right) 
\, \mathcal{Z}^{\mathcal{A}}_4 \, \left(\frac{\partial}{\partial \mathcal{Z}_1^{\mathcal{C}}} \delta_1\right) \, \delta_2 = 0 \,,
\end{equation}
which can be rewritten using (\ref{Ftilde}) and after defining the cross-ratio $v:=\frac{c_{13} c_{24}}{c_{14} c_{23}}$ as 
 \begin{equation}
 \label{diffeq}
 \int \, dc_{14} \, dc_{23} \, dc_{24} \, dv \, c_{13}^{s_2-s_3} \, c_{14}^{-1-s_1-s_2+s_3} \, c_{23}^{-2-s_2}\,
\left(\mathcal{D} f(v)\right) \, \mathcal{Z}^{\mathcal{A}}_4 \, (\partial_{\mathcal{C}}\delta_1) \, \delta_2  = 0 \,,
 \end{equation}
where we have defined
\begin{equation}\label{Ddef}
\mathcal{D} f(v) := v \, (1-v) \, \partial_v f(v) + (s_2 - s_3 + 1 - z - v (2+ s_2))\, f(v) \,.
\end{equation} 
Again recalling the remark made just below \eqref{invariance1}, this simply implies 
\beq\label{Deq}
\mathcal{D} f(v)  = 0\,.
\eeq
It fixes the function $f(v)$ up to an overall constant
\beq\label{firstf}
f(v) = \mathcal{C} \, (1-v)^{-1-s_3-z} \, v^{-1-s_2+s_3+z} \,.
\eeq
Let us now comment on the other term of \eqref{independence}. Repeating the same steps,  we finally obtain a further equation for $f(v)$, similar to, but different from \eqref{Ddef},\eqref{Deq}:
\begin{equation}
v \, (1-v) \, \partial_v f(v) + (s_2 - s_3 + 1 - z - v (2+s_2+s_1-s_3))\, f(v) = 0 \,.
\end{equation} 
Matching the solution to \eqref{firstf}, a further constraint on the representation labels emerges
\beq
\label{equals1}
s_1 = s_3 \,,
\eeq
supplementing our earlier finding \eqref{in=out}. Clearly we then also have
\beq
\label{equals2}
s_2 = s_4 \,.
\eeq
By inserting back the result in (\ref{Ftilde}) and (\ref{kernel1}) we can write down the final expression for the formal kernel of the R-matrix in the Gra\ss mannian description
\begin{equation}
\label{RmatrixGen}
\RR(z) = \int \, \frac{dc_{13}\,dc_{14}\,dc_{23}\,dc_{24}}{c_{13} \, c_{24} \, \mathrm{det}C} \left(-\frac{c_{13}c_{24}}{\mathrm{det}C}\right)^z 
\, c_{24}^{s_1-s_2} (-\mathrm{det}C)^{-s_1} \,
 \, \delta^{N|M}(\mathcal{Z}_1 - \sum_{k=3}^4 c_{1k} \mathcal{Z}_{k}) \, \delta^{N|M}(\mathcal{Z}_2 -  \sum_{k=3}^4 c_{2k} \mathcal{Z}_{k}), 
\end{equation}
with $\mathrm{det}C = (c_{13} c_{24} - c_{14} c_{23})$. Let us remark that the form of the solution is unique up to the usual scalar factor, depending on the spectral parameter $z$, undetermined by the Yang-Baxter equation.

So far we did not fix the set of integration contours in our formal expression \eqref{RmatrixGen}. However, we did make an implicit assumption in our above derivation, which was based on homogeneity assumptions and the imposition of the Yang-Baxter equation. We assumed that all boundary terms vanish when performing partial integrations. There are two cases when this holds. Either all contours are closed and there are no boundaries at all, or else, we have some open contours, for which the boundary terms either vanish or cancel out. The correct choice of contours depends on the precise application of the formula \eqref{RmatrixGen}. In particular, we observed that, in order to compare with the harmonic action result \eqref{Rhop}, we need to put $s_1=s_2$, and then take a combination of these two cases. Namely, for the variables ${c_{13},c_{23},c_{24}}$ we need to take a closed contour encircling 0, while for $v=\frac{c_{13} c_{24}}{c_{14} c_{23}}$ we should take an open contour from 0 to 1. 
If instead we want to make contact with the scattering amplitude problem in $\mathcal{N}=4$ SYM, which will be our main application in the rest of this paper, then we have to take a closed set of contours that contains the support of all delta functions. In this case, the kernel of the R-matrix $\RR(z) $ is a spectral-parameter dependent deformation of the four-point tree-level amplitude, as shown in more detail in the next subsection. It would be very interesting to understand the appropriate contours necessary for these distinct applications from first principles.


\subsection{Deformations of the Four-Point Amplitude in \texorpdfstring{$\mathcal{N}=4$}{} SYM}
\label{Sec:Deformationsoffourpoint}

In the previous section we found a general solution to the Yang-Baxter equation, which is valid for any compact or non-compact oscillator representation of $\mathfrak{gl}(N|M)$. Here we would like to specialize to the case of $\mathfrak{gl}(4|4)$, which is the one relevant to the analysis of scattering amplitudes in $\mathcal{N}=4$ SYM. If we consider particles with physical helicities we should set all representation labels to $s_i=0$. In that case the kernel (\ref{RmatrixGen}) we just derived slightly simplifies to
\begin{equation}
\label{Rmatrix}
\RR(z) = \int \, \frac{dc_{13}\,dc_{14}\,dc_{23}\,dc_{24}}{c_{13} \, c_{24} \, \mathrm{det}C} \left(-\frac{c_{13}c_{24}}{\mathrm{det}C}\right)^z 
 \, \delta^{4|4}(\mathcal{Z}_1 - \sum_{k=3}^4 c_{1k} \mathcal{Z}_{k}) \, \delta^{4|4}(\mathcal{Z}_2 -  \sum_{k=3}^4 c_{2k} \mathcal{Z}_{k})\, .
\end{equation}
In order to establish the relation with amplitudes we should return from the twistor-space to the spinor-helicity formulation, and express
\eqref{Rmatrix} through the $(\lambda^{\alpha},\tilde\lambda^{\dot\alpha}, \eta^A)$ variables. This involves taking a Fourier transform on the $\tilde\mu$ variables\footnote{In our procedure we consider two incoming and two outgoing particles, instead of all incoming as is 
common in the amplitude literature.}
(see appendix \ref{app.Fourier}),
which leads to
\begin{align}
\label{mhv4grass}
\RR(z) =& \int \, \frac{dc_{13}\,dc_{14}\,dc_{23}\,dc_{24}}{c_{13} \, c_{24} \, \mathrm{det}C} \left(-\frac{c_{13}c_{24}}{\mathrm{det}C}\right)^z 
\delta^{4}\left(\eta_1 - \sum_{k=3}^4 c_{1k} \eta_{k}\right)
\delta^{4}\left(\eta_2 - \sum_{k=3}^4 c_{2k} \eta_{k}\right) \nn \\
&\delta^{2}\left(\tilde\lambda_1 - \sum_{k=3}^4 c_{1k} \tilde\lambda_{k}\right)
\delta^{2}\left(\tilde\lambda_2 - \sum_{k=3}^4 c_{2k} \tilde\lambda_{k} \right)
\delta^{2}\left(\lambda_3 - \sum_{k=1}^2 c_{k3} \lambda_{k} \right)\delta^{2}\left(\lambda_4 - \sum_{k=1}^2 c_{k4} \lambda_{k} \right).
\end{align}
As we mentioned in the previous section the integration contours are to be chosen such that the integration localizes on the support of all delta functions. This means that we can perform the integrations on the $c_{ai}$ variables by algebraically solving the system of constraints resulting from the complete set of delta functions. For example, if we solve the delta functions on the $\lambda$-variables we find 
\begin{align}
\lambda_3^\alpha - c_{13} \lambda_1^\alpha - c_{23} \lambda_2^\alpha = 0 \quad \Longrightarrow& \quad c_{13} = \frac{\langle32\rangle}{\langle12\rangle}, \qquad c_{23} = \frac{\langle31\rangle}{\langle21\rangle}, \\ 
\lambda_4^\alpha - c_{14} \lambda_1^\alpha - c_{24} \lambda_2^\alpha = 0 \quad \Longrightarrow& \quad c_{14} = \frac{\langle42\rangle}{\langle12\rangle}, \qquad c_{24} = \frac{\langle41\rangle}{\langle21\rangle} \,,
\end{align}
where  $\langle ij\rangle = \lambda_i^{\alpha} \lambda_{j,\alpha}$.
By substituting these expressions for  the $c_{ai}$ back into (\ref{mhv4grass}),  we finally arrive at
\beq
\label{spectralA}
\RR(z) =  \left(-\frac{\langle23\rangle \langle41\rangle}{\langle12\rangle \langle34\rangle}\right)^z  \frac{\delta^{4}(p) \delta^8(q)}{\langle12\rangle \langle23\rangle \langle34\rangle \langle41\rangle}
= \left(\frac{t}{s}\right)^z \mathcal{A}_4^{\text{MHV}}\,,
\eeq
where $s=(p_1+p_2)^2=(p_3+p_4)^2$ and $t=(p_2-p_3)^2=(p_1-p_4)^2$ are the Mandelstam variables and we have used standard notation for momentum conserving delta functions in the spinor-helicity formalism. Specifically, we have
\beq
p_i^{\alpha\dot\alpha} = \lambda_i^{\alpha}\tilde\lambda_i^{\dot\alpha}\,, \qquad \qquad q_i^{\alpha A} = \lambda_i^{\alpha} \eta_i^A\,, \qquad  \qquad 2 p_i \cdot p_j = \langle ij\rangle [j i] \,,
\eeq 
\beq \label{deltafour}
\delta^{4}(p)= \delta^{4}\left(p_1 + p_2 - p_3 - p_4\right), \qquad  \qquad \delta^{8}(q)= \delta^{8}\left(q_1 + q_2 - q_3 - q_4\right).
\eeq
$\RR(z)$ in (\ref{spectralA}) is the ${\cal N}=4$ SYM four-point tree-level MHV amplitude deformed by a spectral parameter $z$ dependent factor. By construction, it is a solution of the Yang-Baxter equation, and therefore establishes a direct connection between integrability and scattering amplitudes. 



\section{Three-Point Harmonic R-Matrices}
\label{Sec:ch2_ThreePoints}

\subsection{Preliminary Remarks}

As was pointed out and stressed in \cite{ArkaniHamed:2012nw}, in the on-shell approach to Yang-Mills massless scattering the four-point amplitude is not the most elementary building block. It should be replaced by the MHV and $\overline{\mbox{MHV}}$ three-point amplitudes, even though these are not compatible with momentum conservation in $\mathbb{R}^{1,3}$. If this is the case, an interesting and conceptually fundamental question arises in light of our above interpretation of the deformed four-point amplitude of $\mathcal{N}=4$ SYM as the kernel of an R-matrix. Namely, one is led to conjecture that this deformed amplitude should then likewise be composed of deformed three-point MHV and $\overline{\mbox{MHV}}$ amplitudes. Furthermore, the latter should correspond to the kernels of some linear operators one might want to christen ``three-point R-matrices''. In this section we will constructively prove this conjecture, calling the sought deformations $ \mathbf{R}_{\bullet}$ and $ \mathbf{R}_{\circ}$. 

Naively, one might immediately discard the existence of ``three-point R-matrices" from the following argument. In integrable models, particle production and annihilation is forbidden, as the momenta of scattering constituents are individually conserved. However, we should not confuse the ``world-sheet momenta'' of the underlying two-dimensional integrable model with the ``target-space momenta'' of the four-dimensional gauge theory. The former are associated to the spectral parameter, a quantity which will  be related to helicity and central charge, and the latter to the actual on-shell space-time momenta. Furthermore, even in two-dimensional models three-vertices have appeared in the form of bootstrap equations: bound state formation is possible also in two-dimensional integrable models. As the scattering momenta may be complex, an e.g. ``outgoing'' bound state carrying a single momentum label might form from two complex ``incoming'' momenta. Pictorially, we just merge two of the four lines of the four-point R-matrix into one, ending up with a three-point R-matrix.

In the following derivations we focus on  $\mathbf{R}_{\bullet}$, the deformation of the MHV amplitude, but analogous considerations are valid for $ \mathbf{R}_{\circ}$. We will show that the deformations we are looking for can be constructed using two distinct methods. For one, unlike the above four-point case, the requirements of super-Poincar\'e invariance and homogeneity degree preservation are sufficient to nearly, up to a multiplicative factor, fix the form of three-point amplitudes. For another, we will obtain the same results from suitable bootstrap equations. Finally, the interplay between these two methods allows us to establish a relation between the spectral parameters appearing in the bootstrap equations and the central charges of the scattering particles.


\subsection{Bootstrap for Three-Point R-Matrices}
\label{Sec:bootstrap}

There are two ${\cal N}=4$ three-point amplitudes. In Gra\ss mannian language, the  $n=3$, $k=2$ MHV amplitude reads
\begin{equation}
\label{MHVamp}
\mathcal{G}_{3,2}=\mathcal{G}_\bullet = 
\int \frac{dc_{13} dc_{23}}{c_{13} c_{23}}  \,  
\delta^{4|4}(\mathcal{Z}_1- c_{13} \mathcal{Z}_3) \delta^{4|4}(\mathcal{Z}_2- c_{23} \mathcal{Z}_3),
\end{equation}
while the $n=3$, $k=1$ $\overline{\mbox{MHV}}$ amplitude is
\begin{equation}
\label{barMHVamp}
\mathcal{G}_{3,1}=
\mathcal{G}_\circ=
\int \frac{dc_{12} dc_{13}}{c_{12} c_{13}} \, 
\delta^{4|4}(\mathcal{Z}_1- c_{12} \mathcal{Z}_2 -c_{13} \mathcal{Z}_3).
\end{equation}
We are looking for deformations of these three-point amplitudes by modifying the measure in \eqref{MHVamp} and \eqref{barMHVamp} in analogy with the four-point MHV case $n=4$, $k=2$, see \eqref{kernel1}.
Starting with $\mathcal{G}_{3,2}=\mathcal{G}_\bullet$, we thus introduce a general measure factor $F(c_{13},c_{23})$ and make the ansatz
\begin{equation}
\label{RbulletF}
\mathcal{R}_\bullet = \int  dc_{13} dc_{23}  \, F(c_{13}, c_{23})  \delta^{N|M}(\mathcal{Z}_1- c_{13} \mathcal{Z}_3) \delta^{N|M}(\mathcal{Z}_2- c_{23} \mathcal{Z}_3) \,.
\end{equation}
We have also again temporarily generalized from $\mathfrak{gl}(4|4)$ to arbitrary $\mathfrak{gl}(N|M)$, as the calculation is no harder.
Similar to the four-point calculation we assume that the particles corresponding to the first two, trivial columns of the rectangular matrix $C$ in \eqref{C-matrix} are incoming particles, while the third, non-trivial column is related to the outgoing particle. 
One may show by explicit calculation that \eqref{RbulletF} is a $\mathfrak{gl}(N|M)$ invariant quantity. For this one easily checks, in analogy with \eqref{invariance1}, that
\begin{equation}\label{invariance2}
\RR_\bullet \left(\mathcal{Z}^{\mathcal{A}}_1 \frac{\partial}{\partial \mathcal{Z}_1^{\mathcal{C}}}+ 
\mathcal{Z}^{\mathcal{A}}_2 \frac{\partial}{\partial \mathcal{Z}_2^{\mathcal{C}}}\right) -
\mathcal{Z}^{\mathcal{A}}_3 \frac{\partial}{\partial \mathcal{Z}_3^{\mathcal{C}}}
 \RR_\bullet = 0  \,.
\end{equation}
 The homogeneity equations read 
\begin{equation}\label{scaling3}
\mathcal{R}_\bullet \, \mathcal{Z}^{\mathcal{A}}_i \frac{\partial}{\partial \mathcal{Z}_i^{\mathcal{A}}} = s_i \, \mathcal{R}_\bullet\,,\, i=1,2\,, \qquad  \qquad
\mathcal{Z}^{\mathcal{A}}_3 \frac{\partial}{\partial \mathcal{Z}_3^{\mathcal{A}}} \, \mathcal{R}_\bullet  = s_3 \, \mathcal{R}_\bullet \,.
\end{equation}
These equations are naturally solved with the technique of separation of variables. We take
\begin{equation}
 F(c_{13}, c_{23}) = f_{13}(c_{13}) \, f_{23}(c_{23}) \,.
\end{equation}
Proceeding as in the four-point case, we find that the first two equations in \eqref{scaling3} fix the functions $f_{13}$ and $f_{23}$, while the third one implies the conservation of representation labels of the three-point vertex
\begin{equation}
f_{13}(c_{13})=c_{13}^{-s_1-1}, \qquad  \qquad f_{23}(c_{23})=c_{23}^{-s_2-1},  \qquad  \qquad s_3 = s_1 + s_2 \,.
\end{equation}
This fixes the three-point deformed MHV amplitude up to a multiplicative scalar factor.

\begin{figure}[t]
\begin{center}
\scalebox{0.25}{\input{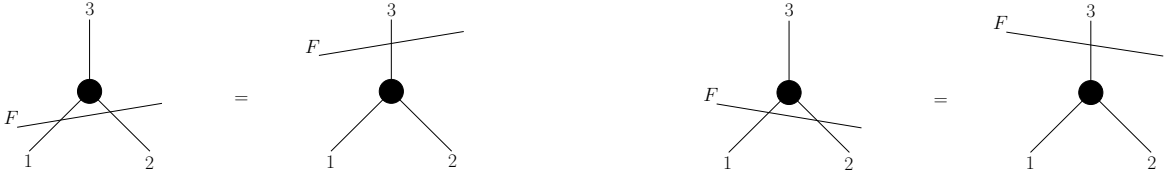}}
  \caption{Bootstrap equations for the deformed three-point MHV amplitude $\mathcal{R}_{\bullet}$. $F$ stands for the fundamental representation of $\mathfrak{gl}(N|M)$.}
 \label{Fig:bootstrapblack}
 \end{center}
 \end{figure}

There exists an alternative method to derive the same result. It takes its origin from integrable models. We assume that a proper deformation of the three-point amplitude is given as the common solution to the two bootstrap equations given in figure \ref{Fig:bootstrapblack}.
One can think about these bootstrap equations as a degenerate version of the Yang-Baxter equation, where the two particles emerging from the four-vertex form a bound-state; i.e.~the two corresponding lines are merged into a single line. In our calculation, however, we treat all three particles on an equal footing, i.e.~we ignore the fact that one of them may be thought of as a bound state. In the kernel form, the bootstrap equations may then be written as\footnote{As opposed to the Yang-Baxter equation, where the overall normalizations of the R-matrices do not play a role, for the bootstrap equation the normalizations of the intertwiners \eqref{Ri3} modify the equation. Our, admittedly somewhat ad-hoc, choice leads to the proper building blocks for deformed amplitudes.}
\begin{align}
\label{eq.bootstrap}
\left(\delta^{\mathcal{A}}_{\mathcal{C}} +(-1)^{\mathcal{C}} \frac{(J_3)^{\mathcal{A}}_{\mathcal{C}}}{\tilde{z}_1}\right) \,\mathcal{R}_{\bullet} &= \mathcal{R}_{\bullet} \,\left(\delta^{\mathcal{A}}_{\mathcal{B}}+ (-1)^{\mathcal{B}}\frac{(J_1)^{\mathcal{A}}_{\mathcal{B}}}{\tilde{z}_1}\right) \left(\delta^{\mathcal{B}}_{\mathcal{C}} +(-1)^{\mathcal{C}} \frac{(J_2)^{\mathcal{B}}_{\mathcal{C}}}{\tilde{z}_2}\right),\nn \\
\left(\delta^{\mathcal{A}}_{\mathcal{C}} +(-1)^{\mathcal{C}} \frac{(J_3)^{\mathcal{A}}_{\mathcal{C}}}{\tilde{z}_3}\right) \,\mathcal{R}_{\bullet} &= \mathcal{R}_{\bullet} \,\left(\delta^{\mathcal{A}}_{\mathcal{B}}+(-1)^{\mathcal{B}} \frac{(J_2)^{\mathcal{A}}_{\mathcal{B}}}{\tilde{z}_3}\right) \left(\delta^{\mathcal{B}}_{\mathcal{C}}+(-1)^{\mathcal{C}} \frac{(J_1)^{\mathcal{B}}_{\mathcal{C}}}{\tilde{z}_4}\right),
\end{align}
where we used the explicit, standard form for the R-matrices intertwining the fundamental representation with any representation of $\mathfrak{gl}(N|M)$, {\it cf} \eqref{Ri3}, and we replaced $J_{i\,C} ^{\mathcal{A}}$ by $(J_{i})^{\mathcal{A}}_{\mathcal{C}}$ for clarity. As for the Yang-Baxter equation, by expanding in the difference of spectral parameters, we may derive, a priori, two conditions from each of the two bootstrap equations. The first two actually coincide, and again simply express the $\mathfrak{gl}(N|M)$ invariance of $\mathcal{R}_{\bullet}$:
\begin{equation}
\label{eq.invgl}
(J_3)^{\mathcal{A}}_{\mathcal{B}} \, \mathcal{R}_{\bullet} =  \mathcal{R}_{\bullet} \, ((J_1)^{\mathcal{A}}_{\mathcal{B}} + (J_2)^{\mathcal{A}}_{\mathcal{B}}) \,.
\end{equation}
The second two conditions are indeed distinct, and are both quadratic in the generators $J$. Using \eqref{Ri3}, we may succinctly write them in the following form
\begin{align}
\label{3point}
\nonumber
z_1
\mathbf{L}_3(0)\,\mathcal{R}_{\bullet} &= \mathcal{R}_{\bullet}\,
\mathbf{L}_1(0)\,\mathbf{L}_2(z_1)\,,\\ 
z_2
\mathbf{L}_3(0)\,\mathcal{R}_{\bullet} &= \mathcal{R}_{\bullet}\,
\mathbf{L}_2(0)\,\mathbf{L}_1(z_2)\,,
\end{align}
where we have defined $z_1 \equiv \tilde{z}_2-\tilde{z}_1$ and  $z_2 \equiv \tilde{z}_3-\tilde{z}_4$.
The $\mathfrak{gl}(N|M)$ invariance \eqref{eq.invgl} suggests the Gra\ss mannian form \eqref{RbulletF} of $\mathcal{R}_{\bullet}$. When we solve the two further conditions \eqref{3point} we find, in generalization of \eqref{MHVamp},
\begin{equation}\label{kernels3ptblack}
\mathcal{R}_\bullet=\int \frac{dc_{13} dc_{23}}{c_{13} c_{23}}{
\frac{1}{c_{13}^{z_1} c_{23}^{z_2}}}\delta^{N|M}(\mathcal{Z}_1- c_{13} \mathcal{Z}_3) \delta^{N|M}(\mathcal{Z}_2- c_{23} \mathcal{Z}_3).
\end{equation}
It is exactly the same formula as derived above from the homogeneity properties if we identify
\begin{equation} 
z_1 \equiv s_1 \qquad  {\rm and} \qquad z_2 \equiv s_2  \,.
\end{equation}
This relates the spectral parameters to the representation labels.

We are left with also finding the deformed $\overline{\mbox{MHV}}$ amplitudes, i.e. the $k=1$ case above. Solving once more either the appropriate homogeneity equations or, alternatively, the correct bootstrap equations as presented in figure \ref{Fig:bootstrapwhite}, one ends up, in generalization of \eqref{barMHVamp}, with
\begin{figure}[t]
\begin{center}
\scalebox{0.25}{\input{bootstrapwhite.pstex_t}}
  \caption{Bootstrap equations for the deformed three-point $\overline{\mbox{MHV}}$ amplitude $\mathcal{R}_{\circ}$. $F$ stands for the fundamental representation of $\mathfrak{gl}(N|M)$.}
 \label{Fig:bootstrapwhite}
 \end{center}
 \end{figure}
\begin{equation}
\label{kernels3ptwhite}
\mathcal{R}_\circ=\int \frac{dc_{12} dc_{13}}{c_{12}
c_{13}}{\frac{1}{c_{12}^{z_2} c_{13}^{z_3}}}\delta^{N|M}(\mathcal{Z}_1- c_{12} \mathcal{Z}_2 - c_{13} \mathcal{Z}_3),
\end{equation}
where the spectral parameters $z_2=s_2$, $z_3=s_3$ are related to the representations labels $s_2$, $s_3$.
Recall that the solutions to the bootstrap equations \eqref{kernels3ptblack} and \eqref{kernels3ptwhite} we derived are valid for arbitrary $\mathfrak{gl}(N|M)$. In order to establish a direct link with the scattering amplitudes of $\mathcal{N}=4$ SYM, we return to the special case $N=M=4$.
We may then once more translate these expressions to super-spinor-helicity space. After integration over the complex parameters on the joint support of all delta functions, the three-point R-matrix kernels take the form\footnote{In order to render \eqref{Rz3} more symmetric, we introduced a third spectral parameter $z_3$, dropped an overall scalar function, and considered all particles incoming.}
\begin{align}
\label{Rz3}
\mathcal{R}_\bullet & = \frac{\delta^{4}(p^{\alpha\dot\alpha}) \delta^{8}(q^{\alpha A}) }{\langle 1\,2\rangle^{1+z_3} \langle 2\,3\rangle^{1+z_1} \langle 3\,1\rangle^{1+z_2}} \,,\nn \\
\mathcal{R}_\circ & = \frac{\delta^{4}(p^{\alpha\dot\alpha}) \delta^{4}(\tilde q^A)}{ [1\,2]^{1-z_3} [2\,3]^{1-z_1} [3\,1]^{1-z_2}}  \,,
\end{align}
where $\tilde q^A =  [23] \eta_1^A + [31] \eta_2^A + [12] \eta_3^A $, and with a constraint $z_1+z_2+z_3=0$ on the three spectral parameters.
Remarkably, the expressions in \eqref{Rz3}  for the ${\cal N}=4$ deformed three-particle amplitudes turn out to be identical, apart from the delta functions in the numerators, to standard conformal field theory three-point correlators. 


\subsection{Four-Point R-Matrix Kernel from Building Blocks}
\label{Sec:gluing.fourpt}
\begin{figure}[ht]
\begin{center}
\scalebox{0.35}{\input{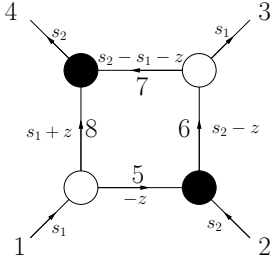}}
\end{center}
\caption{The R-matrix kernel from an on-shell diagram. The black and white vertices are the deformed three-point MHV and $\overline{\mbox{MHV}}$ amplitudes $\mathcal{R}_{\bullet}$ and $\mathcal{R}_{\circ}$, respectively. We labelled the edges with the deformed particle central charges.}
\label{Fig:Fourglued}
\end{figure}

Having constructed the proper deformations of the three-point amplitudes, we next show that one can reconstruct the R-matrix kernel \eqref{RmatrixGen} using the on-shell diagrams prescription given in \cite{ArkaniHamed:2012nw}. The relevant diagram for the four-point MHV amplitude is given in figure \ref{Fig:Fourglued}.
 In order to evaluate this diagram one identifies black and white vertices with, respectively, deformed MHV and $\overline{\mbox{MHV}}$ three-particle amplitudes and subsequently combines two of each kind in an appropriate fashion. One out of many possible deformation choices for all internal and external particles is depicted in the picture. Each of the vertices comes with two complex parameters, and we have to integrate over all four internal lines. One gets 
\begin{align}
\label{Rgluing}
\mathcal{R}_{\mathrm{glued}} &= 
\int \prod_{i=5}^8 \, d^{N|M}\mathcal{Z}_{i} \, \int 
\frac{ dc_{15}\,dc_{18}\,dc_{26}\,dc_{56}\,dc_{63}\,dc_{67}\,dc_{74}\,dc_{84}}{c_{15}^{1-z}\,c_{18}^{1+z+s_1}\,c_{26}^{1+s_2}\,c_{56}^{1-z}\,c_{63}^{1+s_1}\,c_{67}^{1-z+s_2-s_1}\,c_{74}^{1-z+s_2-s_1}\,c_{84}^{1+z+s_1}} 
 \delta(\mathcal{Z}_5-c_{56}\mathcal{Z}_6)\nn \\
 &\hspace{0.3cm}
\delta(\mathcal{Z}_{2}-c_{26}\mathcal{Z}_6)\,\delta(\mathcal{Z}_6-c_{63}\mathcal{Z}_3-c_{67}\mathcal{Z}_7)\,\delta(\mathcal{Z}_7-c_{74}\mathcal{Z}_4)\,\delta(\mathcal{Z}_8-c_{84}\mathcal{Z}_4)\,\delta(\mathcal{Z}_1 - c_{15} \mathcal{Z}_{5}-c_{18}\mathcal{Z}_8) ,
\end{align}
where we used the shorthand notation $\delta=\delta^{N|M}$.
By performing the integration over $(\mathcal{Z}_{5}, \mathcal{Z}_{6}, \mathcal{Z}_{7}, \mathcal{Z}_{8})$, and after the following change of variables
\beq
\label{matrixC}
C=
\left(
\begin{array}{cccc}
 1 & 0 & -c_{15}c_{56}c_{63}  & -(c_{18}c_{84}+c_{15}c_{56}c_{67}c_{74})  \\
 0 & 1 & -c_{26}c_{63}  & -c_{26}c_{67}c_{74}  
\end{array}
\right) 
=
\left(
\begin{array}{cccc}
 1 & 0 & -c_{13}  & -c_{14}  \\
 0 & 1 & -c_{23} & -c_{24}  
\end{array}
\right),
\ee
we finally arrive at 
\begin{align}
\label{Rglued}
\mathcal{R}_{\mathrm{glued}} =& \left( \int \frac{dc_{56} \,dc_{63} \,dc_{74} \,dc_{84}}{c_{56} c_{63} c_{74} c_{84}} \right) 
\int \frac{dc_{13}\,dc_{24}\,dc_{14}\,dc_{23}}{c_{13} \, c_{24} \, \mathrm{det}C} \left(-\frac{c_{13}c_{24}}{\mathrm{det}C}\right)^{z} (-\mathrm{det}C)^{-s_1} c_{24}^{s_1-s_2} \, \nn\\
& \, \delta^{N|M}(\mathcal{Z}_1 - \sum_{k=3}^4 c_{1k} \mathcal{Z}_{k}) \, \delta^{N|M}(\mathcal{Z}_2 -  \sum_{k=3}^4 c_{2k} \mathcal{Z}_{k})\, ,
\end{align}
where we have dropped an irrelevant constant factor.
We notice that four of the original complex variables decouple completely. We may thus integrate these variables over contours encircling 0, leading to a numerical factor, which may again be dropped.
Excitingly, \eqref{Rglued} then turns exactly into \eqref{RmatrixGen}.



\section{Deformation of Generic On-Shell Diagrams}
\label{Sec:ch3_ManyPoint}

\subsection{Preliminary Remarks}
\label{Sec:ch3_preliminary}
In the preceding sections we derived the deformations of the three- and four-point tree-level amplitudes \eqref{Rz3} and \eqref{spectralA}, respectively. 
We also showed in section \ref{Sec:gluing.fourpt} that a specific combination of deformed MHV and $\overline{\mbox{MHV}}$ three-point vertices, through an appropriate procedure of integration, reconstructs the R-matrix kernel. Pictorially the steps are encoded in the on-shell diagram of figure~\ref{Fig:Fourglued}.
Our next goal is to generalize our procedure to provide the deformations for general on-shell diagrams, which may be related to tree- and loop-level amplitudes. 
Indeed, any amplitude at any loop order was claimed to arise from the conjectured all-loop BCFW recursion relation \cite{ArkaniHamed:2010kv,Boels:2010nw}. For given loop level, number of particles $n$, and  helicity $n-2k$, the amplitude is then expressed as a sum of on-shell diagrams as shown in  \cite{ArkaniHamed:2012nw}. Every on-shell diagram is a planar graph made from two types of cubic (=trivalent) vertices, representing the ``black'' three-point  MHV and the ``white'' three-point $\overline{\mbox{MHV}}$ amplitudes. The amplitude then arises via integration of the
on-shell degrees of freedom for all internal particle lines.
It is then natural to ask how to spectrally deform any such on-shell diagram and whether there exists a formula valid for all of them. As we will show, the answer is positive at least for
the $n$-point MHV and $\overline{\mbox{MHV}}$ amplitudes, beautifully generalizing the undeformed versions. 

Some remarks are in order to avoid confusion about the relation between our deformed on-shell diagrams and scattering amplitudes.
First of all let us recapitulate our deformation procedure for three- and four-points at tree level. First, we multiplicatively deformed the measure in \eqref{ACCK} by some function $F$, {\it cf} \eqref{RbulletF}, \eqref{kernel1}, which was then fixed by imposing bootstrap and Yang-Baxter equations, respectively. The results, written as Gra\ss mannian integrals in \eqref{kernels3ptblack}, \eqref{kernels3ptwhite}, \eqref{Rmatrix}, were then translated to spinor-helicity variables in \eqref{Rz3} and \eqref{spectralA}. This was possible because all integrations localized on the joint support of all delta functions. The same property holds for the general MHV and $\overline{\mbox{MHV}}$ cases, where $k=2$ and $k=n-2$ respectively. For these we will shortly derive an explicit expression for their deformations for general $n$, using the on-shell diagrams built from the three-point deformed vertices. However, moving beyond MHV and $\overline{\mbox{MHV}}$ amplitudes, the problem of determining the appropriate deformation arises. The reason is that in the formula \eqref{ACCK} for $2< k< n-2$ not all integrations are saturated by delta functions and we are left with non-trivial integrations in the Gra\ss mannian space. Since the undeformed 
measure is a meromorphic function in these variables, one can find the final expression for any tree-level amplitude by evaluating a particular sum over residues at multidimensional poles of \eqref{ACCK}. The ``right'' sum (or choice of integration contours) appears to be 
 given by the BCFW recursion relation. In \cite{ArkaniHamed:2012nw} the hierarchy of residues was mapped to the cell decomposition of the positive part of the Gra\ss mannian $G(n,k)$. In that formalism, the expression \eqref{ACCK} is related to the so-called {\em top cell} of the positive Gra\ss mannian, while each of the residues is related to a particular lower cell, which belongs to the boundary of the positive part. Each element of the cell decomposition is in one-to-one correspondence with some Gra\ss mannian integral, which can be graphically described by an on-shell diagram. 
In the sequel we will present the procedure to find the deformation for any 
such on-shell diagram. This seemingly 
enables us to derive an arbitrary deformed non-MHV amplitude, but there is a caveat:
 Since we are currently lacking a deformed version of the BCFW recursion relation, we are not  able to recombine these deformed on-shell diagrams into what should then be called deformed non-MHV scattering amplitudes. At the same time, we are convinced that a proper principle on how to do this should exist. The most elegant way would presumably be a deformed BCFW relation. This is left to future work.

In the following we sketch how to find the deformation for any of such diagram, drawing heavily on the rather elaborate formalism of \cite{ArkaniHamed:2012nw}, and mainly pointing out the necessary generalizations when applying our deformation procedure. In other words, in this section our presentation is not fully self-contained, and requires some familiarity with \cite{ArkaniHamed:2010kv,ArkaniHamed:2010gh,ArkaniHamed:2012nw}. The main message is that the on-shell diagrams may indeed be naturally deformed (section \ref{Sec:ch3_Deformations}), and that the graphic rules of (un)merging, flipping and square-moving of \cite{ArkaniHamed:2012nw} are preserved under a subclass of our general multi-parameter deformation (section \ref{Sec:ch3_Moves}).


\subsection{Undeformed On-Shell Formalism}
\label{Sec:ch3_Undeformed}
\begin{figure}[ht]
\begin{center}
\scalebox{0.35}{\input{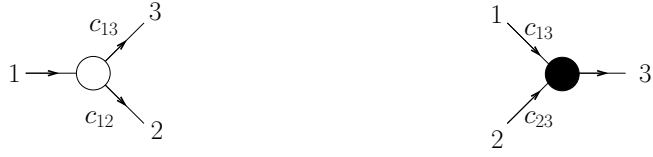}}
\end{center}
\caption{Three-point amplitudes dressed with arrows and weights.}
\label{Fig:threewitharrows}
\end{figure}

In this section we discuss some of the structure presented in \cite{ArkaniHamed:2012nw} in order to set up the background needed for the next section.  
Let us start by recalling the notion of perfect orientation for on-shell diagrams, which will be helpful in extracting important mathematical structure directly from a given graph.
Each diagram is said to admit a {\em perfect orientation} if its edges can be decorated with arrows such that for each white vertex there is one incoming arrow, while for each black vertex there are two incoming arrows. It is claimed in \cite{ArkaniHamed:2012nw} that ``all on-shell diagrams relevant to physics can be given a perfect orientation'', and therefore be composed from the two vertices in figure \ref{Fig:threewitharrows}. 
For the purposes of our current study, it suffices to restrict to perfect orientations without cycles.
We may relate the ensuing orientation pattern at each black or white vertex to its respective associated matrix in \eqref{C-matrix}, which read
$C_\bullet=\bigl( \begin{smallmatrix}
  1&0&-c_{13}\\ 0&1&-c_{23}
\end{smallmatrix} \bigr)
$
and
$C_\circ=\bigl( \begin{smallmatrix}
  1&-c_{12}&-c_{13}
\end{smallmatrix} \bigr).
$
Recall that the vertices stand for the Gra{\ss}mannian integrals $\mathcal{G}_\bullet$ in \eqref{MHVamp} and $\mathcal{G}_\circ$ in \eqref{barMHVamp}, and we accordingly attach the integration variables $c_{13},c_{23}$ and $c_{12},c_{13}$ to the edges as in figure~\ref{Fig:threewitharrows}.

To retrieve the formula encoded by an on-shell diagram one has to follow a gluing procedure as
follows: First one multiplies all three-point vertices $\mathcal{G}_\bullet$ and $\mathcal{G}_\circ$ in the Gra{\ss}mannian integral representation with edge-variables $c_{ij}$. Then one  
integrates over the on-shell degrees of freedom on all internal edges, i.e.~the
super-twistors associated to these edges. 
It has been shown in \cite{ArkaniHamed:2012nw} that this gluing procedure associates each 
on-shell diagram with a $k \times n$ matrix $C$  representing an element in the Gra{\ss}mannian variety $G(n,k)$ as in \eqref{C-matrix}. This matrix is constructed from the edge-variables $c_{ij}$ of all trivalent vertices comprising the graph. 
The gluing also gives rise to an $(n_F - 1)$-dimensional measure of integration $dC$, where $n_F$ is the number of faces of the given diagram.
We already encountered an example in the previous section, albeit in the deformed case: 
Four three-point vertices were glued to yield the deformed four-point amplitude. 
See in particular the matrix \eqref{matrixC}. 
In fact we shall not be more explicit here as there exists a set of variables known as 
face variables $f_i$, where the result simplifies and to which we now turn.

\begin{figure}[t]
\begin{center}
\scalebox{0.35}{\input{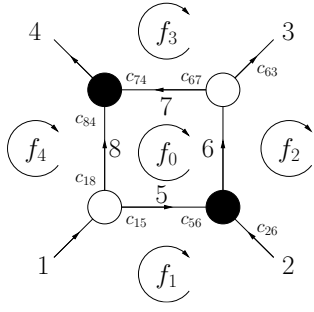}}
\end{center}
\caption{Face variables.}
\label{Fig:facevariable}
\end{figure}

As the name indicates these variables are associated to the faces of a given on-shell diagram. 
The face-variables $f_{i}$ are functions of the edge-variables $c_{ij}$ along the face boundaries. Namely, for a given face we take the product of all edge-variables on the clockwise-aligned edges and divide by the product of all edge-variables on the anti-clockwise-aligned edges.
Let us demonstrate the gluing construction and the change to face variables for 
the MHV four-point example, compare 
figure~\ref{Fig:facevariable}. 
Gluing the undeformed three-point vertices according to the nomenclature of 
figure~\ref{Fig:facevariable} yields 
\begin{align}
\mathcal{A}_{\mathrm{glued}} =& 
\int \prod_{i=5}^8 \, d^{N|M}\mathcal{Z}_{i} \, \int 
\frac{ dc_{15}\,dc_{18}\,dc_{26}\,dc_{56}\,dc_{63}\,dc_{67}\,dc_{74}\,dc_{84}}{c_{15}\,c_{18}\,c_{26}\,c_{56}\,c_{63}\,c_{67}\,c_{74}\,c_{84}} 
\, \delta(\mathcal{Z}_5-c_{56}\mathcal{Z}_6)\,\delta(\mathcal{Z}_{2}-c_{26}\mathcal{Z}_6)\nn \\
 &\hspace{0.3cm}
\delta(\mathcal{Z}_6-c_{63}\mathcal{Z}_3-c_{67}\mathcal{Z}_7)\,\delta(\mathcal{Z}_7-c_{74}\mathcal{Z}_4)\,\delta(\mathcal{Z}_8-c_{84}\mathcal{Z}_4)\,\delta(\mathcal{Z}_1 - c_{15} \mathcal{Z}_{5}-c_{18}\mathcal{Z}_8) ,
\label{Aglued1}
\end{align}
which is nothing but the undeformed version of \eqref{Rgluing}. We now make the change from
edge to 
face variables. Following the composition rule introduced above we define the face variables 
\begin{equation}
\label{face.variables}
f_0 = \frac{c_{18}c_{84}}{c_{74}c_{67}c_{15}c_{56}} \,,  \,\,f_1= \frac{c_{15}c_{56}}{c_{26}}\,, \,\, f_2= c_{26}c_{63}\,, \,\, f_3=\frac{c_{67}c_{74}}{c_{63}} \,, \,\, f_4=\frac{1}{c_{18}c_{84}} \,.
\end{equation}
putting the clock-wise (anti-clock wise) oriented edge-variables in the numerator (denominator).
Applying this change of variables to the expression \eqref{Aglued1} and performing the four super-twistor integrals via the delta functions yields 
\begin{equation}
\mathcal{A}_{\mathrm{glued}} = \int \prod_{i=1}^4 \frac{df_i}{f_i}
 \delta^{4|4}(\mathcal{Z}_1 -  f_1 f_2 \, \mathcal{Z}_{3} - (1+f_0)f_1 f_2 f_3 \, \mathcal{Z}_{4}) \, \delta^{4|4}(\mathcal{Z}_2 - f_2 \, \mathcal{Z}_3 - f_2 f_3 \,\mathcal{Z}_4)\, ,
\end{equation}
where we have omitted a trivial constant arising from the factorization of four of the edge-variable
integrals (\emph{cf} the case of $\mathcal{R}_{\mathrm{glued}}$ in \eqref{Rglued}). In fact not only the measure can be derived directly by looking at the graph, but also the factors inside the delta functions! The matrix determining the arguments of the delta functions, introduced in the most general case in \eqref{C-matrix}, is
\beq
C(f)=
\left(
\begin{array}{cccc}
 1 & 0 & -f_1 f_2  & -(1+f_0)f_1 f_2 f_3  \\
 0 & 1 & -f_2  & -f_2 f_3   
\end{array}
\right) 
=
\left(
\begin{array}{cccc}
 1 & 0 & -c_{13}(f)  & -c_{14}(f)  \\
 0 & 1 & -c_{23}(f) & -c_{24}(f)  
\end{array}
\right).
\ee
We note that each $c_{ai}(f)$ is given by the product of the face variables which are on the right of the path $a\rightarrow i$ following the arrows. In case of multiple paths, one has to sum the respective partial results.
This is the so-called boundary measurement, see \cite{Postnikov}.

One can now generalize this result. For a general on-shell diagram with $n_F$ faces, the formula will read
\begin{equation}\label{undeformed.amplitudes}
\int \prod_{i=1}^{n_F-1} \frac{df_i}{f_i}\prod_{a=1}^k \delta^{4|4}(\mathcal{Z}_a-\sum_{i=k+1}^n c_{ai}(f) \mathcal{Z}_i) \,,
\end{equation}   
where $C(f)$ can again be read off from the boundary measurement.

The careful reader may have noticed that  the product in \eqref{undeformed.amplitudes} only runs over $n_F-1$ faces. As was shown in \cite{ArkaniHamed:2012nw}, this is indeed the proper dimension of the matrix $C(f)$ related to any on-shell diagram with $n_F$ faces. The remaining face variable can be always related to all the others. This is due to the fact that since we chose the same orientation for all faces in the diagram every edge-variable appears once in a numerator and once in a denominator. Hence it is easily seen that
\begin{equation}
\prod_{i=1}^{n_F}f_i=1 \,.
\end{equation}
One may check this for the face variables of our example presented in \eqref{face.variables}.


\subsection{Deformed On-Shell Formalism}
\label{Sec:ch3_Deformations}

After having sketched the face-variable formalism for the undeformed case, let us investigate how the formulas presented get modified once we replace three-point amplitudes by their deformed counterparts $\mathcal{R}_\bullet$ and $\mathcal{R}_\circ$ as constructed in section \ref{Sec:bootstrap}. 

In fact we would like to stress here
that the somewhat \emph{ad hoc} orientation for three-point vertices
introduced in figure~\ref{Fig:threewitharrows} naturally arises from our spectral parameter
deformation point of view: The bootstrap and Yang-Baxter equations discussed in previous sections carry 
the notion of ``incoming" and ``outgoing" particles, corresponding to the arrows in
figure~\ref{Fig:threewitharrows}. Indeed this is also true for the Yang-Baxter equation and  
will reappear in the case of a generalized Yang-Baxter equation to be 
discussed in a later section. 

Let us then start with a given perfectly oriented on-shell diagram.  
We interpret all three-point vertices of the graph as deformed three-point amplitudes. The deformation introduces non-physical helicities (resulting in non-vanishing central charges) for each particle in the diagram. Remembering that there is a conservation of central charge at each vertex, we can introduce region central charges $\zeta_i$, one for each face of the on-shell diagram. We call these newly introduced quantities {\em face spectral parameters}. 
Moreover we understand a {\em dressed on-shell diagram} as an on-shell diagram with face spectral parameters attached to all faces. There is a way to read off the helicities (or central charges) 
of particles running along an edge from the two neighboring face spectral parameters. For each oriented edge of the on-shell diagram,  one has to take the difference of the right minus left face spectral parameter. Since only differences of face spectral parameters have physical relevance, there is always the possibility of redefining them by adding a common number, this does not change central charges of particles. This allows to always fix one of the face spectral parameters to be zero.

We are now in a position to translate dressed on-shell diagrams into integrals of Gra\ss mannian type. We adapt the gluing procedure we have reviewed in the above subsection to our deformed version. This implies that for a given on-shell diagram we simply multiply all deformed three-point vertices and integrate over the on-shell super-twistors of all internal particles. The 
$\zeta$-independent part recombines again into \eqref{undeformed.amplitudes} and we need to only focus on the part depending on the face spectral parameters. For every edge-variable $c_{ij}$ of the deformed three-point amplitude we have an additional contribution to the integration measure of the form $c_{ij}^{\zeta_{\mbox{\tiny L}}-\zeta_{\mbox{\tiny R}}}$, where $\zeta_{\mbox{\tiny R}}$ ($\zeta_{\mbox{\tiny L}}$) is the face spectral parameter attached to the face on the right (left) of the edge of $c_{ij}$. Collecting all edge-variables with a common exponent of a face spectral parameter $\zeta_i$ nicely combines into the face-variable factor $f_i^{-\zeta_{i}}$. Then the final result simply is
\begin{equation}\label{deformed.amplitudes}
\int \prod_{i=1}^{n_F-1} \frac{df_i}{f_i^{1+\zeta_i}}\prod_{a=1}^k \delta^{4|4}(\mathcal{Z}_a-\sum_{i=k+1}^n c_{ai}(f) \mathcal{Z}_i)  .
\end{equation}   
As we can see the only difference of \eqref{deformed.amplitudes} compared to the undeformed case \eqref{undeformed.amplitudes} is in a different measure of integration
\begin{equation}\label{deformedmeasure}
\prod_{i=1}^{n_F-1}\frac{df_i}{f_i}\longrightarrow \prod_{i=1}^{n_F-1}\frac{df_i}{f_i^{1+\zeta_i}} \,.
\end{equation}


\subsection{MHV \texorpdfstring{$n$}{}-Point Example}

\begin{figure}[t]
\begin{center}
\scalebox{0.35}{\input{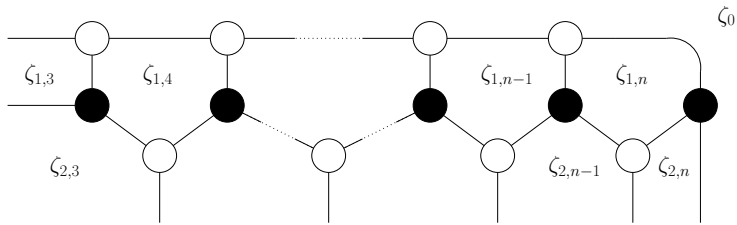}}
\end{center}
\caption{MHV $n$-point on-shell diagram with assignment of face spectral parameters.}
\label{Fig:MHVnex}
\end{figure}

As an example for the use of \eqref{deformed.amplitudes} let us now give the explicit form of all tree-level MHV deformed amplitudes. For this purpose, we translate \eqref{deformed.amplitudes} into the spinor-helicity language as in appendix~\ref{app.Fourier}. In the case of $k=2$, the integration on the $c_{ai}$ variables in \eqref{deformed.amplitudes} can be performed for any number of particles $n$ by algebraically solving the system of constraints from the complete set of delta functions. These constraints read
\begin{equation}
\lambda_{i}^{\alpha}-c_{1i}\lambda_{1}^{\alpha}-c_{2i}\lambda_{2}^{\alpha}\,, \qquad \mbox{ for }\alpha=1,2 \mbox{ and } i=3,\ldots,n\,,
\end{equation}  
and the variables $c_{ai}$ take the  values
\begin{equation}
c_{1i}=\frac{\langle i2 \rangle}{\langle 12\rangle}\,,\qquad c_{2i}=\frac{\langle i1\rangle }{\langle 21 \rangle}\,.
\end{equation}
We use formula \eqref{MHVFC}, which encodes the integral measure of the deformed Gra\ss mannian integral for $k=2$, together with the enumeration of faces we introduced in appendix~\ref{app.quivers}. Upon performing integrals and using the explicit form of the minor
\begin{equation}\label{roundbracket}
(ij)=c_{1i}c_{2j}-c_{2i}c_{1j}=\frac{\langle ij\rangle}{\langle 12\rangle}\,,
\end{equation}
we find the general spectral parameter deformation of the $n$-point MHV amplitudes
\begin{equation}
\label{Rn2deform}
\mathcal{R}_{n,2}=\mathcal{A}_{n,2}\  \left(\frac{\langle n-1 \, n\rangle }{\langle 1n-1\rangle}\right)^{\zeta_{0}}\left(-\frac{\langle 13\rangle }{\langle 23\rangle}\right)^{\zeta_{1,3}} \left(-\frac{\langle 12\rangle }{\langle 13\rangle}\right)^{\zeta_{2,3}}\prod_{i=4}^n \left(\frac{\langle i-2\,\, i-1\rangle \langle1\, i\rangle}{\langle i-1\, i\rangle \langle 1\, i-2\rangle}\right)^{\zeta_{1, i}}\left(\frac{\langle 1\, i-1\rangle}{\langle 1\, i\rangle}\right)^{\zeta_{2, i}}   \, . 
\end{equation}
In the above the face spectral parameters $\zeta_{i,j}$ are associated to the faces of the on-shell
diagram given in figure~\ref{Fig:MHVnex}.
We consider here the case when all external particles have non-physical helicities. A similar calculation can be done also for $\overline{\mbox{MHV}}$ amplitudes and one obtains analogous formula for a general deformation.


\subsection{Moves and Reduction} 
\label{Sec:ch3_Moves}

In \cite{Postnikov} a relation between on-shell diagrams and elements of the cell decomposition of the positive Gra\ss mannian was presented. These cells are parametrized by the face-variable matrices $C(f)$ associated to a given diagram as discussed in our example in \eqref{deformed.amplitudes}. A natural question is which class of diagrams parametrizes the same cell. The answer is that two diagrams are in the same equivalence class if they are related to each other by a set of basic operations: flip move, square move and reduction. In this section we study how the deformed measure \eqref{deformedmeasure} transforms under such operations. We shall be particularly interested in the case when the measure is invariant under such transformations.

\begin{figure}[ht]
\begin{center}
\scalebox{0.25}{\input{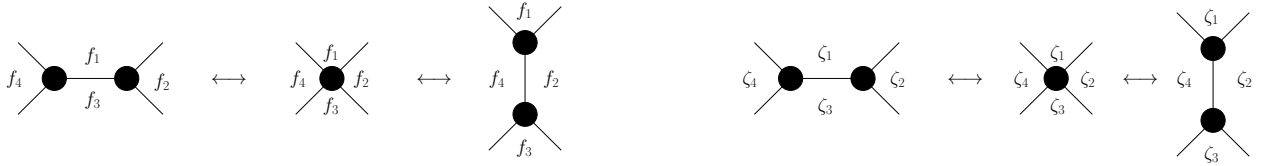}}
\end{center}
\caption{Transformation rules for dressed on-shell diagrams under the flip move, as composition of merge-unmerge transformations. The left picture describes the transformation of face variables while the right one encodes face spectral parameters. We can refrain from assigning arrows to the graphs as these relations hold for all possible orientations.}
\label{Fig:Flipmove}
\end{figure}

\begin{figure}[ht]
\begin{center}
\scalebox{0.25}{\input{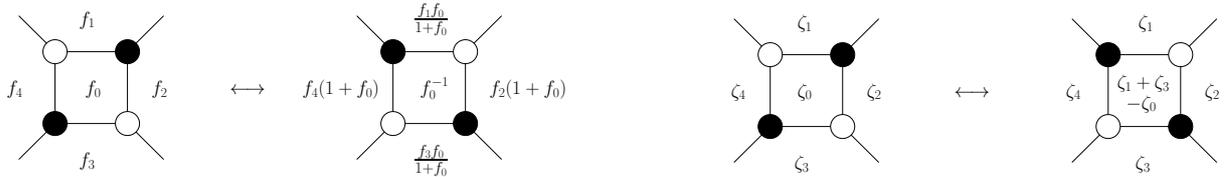}}
\end{center}
\caption{Transformation rules for on-shell diagrams under the square move. The left picture describes the transformation of the face-variables while the right one desrcibes the face spectral parameter transformation. The move leaves the integration mesure invariant only if 
$\zeta_1+\zeta_3=\zeta_2+\zeta_4$ holds. The relation is true for all arrow orientations.}
\label{Fig:Squaremove}
\end{figure}

\begin{figure}[p!]
\begin{center}
\scalebox{0.184}{\input{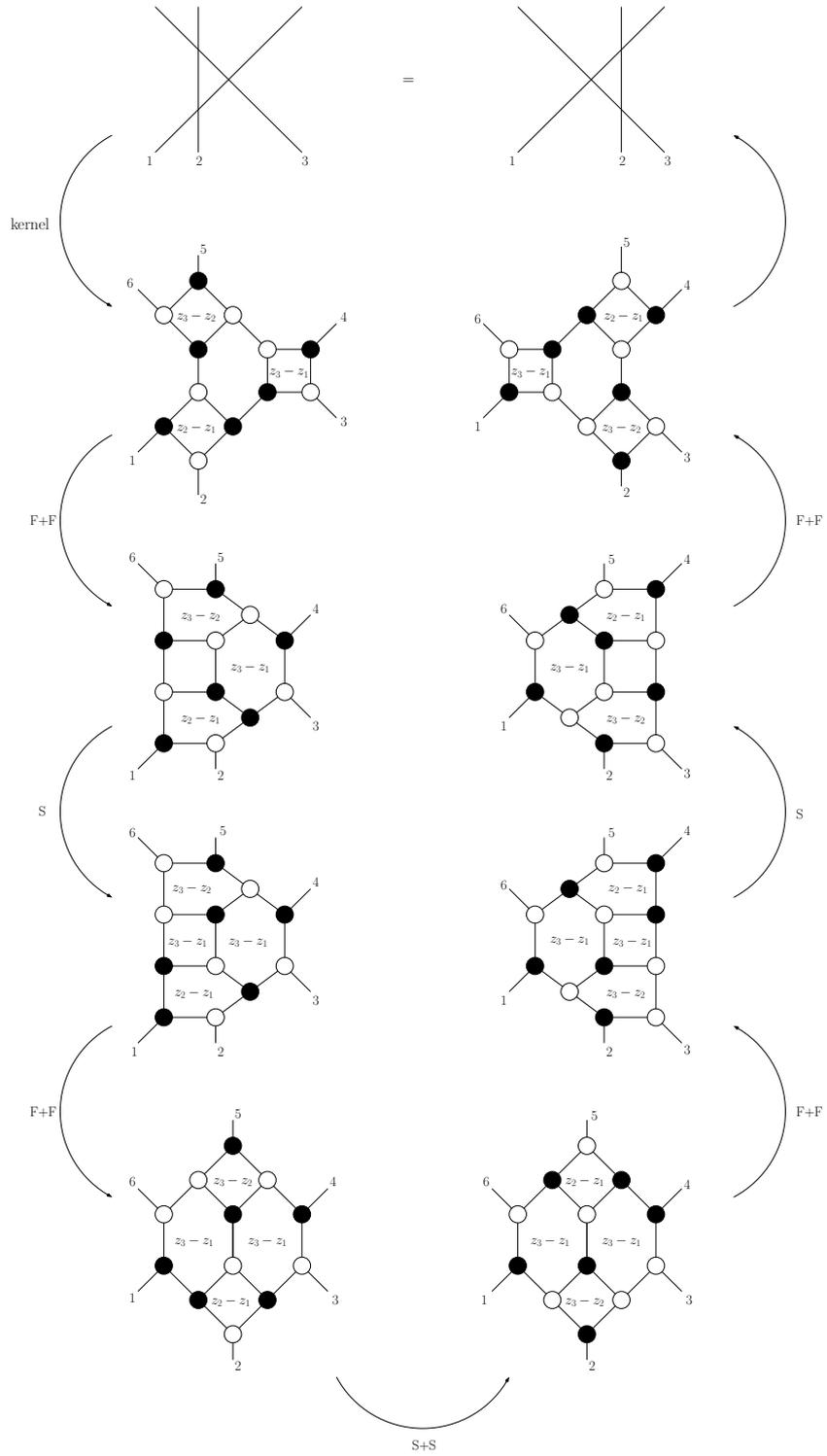}}
\end{center}
\caption{A diagrammatic derivation of the Yang-Baxter equation. At each step we marked whether the square move (S) or flip move (F) was performed.}
\label{Fig:YBEFromMoves}
\end{figure}

Let us start from the description of the flip move, which is a combination of a merge and unmerge  transformation, see figure~\ref{Fig:Flipmove}. One sees that the flip move,  performed locally in an on-shell diagram, does not change the matrix $C(f)$ and leaves the integration measure invariant. This entails the invariance under the flip move of the deformed graph  for the face spectral parameter dependence.
The same transformation is possible with white vertices instead of black ones. For the square move the prescription is a bit more involved. The first observation is that performing the square move not always leaves the measure invariant. If we require the measure to be invariant we have to impose the following relation on the face spectral parameters
\begin{equation}
\zeta_1+\zeta_3=\zeta_2+\zeta_4\,.
\end{equation}   
In that case we can encode the transformation rules in the two pictures presented in figure \ref{Fig:Squaremove}. 
We see that the face variables are altered in the same way as for the square move relevant to the undeformed diagrams, while the spectral parameters get modified only in the face for which the square move was performed. Using the flip and square moves  described above, one can give a very nice diagrammatic derivation of the Yang-Baxter equation (see figure \ref{Fig:YBEFromMoves}).

All possible transformations of the on-shell diagrams, which are compositions of the square and flip moves, can be understood as {\em mutations} in the language of cluster algebras. Since the definition of these algebras is very involved and not necessary in full generality to the purpose of this work, we will not present it here. We invite the interested reader to consult\cite{ClusterAlgebras} for an introduction. As every on-shell diagram can be transformed into a bipartite graph using the merge transformations as in figure \ref{Fig:Flipmove}, its dual graph is also oriented and we will refer to it as a {\em quiver}. One can indeed assign an orientation to the edges of the dual graph by demanding that e.g.~white vertices are always on the right of a given edge. To each node $i$ of the quiver diagram, which corresponds to a face of the original on-shell diagram, we associate a face variable $f_i$. Additionally, for dressed diagrams encoding deformed amplitudes,  we associate face spectral parameters $\zeta_i$. Given a quiver diagram we can construct a new quiver by mutating it at any vertex $j$. The mutated quiver is obtained by applying the following operations to the original quiver: for each path $i\to j\to k$ we add an arrow $i\to k$, we reverse all arrows incident to $j$ and we remove all two-cycles from the quiver we get. The mutation rules for face variables, which are generalizations of the formulas in figure~\ref{Fig:Squaremove}, are
\begin{equation}
f_i'=\begin{cases}f_j^{-1}\,,& \mbox{if }i=j\,,\\
f_i(1+f_{j})\,,& \mbox{if }i\to j\,,\\
f_i f_j(1+f_j)^{-1}\,,&\mbox{if }j\to i\,,\end{cases}
\end{equation}
where by $i\to j$ we denote that there existed an arrow from the node $i$ to the node $j$ in the original quiver, and analogously for $j\to i$. In the case of dressed quiver diagrams we have to additionally give transformation rules for face spectral parameters. In that case, similar to the square move, the mutation does not always leave the measure invariant. The requirement is that
\begin{equation}\label{zrestrictions}
\sum_{i\to j}\zeta_i-\sum_{j\to i}\zeta_i=0  \,.
\end{equation}
With this restriction the mutation rules for face spectral parameters are
\begin{equation}
\zeta_i'=\begin{cases} -\zeta_j+\sum_{i\to j}\zeta_i&\mbox{for }i=j \\ \zeta_i&\mbox{for }i\neq j\end{cases}
\end{equation}
in generalization to the formula given in figure~\ref{Fig:Squaremove}.
We will discuss more on mutations in the following sections, when we will check the Yangian symmetry of the dressed on-shell diagrams. It turns out that the restrictions \eqref{zrestrictions} will play a crucial role there. It will be also important for the one-loop example.  

\begin{figure}[ht]
\begin{center}
\scalebox{0.25}{\input{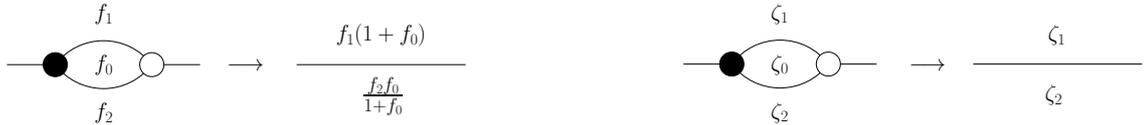}}
\end{center}
\caption{Transformation rules for on-shell diagrams under the reduction. The left picture carries the face-variables while the right picture carries the face spectral parameters of the deformation. Again we did not assign arrows as the relations hold for all orientations.}
\label{Fig:reduction}
\end{figure}

The last transformation we consider is the reduction.
As opposed to the square and flip moves this transformation reduces the number of faces of a given diagram by one, see figure~\ref{Fig:reduction}.

\noindent The measure associated to the three faces involved in the operation is
\begin{equation}
 \frac{df_0}{f^{1-\zeta_0}_0}\,\frac{df_1}{f^{1-\zeta_1}_1}\,\frac{df_2}{f^{1-\zeta_2}_2}\,=\,
\left(\frac{df_0}{f^{1-\zeta_0}_0}\,\frac{\left(1+f_0^{-1}\right)^{\zeta_2}}{\left(1+f_0\right)^{\zeta_1}}\,\right)\,
\frac{df'_1}{\left(f'_1\right)^{1-\zeta_1}}\,\frac{df'_2}{\left(f'_2\right)^{1-\zeta_2}}\,,
\end{equation}
where $f'_0\,=\,f_0$. We will not make use of this transformation in the present work. However, it was claimed in \cite{ArkaniHamed:2012nw} that the reduction is directly relevant to the calculation of loop amplitudes.



\section{Symmetries of Deformations}
\label{Sec:ch4_Symmetries}

\subsection{Preliminary Remarks}
Having established a spectral parameter deformation of on-shell diagrams, which was motivated by their relation to scattering amplitudes,
we should now clarify their symmetry properties under the action of the superconformal
algebra $\alg{su}(2,2|4)$ and its Yangian extension $Y[\alg{su}(2,2|4)]$. For the latter, it is sufficient to consider the action of the level-zero and level-one generators $J^{(0)}$ and $J^{(1)}$, respectively. Through the commutation relations the invariance under higher generators is then manifest. Here we assume the super Serre relations to hold, a
property which was analyzed in \cite{Dolan:2004ps} for the on-shell supermultiplet
representation we consider.
The standard definition of level-zero generators acting on scattering amplitudes is
\begin{equation}\label{level_zero}
J^{(0)\, \mathcal{A}}{}_{\mathcal{B}}=\sum_{i=1}^n J^{(0) \mathcal{A}}_{i \quad\,\, \mathcal{B}}=\sum_{i=1}^n\left(J^{\mathcal{A}}_{i\,\mathcal{B}}-\frac{1}{8}(-1)^{\mathcal{B}}\delta^{\mathcal{A}}_{\mathcal{B}} \sum_{\mathcal{C}} (-1)^{\mathcal{C}}J^{\mathcal{C}}_{i\,\mathcal{C}}\right),
\end{equation}
where we removed the super trace from the generators \eqref{generators} and summed over all particles. Indeed the invariance of the four-point amplitude under the action of \eqref{level_zero} follows from \eqref{gl_inv} for $z=0$. In turn, the level-one generators  are given by  bilocal formula and take
a very compact form given in \cite{Drummond:2009fd}
\be
\label{levelonetwistor}
J^{(1)\, \mathcal{A}}{}_{\mathcal{B}}= \sum_{i=1}^n \alpha_{i}\, J^{(0) \mathcal{A}}_{i \quad\,\, \mathcal{B}}-
\sum_{i>j} (-1)^{\mathcal{C}} \left (J^{(0) \mathcal{A}}_{i \quad\,\, \mathcal{C}}\, J^{(0) \mathcal{C}}_{j \quad\, \mathcal{B}}
- i\leftrightarrow j \right ),
\ee
where the $\alpha_{i}$ are a priori arbitrary local parameters. Actually, for the undeformed
super-amplitudes, the invariance under the level-one generators \eqref{levelonetwistor} holds for $\alpha_{i}=0$. We will see shortly that our deformation ``turns on'' these parameters $\alpha_{i}$.

We will start by establishing the symmetry properties of the spectrally deformed
four-point amplitude with external physical helicities \eqn{spectralA} 
\be
\label{RRz}
\RR_{4,2} =  \left(-\frac{\langle23\rangle \langle41\rangle}{\langle12\rangle \langle34\rangle}\right)^z  \frac{\delta^{4}(p) \delta^8(q)}{\langle12\rangle \langle23\rangle \langle34\rangle \langle41\rangle}
= \left(-\frac{\langle23\rangle \langle41\rangle}{\langle12\rangle \langle34\rangle}\right)^z  \, 
\mathcal{A}^{\text{MHV}}_{4}
\, .
\ee
For this particularly simple case it is straightforward to use the super-spinor-helicity formalism.  Moving to higher points and helicities, the analogous calculations become very involved. The Gra{\ss}mannian representation will then be more suitable in establishing the sought results on the symmetry properties.

\subsection{Superconformal Symmetries}

Let us begin with the superconformal symmetry of the four-point R-matrix kernel $\RR_{4,2}$.
In the super-spinor-helicity space  
the generators of the $\alg{su}(2,2|4)$  algebra, \emph{i.e.}~the Poincar\'e, dilatation, special-conformal
transformations and their super-partners, have the following
single-particle representation
as differential operators of degree zero, one and two
\begin{align}
\label{scgen}
p_{i\, \a\da}&=\la_{i\, \a}\,\tla_{i\, \da}\, ,\qquad q^{A\,\a}_{i}=\la_{i}^{\a}\, \eta^{A}_{i}\, , 
\qquad \bar q_{i\, A}^{\da}= \tla^{\da}_{i}\,\partial_{i\, A}
\nn\\
m_{i\, \a\b}&= \la_{i(\a}\, \partial_{i\, |\b)}\, , \qquad {\bar m}_{i\, \da\db}= \tla_{i(\da}\, 
\partial_{i\, |\db)}\, , \qquad   d_{i}=\half\la^{\a}_{i}\partial_{i\, \a} +
\half\tla^{\da}_{i}\partial_{i\, \da} +
\eta^{A}_{i}\partial_{i\, A} + 1\, , \nn \\
k_{i\, \a\da}&= \partial_{i\, \a}\, \partial_{i\, \da}\, ,\qquad
s_{i\, A\a}= \partial_{i\, \a}\, \partial_{i\, A}\, , \qquad
{\bar s}^{A}_{i\, \da}= \eta^{A}_{i}\, \partial_{i\, \da}\, , \nn\\
r^{A}_{j\, B}& = -\eta^{A}_{i}\partial_{i\, B}+ \quarter \delta^{A}_{B}\, \eta^{C}_i\, \partial_{i\, C}\, , \qquad c_{i}=1 + \half \la^{\a}_{i}\partial_{i\, \a} - \half \tla^{\da}_{i}
\partial_{i\, \da} - \half \eta^{A}_{i}
\partial_{i\, A} 
\, .
\end{align}
As the amplitude $\mathcal{A}^{\text{MHV}}_{4}$ is superconformally invariant, the following symmetries of the deformed four-point amplitude 
\be
\{p, \, m, \, \bar m, \, d, \, q,\, \bar q,\, \bar s\}\, \RR_{4,2} =0\, ,
\ee
are manifest due to the fact that the deformation factor 
\be
\mathcal{K}:=\left(-\frac{\langle23\rangle \langle41\rangle}{\langle12\rangle \langle34\rangle}\right)^z 
\ee
is clearly scale- and Poincar\'e-invariant,
and does not depend on the $\tla_{i}$ or $\eta_{i}$. Therefore the only non-manifest symmetries
of $\RR_{4,2}$ are the special conformal $k_{\a\da}$ and the superconformal $s_{A\alpha}$ 
transformations. Due to
the structure of the superconformal algebra it suffices to consider only 
one of them, say $k_{\a\da}$:
\be
k_{\a\da}\RR_{4,2}=\sum_{i=1}^{4} \left (\partial_{i\, \da}\, \mathcal{A}_{4}^{\text{MHV}} \right )\,
\partial_{i\, \a}\, \mathcal{K} 
= \left(\frac{\partial}{\partial p^{\b\da}}\, \delta^{4}(p)\right)\,
\frac{\delta^{8}(q)}{\langle12\rangle \langle23\rangle \langle34\rangle \langle41\rangle}\, 
\sum_{i=1}^{4}\, \la_{i}^{\b}\, \partial_{i\, \a}\, 
\mathcal{K} \, .
\ee
Noting that 
\be
\sum_{i=1}^{n}\la_{i\, \b}\, \partial_{i\, \a}= m_{\a\b}- \half \epsilon_{\a\b}\, \sum_{i=1}^{n}
\la^{\g}_{i}\, \partial_{i\, \g}\,,
\ee
and using the manifest invariance of $\mathcal{K}$ under $m_{\a\b}$ and scale transformations $\sum_{i}\la^{\g}_{i}\, \partial_{i\, \g}$
in the $\la_{i}$ spinors, we conclude
\be
k_{\a\da}\RR_{4,2}= 0 \, .
\ee

These arguments easily lift to the most general $n$-point MHV
amplitudes with continuously deformed helicity assignments on the external legs. From \eqn{Rn2deform},
the deformation of $\RR_{n,2}$ away from the MHV $n$-point super-amplitude 
$\mathcal{A}^{\text{MHV}}_{n}$ is holomorphic, i.e.~comprised purely of helicity spinors $\la_{i}^{\a}$. Superconformal symmetry is hence manifest as soon as we have global scaling symmetry of the correction
term in the $\la_{i}$ spinors. One has, {\it cf}
\eqref{Rn2deform},
\be
\label{In2}
\mathcal{F}_{n,2}= \left(\frac{\langle n-1 \, n\rangle }{\langle 1n-1\rangle}\right)^{\zeta_{0}}\left(-\frac{\langle 13\rangle }{\langle 23\rangle}\right)^{\zeta_{1,3}} \left(-\frac{\langle 12\rangle }{\langle 13\rangle}\right)^{\zeta_{2,3}}\prod_{i=4}^n \left(\frac{\langle i-2\,\, i-1\rangle \langle1\, i\rangle}{\langle i-1\, i\rangle \langle 1\, i-2\rangle}\right)^{\zeta_{1, i}}\left(\frac{\langle 1\, i-1\rangle}{\langle 1\, i\rangle}\right)^{\zeta_{2, i}}.
\ee
Invariance under a global scaling in $\lambda_{i}$ is obvious, and we conclude 
\be
\{m,\, {\bar m},\, p,\, q,\,
{\bar q};\, d,\, k, s, {\bar s};\,
c,\, r\}\, \RR_{n,2}=0 \,  ,
\ee
i.e.~the deformed $n$-point MHV amplitudes are superconformally invariant. Note that only
the total central charge $c$ is conserved, in contradistinction to undeformed amplitudes, for which the central charge individually vanishes for all particles.

Actually, one can easily infer superconformal invariance of the most general deformed on-shell diagram by considering its Gra\ss mannian formulation in super-twistor space. 
Indeed, independently of the integration measure, the delta functions appearing in \eqref{deformed.amplitudes} are superconformally invariant. This can be verified straightforwardly by acting with
 the level-zero generators \eqref{level_zero}
in the super-twistor representation \eqref{generators}.

\subsection{Yangian Symmetries}

Let us now turn to the Yangian symmetries of the deformed on-shell diagrams. The level one generators of the Yangian algebra $Y[\alg{su}(2,2|4)]$ are given by \eqref{levelonetwistor}. 
Once again we start with the
deformed amplitude $\RR_{4,2}$. Focusing on the bi-local term, it is advantageous to consider the action of the level-one supersymmetry generator
$q^{(1)\, A}_{\a}$ in the super-helicity spinor representation \cite{Drummond:2009fd}
\be
q^{(1)\, A}_{\a} = \sum_{i>j} \left \{\, m^{\g}_{i\, \a}\, q_{j\, \g}^{A}- \half
\, (d_{i}+c_{i})\, q_{j\, \a}^{A}+p_{i\, \a}^{\db}\, {\bar s}^{A}_{j\, \db} +q^{B}_{i\, \a}
\, r_{j\, B}^{A} - i\leftrightarrow j \right \} \,.
\ee
We see from \eqn{scgen}
that $q^{(1)\, A}_{\a}$ is a first order differential operator. Consider
its action on $\RR_{4,2}= \mathcal{A}^{\text{MHV}}_{4}\, \mathcal{K}$.
From the invariance of the undeformed amplitude $q^{(1)\, A}_{\a}\, \mathcal{A}^{\text{MHV}}_{4}=0$, together with the $\tla^{\da}_{i}$ and $\eta^{A}_{i}$ independence of the deformation term $\mathcal{K}$ of \eqn{In2}, it follows that
\begin{align}
q^{(1)\, A}_{\a}\, \RR_{4,2}& = \mathcal{A}^{\text{MHV}}_{4}\,
\sum_{i>j}\left\{\, ( -m_{i\, \g\a}-\half \epsilon_{\a\g}\, 
\la_{i}^{\delta}\partial_{i\, \delta})\,
q^{A\g}_{j} - i\leftrightarrow j \right \}\, \mathcal{K} \nn\\
& = - \mathcal{A}^{\text{MHV}}_{4}\,
\sum_{i>j}\left\{\, \la_{j}^{\g}\, \eta^{A}_{j}\, \la_{i\, \a}\, \partial_{i\, \gamma}
- i\leftrightarrow j \right \}\, \mathcal{K}\, .
\end{align}
We now add, by virtue of $q\,\delta(q)=0$, a suitable vanishing contribution, namely the first term in the bracket 
summed freely over all indices $i$ and $j$, and find
\be
q^{(1)\, A}_{\a}\, \RR_{4,2} =  2z \, \RR_{4,2}\sum_{i>j} \eta_{j}^{A}\, \la_{i\,\a}\,
 \Bigl [\delta_{i,2}\,\left(\frac{\vev{ij}}{\vev{1i}} - \frac{\vev{3j}}{\vev{3i}}\right)
 +\delta_{i,3}\,\left(\frac{\vev{4j}}{\vev{4i}} - \frac{\vev{2j}}{\vev{2i}}\right)
 +\delta_{i,4}\,\left(\frac{\vev{3j}}{\vev{3i}} - \frac{\vev{1j}}{\vev{1i}}\right)
    \Bigr ] \, .
\ee
Further manipulating this expression using total $q$-conservation, one may reduce
it to the compact form
\be
q^{(1)\, A}_{\a}\, \RR_{4,2} =  2z \, 
\Bigl ( q_{2}^{A\a} + q_{4}^{A\a} \Bigr ) \, \RR_{4,2}\, .
\ee
We conclude that indeed $\RR_{4,2}$ is Yangian invariant with a locally $z$-deformed 
level-one generator in the sense of \eqn{levelonetwistor}
\be
\label{alphas4}
J^{(1)}\, \RR_{4,2} = 0 \, \qquad \text{with}\quad
\alpha_{i}= 2z\{0,1,0,1\}\, .
\ee
Of course the $\alpha_{i}$ are only determined up to an overall constant shift, in view of the
level zero symmetry $J^{(0)}\, \RR_{4,2}=0$. Here we
used this freedom to put $\alpha_{1}=0$.


Let us now consider the general deformed on-shell diagrams describing the top cell, see section \ref{Sec:ch3_ManyPoint}. To this purpose we move to the Gra\ss mannian formulation.
For later convenience we write
 \begin{equation}
 \label{Rtop}
 \mathcal{R}^{\mathrm{top}}_{n,k} = \int   \prod_{a=1}^{k} \prod_{i=k+1}^{n} dc_{ai}  \, F(C,\{\zeta\}) \,   \delta_a\,, \qquad \qquad \delta_a \equiv \delta^{4|4}\left( \mathcal{Z}^{\mathcal{A}}_a - \sum_{i=k+1}^n c_{ai} \mathcal{Z}^{\mathcal{A}}_{i}\right),
\end{equation}  
where with $F(C,\{\zeta\})$  we indicate  the Gra\ss mannian measure, which depends on the variables $c_{ai}$ of the matrix $C$  \eqref{C-matrix} and the face spectral parameters of the dressed on-shell diagram under consideration. 
The level-one generators \eqref{levelonetwistor} 
acting on $\RR_{n,k}^{\mathrm{top}}$ should yield
\begin{equation}
J^{(1)\, \mathcal{A}}{}_{\mathcal{B}}\, \RR^{\mathrm{top}}_{n,k}  = 0 \,.
\end{equation}
We transform this condition into a differential equation for the function $F(C,\{\zeta\})$, similar to the one obtained in \cite{Drummond:2010uq}.
The actual tricks that have to be performed were already applied in section \ref{Sec:ch1_RmatrixGrass}. More specifically, one has to use the commutation relations for the oscillators, the generalization of \eqref{trade} to any $n$ and $k$, and integration by parts.
We then arrive at an equation of the type
\begin{equation}
\label{eq.diff_yangian}
\sum_{a=1}^k \sum_{i=k+1}^n \int  \prod_{b=1}^{k} \prod_{j=k+1}^{n} dc_{bj} \,(\mathcal{DF}_{a,i})\,  c_{ai} \, \mathcal{Z}^{\mathcal{A}}_i \, \delta_1 \dots (\partial_{\mathcal{B}}\delta_a) \dots \delta_k  = 0\,,
\end{equation}
where we have defined
 \begin{equation}
\mathcal{DF}_{a,i} = -(n-2i+2a + \alpha_i - \alpha_a) F(C,\{\zeta\}) + \sum_{b\neq a =1}^k \mathrm{sign}(b - a) F_{ab}  \frac{c_{bi}}{c_{ai}}+ \sum_{j\neq i =k+1}^n \mathrm{sign}(i - j) F_{ij}  \frac{c_{aj}}{c_{ai}}  \,.
 \end{equation}
The $F_{ab} $ and $F_{ij}$ are derivatives of the function $F(C,\{\zeta\})$ with respect to the variables $c_{ai}$ and take the form 
\begin{eqnarray}
F_{ab}   &=&
    \sum_{i=k+1}^n c_{ai} \frac{\partial}{\partial c_{bi}} F(C,\{\zeta\})  \,,\\
   F_{ij}&=&    \sum_{a=1}^k c_{ai} \frac{\partial}{\partial c_{aj}} F(C,\{\zeta\})   \,.
\end{eqnarray}
 Similar to \eqref{independence} we have that all $\mathcal{Z}^{\mathcal{A}}_{i}(\partial_{\mathcal{B}}\delta_a)$ are linearly independent. Thus, insisting on \eqref{eq.diff_yangian}, we have to set for all $a=1,\ldots, k$ and $i=k+1,\ldots, n$
  \begin{equation}
 \label{eq.YangianF}
\mathcal{DF}_{a,i} = 0 \,.
 \end{equation} 
 This gives a set of differential equations for any $n$ and $k$, which could be solved in principle. Instead, we use the results we already obtained for the deformed diagrams, and check if they satisfy the conditions \eqref{eq.YangianF}. Let us first focus on the case of $k=2$ given by \eqref{Rn2deform}, which is relevant to the deformations of the MHV amplitudes. We observe that for generic values of the face spectral parameters $\zeta_{i, j}$ the equations \eqref{eq.YangianF} do not hold. However, they {\it are} satisfied if we additionally impose some constraints on the $\zeta_{i, j}$. Explicitly we have
\begin{equation}
\label{Yangianinvariance}
\zeta_{2,i} + \zeta_{1,i-1} - \zeta_{2,i-1} - \zeta_{1,i+1} = 0\,,  \qquad \mathrm{for} \qquad \, i = 4, \dots, n  \,,
\end{equation}
with $\zeta_{1,n+1}=\zeta_0$, 
 see again appendix \ref{app.quivers}.
Interestingly, we recognize these constraints to be identical to the ones we derived in section \ref{Sec:ch3_Moves} when insisting on the validity of the square moves! To be more precise, the requirement of Yangian invariance imposes precisely the same conditions on the spectral parameters as the ones necessary to ensure invariance of the measure under all possible cluster mutations! We observed this to hold true for all top cells that we checked. However, we are currently lacking a general proof of this observation. It turns out that the number of independent face spectral parameters, after solving all constraints \eqref{Yangianinvariance}, equals $n-1$. In appendix \ref{app.alphas} we collect the values of the parameters $\alpha_i$, for which \eqref{eq.YangianF} holds, for various MHV deformed amplitudes.


\subsection{Generalized Yang-Baxter Equation}

\begin{figure}[ht]
\begin{center}
\scalebox{0.35}{\input{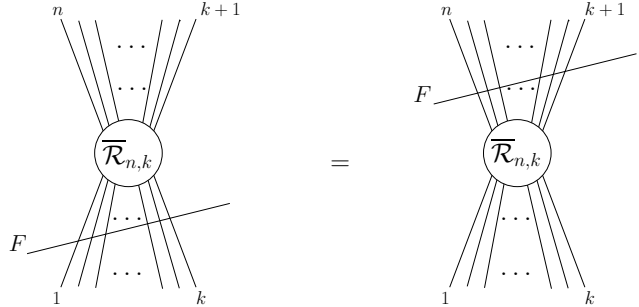}}
\end{center}
\caption{Generalized Yang-Baxter equation. $F$ stands for the fundamental representation of $\mathfrak{gl}(N|M)$.}
\label{Fig:GeneralisedYBE}
\end{figure}

There exists a further method for deriving the constraints on the face spectral parameters imposed by Yangian symmetry. Let us return to the graphical form of the Yang-Baxter equation and bootstrap equations investigated in sections \ref{Sec:ch1_FourPoint} and \ref{Sec:ch2_ThreePoints}, respectively. In both cases we can describe these equations by taking a ``fundamental particle'' and shifting its corresponding line through the four- or three-point vertices, $\mathcal{R}$ and $\mathcal{R}_{\circ}$, $\mathcal{R}_{\bullet}$, respectively. One may generalize these pictures and write down a general equation for an object with $k$ incoming and $n-k$ outgoing particles, see figure \ref{Fig:GeneralisedYBE}.
We will call the equation associated with this picture the {\em generalized Yang-Baxter equation}. In algebraic form we may write it as
\begin{equation}\label{gen.YBE}
\overline{\mathcal{R}}_{n,k}\,\mathbf{L}_{k}(z_k)\,\mathbf{L}_{k-1}(z_{k-1})\ldots \mathbf{L}_{1}(z_1)=\mathbf{L}_{k+1}(z_{k+1})  \ldots \mathbf{L}_{n-1 }(z_{n-1})\, \mathbf{L}_{n }(z_n)\,\overline{\mathcal{R}}_{n,k}\,,
\end{equation}
where $\mathbf{L}_{i}(z_i)$ are again the R-matrices given by  \eqref{Ri3}, which intertwine the fundamental representation with a general representation. For any $n$ and $k$, \eqref{gen.YBE} is a linear equation for $\overline{\mathcal{R}}_{n,k}$. It defines $\overline{\mathcal{R}}_{n,k}$ as  an intertwiner of two representations of the Yangian. For the moment, let us keep the spectral parameters $z_i$ for $i=1,\ldots,n$ unspecified, i.~e.~they can be any complex number. 

The expressions \eqref{deformed.amplitudes} for the deformed on-shell diagrams are natural candidate solutions to \eqref{gen.YBE}. Namely, since they may be encoded by on-shell diagrams, we can use the bootstrap equations to shift the fundamental line step by step through all vertices of the on-shell diagram. However, some care is needed, as these bootstrap equations are not all independent. Note that outgoing particles from one vertex can be incoming particles for another vertex. This means that spectral parameters are ``propagating'' during the shifting process. Actually, this is precisely what is needed, as it yields once more the constraints on the face spectral parameters we observed earlier. In particular, the spectral parameters $z_i$ in \eqref{gen.YBE} are related to each other, in generalization of the mechanism we noticed in the case of the Yang-Baxter as well as bootstrap equations. In e.g.~the former case, we found $z_3=z_1$ and $z_4=z_2$. A general prescription on how to fix spectral parameters in \eqref{gen.YBE} can be encoded into the form of Bethe equations, and will be presented in \cite{NRYMtoappear} based on considerations from the Algebraic Bethe Ansatz approach. 

We studied examples of on-shell diagrams related to top cells of $G(n,k)$ and found that the relations between face spectral parameters stemming from the generalized Yang-Baxter equation are indeed identical to the ones resulting from directly demanding Yangian invariance. This statement is far from trivial, and it would be very instructive to find a general proof of this fact. Interestingly, the generalized Yang-Baxter equation imposes the same kind of relations also for all lower cell examples we studied. However, for the moment we cannot comment on the Yangian invariance for lower cells, as we lack an equivalent of \eqref{eq.YangianF}.



\section{Spectral Regularization of Loop Amplitudes}
\label{Sec:ch5_OneloopReg}

Clearly there is no compelling reason to deform the tree-level amplitudes by spectral parameters. While mathematically interesting, it even appears to be ``physically wrong'' to deform the helicities of physical particles. In this section we would like to demonstrate that the situation drastically changes when taking into account radiative corrections. 

The undeformed on-shell three-point vertices have been used in \cite{ArkaniHamed:2012nw} to also construct the formal loop integrands of $\mathcal{N}=4$ SYM. We will sketch this rather intricate procedure below. In the approach of  \cite{ArkaniHamed:2012nw}, the subsequent step \emph{integrand}$\to$\emph{integral} is highly non-trivial, and, to our understanding, in some sense even a priori ill-defined at generic loop order. The reason is that on-shell loop amplitudes in massless gauge theories show infrared divergences when loop momenta become soft and/or collinear to some external momentum.  An efficient and consistent regularization of loop momentum integration is required. Many different methods have been studied in the past decades, all having
certain advantages as well as drawbacks. As for the latter, a common feature is the breaking or modification of some symmetry of the quantum field theory. Dimensional regularization, in its various versions
\cite{'tHooft:1972fi,Bollini:1972ui,Ashmore:1972uj,Cicuta:1972jf,Ellis:1985er,Bern:1991aq,Bern:2002zk} 
is by now the most frequently used method to render divergent integrals finite. It is often used in conjunction with dimensional reduction, which preserves space-time
supersymmetry \cite{Siegel:1979wq}. 
Some years ago, an AdS-inspired mass regularization was proposed in \cite{Alday:2009zm}, which maintains \emph{extended} dual conformal
symmetry. Very recently, a novel regularization that manifestly preserves dual conformal symmetry has been introduced in \cite{Bourjaily:2013mma}.
Interestingly, however, in all these schemes conventional conformal symmetry is broken. 

As a first example, let us consider in this section the simplest infrared-divergent case, namely the one-loop four-point amplitude. In the $\mathcal{N}=4$ theory it factorizes into the
tree-level amplitude times the so-called scalar box integral $I^{\mbox{\tiny box}}_4$
\begin{equation}
\label{oneloop}
\mathcal{A}_{4,2}^{\mbox{\tiny 1-loop}}=\mathcal{A}_{4,2}^{\mbox{\tiny
tree}}\, I^{\mbox{\tiny box}}_4 \,,
\end{equation}
where the latter reads
\begin{equation}
\label{Ibox4}
I^{\mbox{\tiny box}}_4 = s_{12}\, s_{23} \, \int d^4 q \frac{1}{q^2 (q+p_1)^2
(q+p_1+p_2)^2 (q-p_4)^2} \,.
\end{equation}
Here $s_{12}=(p_1+p_2)^2$ and $s_{23}=(p_2+p_3)^2$ are Mandelstam variables. This integral shows IR divergences, which require special attention.
In the dimensional regularization scheme the number of space-time dimensions is formally modified to $D=4-2\epsilon_{\mbox{\tiny IR}}$ with
$\epsilon_{\mbox{\tiny IR}} < 0$. The infrared singularities then show up as poles in the parameter
$\epsilon_{\mbox{\tiny IR}}$ as $\epsilon_{\mbox{\tiny IR}}\to 0$.
The result for the box integral is well-known and reads
\begin{equation}\label{dimeregbox}
I^{\mbox{\tiny box}}_4 =  \frac{2}{\epsilon_{\mbox{\tiny IR}}^2}\left[
\left(\frac{-s_{12}}{\mu^2}\right)^{-\epsilon_{\mbox{\tiny IR}}}+
\left(\frac{-s_{23}}{\mu^2}\right)^{-\epsilon_{\mbox{\tiny IR}}}\right]-
\log^2 \left(\frac{s_{12}}{s_{23}}\right)-\frac{4}{3}\pi^2 +
\mathcal{O}(\epsilon_{\mbox{\tiny IR}}).
\end{equation}
It depends on the Mandelstam variables as well as on a regularization scale $\mu$.
Conformal as well as dual-conformal symmetry are manifestly broken due to the explicit appearance of the scale $\mu$.

Our main objective in this section is to avoid this and similar symmetry-breaking schemes, and to present instead a novel, natural way to regulate loop integral while respecting superconformal symmetry.
We will show in the following that the spectral parameter can be used to this purpose, at least at the one-loop level. We call the new scheme spectral regularization.
As we shall demonstrate our spectral regularization scheme introduces a self setting
dynamical scale, set by the kinematical data of the amplitude, to regulate the IR-divergent 
integrals. It is akin to the analytical regularization \cite{AnalyticReg1, AnalyticReg2}
and respects conventional superconformal
symmetry.

Let us start by specifying the setup in which we will be working in the following.
For sake of mathematical precision we find it more convenient to abandon the Gra\ss mannian formalism and 
use instead the generalization of the Parke-Taylor formulas encoding the deformed three-point vertices of on-shell diagrams \eqref{Rz3}. 
The gluing procedure becomes quite subtle, since we need to perform on-shell integrations over massless particles. 
The usual parametrization of massless momenta in terms of spinor helicity variables
 \begin{equation}\label{massless}
p^{\alpha \dot\alpha}=\lambda^{\alpha}\tilde\lambda^{\dot \alpha}
\end{equation}
is not suitable here because of its $GL(1)$ invariance
\begin{equation}
\lambda^{\alpha}\to \beta \lambda^{\alpha} \,, \mbox{ and } \tilde\lambda^{\dot\alpha}\to\beta^{-1}\tilde\lambda^{\dot \alpha} \,.
\end{equation}
In order to avoid this redundancy, we express the spinor helicity variables in terms of three independent quantities $t, x, y$ in the following manner
\begin{equation}
\lambda^{\alpha}=\left(\begin{tabular}{c}$t$\\$t\, x$\end{tabular}\right)\,,\qquad\qquad \tilde\lambda^{\dot\alpha}=\left(1\,\,y \right) \,.
\end{equation}
With this parametrization the massless momentum \eqref{massless} takes the form
\begin{equation}
p^{\alpha\dot\alpha}=\left( \begin{tabular}{cc}$t$&$t \,y$\\$t\,x$&$t\,x\,y$\end{tabular}\right) \,.
\end{equation}
We will not impose any additional constraints on the variables $t$, $x$ and $y$ and allow them to be any complex numbers, 
leading to a complexified Minkowski momentum space. The three-dimensional on-shell integration could then formally be written as
\begin{equation}\label{onshellmeasure}
\int \frac{d^3 p}{2 p_0}=\int \frac{t}{4}\,dt \, dx\, dy \,,
\end{equation}
where we did not specify the precise domain of integration. Indeed, in the following we will evaluate all integrals formally, assuming that the integrations localize on the support of delta functions.
At the very end of the calculation, having saturated all delta functions, we will rewrite the final result as an integral over an off-shell, real 4-momentum with measure $\int d^4 p$.

Let us remark that one could also go to (2,2) signature and parametrize momenta with real $(t,x,y)$. 
In this case all integrals would be over intervals on the real line. After saturating all delta functions one would end up with an integral over real off-shell 
momentum in (2,2) signature. A Wick rotation would translate the result to Minkowski space. 
Both methods should give the same final integral. However, the calculation in (2,2) signature forces us to deal with very many absolute values. E.~g.~in \eqref{onshellmeasure}, we should replace $t\to |t|$. 
This is technically much more involved. We will therefore formally perform the calculation in complexified Minkowski space, and in particular drop all absolute values for the integration variables. This leads to sensible results in all cases we have investigated to date. However, it would be important to gain a deeper understanding of this rather ad-hoc procedure.

Similarly, there is a simple way to parametrize any off-shell momentum using the $(t,x,y)$ variables. 
In order to do so we introduce a reference on-shell momentum $(\tilde t,\tilde x,\tilde y)$ and some parameter $\tau$. 
At the end of any calculation, the result should not depend on the particular reference momentum we choose.
Then the off-shell momentum can be rewritten as
\begin{equation}\label{ch4.off.par}
q^{\alpha\dot\alpha}=p^{\alpha\dot\alpha}+\tau\, \tilde p^{\alpha\dot\alpha}=
\left( \begin{tabular}{cc}$t+\tau \tilde t$&$t \,y+\tau \, \tilde t \,\tilde y$\\$t\,x+\tau \, \tilde t \,\tilde x$&$t\,x\,y+\tau\, \tilde t\, \tilde x\,\tilde y$\end{tabular}\right).
\end{equation}
Note that 
\begin{equation}
q^2=\det q^{\alpha\dot\alpha}=t\,\tilde t\, \tau(x-\tilde x)(y-\tilde y) \,.
\end{equation}
The off-shell integration that we will encounter in the final result of the loop calculation, dropping constants and absolute values, then becomes 
\begin{equation}
\label{onshellQ}
\int \frac{d^4 q}{q^2}=\int t \, dt\, dx\, dy\, \frac{d\tau}{\tau} \,.
\end{equation}
Before starting any calculation of on-shell diagrams, let us present some more details on the ingredients we will need, expressing them in the new variables.
Specifically, using $(t,x,y)$, the momentum-conservation delta functions take a particularly simple form 
\begin{equation}\label{ch4.cons.mom}
\delta^4(p^\mu)=\delta^4(p^{\alpha\dot\alpha})=\delta\left(\sum_i t_i\right) 
\delta\left(\sum_i t_i\, x_i\right)\delta\left(\sum_i t_i\,y_i\right)\delta\left(\sum_i t_i\,x_i\,y_i\right) \,.
\end{equation}
Furthermore, the angle and square brackets turn into
\begin{eqnarray}
\label{AngleSquare}
\langle i\,j\rangle =t_i\, t_j (x_j-x_i) \nonumber \\
\,[ i\,j ]=(y_j-y_i) \,.
\end{eqnarray}

We are now ready to use the  variables introduced above to evaluate amplitudes given by on-shell diagrams. 
As a warm up, consider the deformed tree-level four-point amplitude.  
The steps of this calculation will not differ too much from the ones of more complicated on-shell diagrams in the following sections. 

\begin{figure}[ht]
\begin{center}
\scalebox{0.35}{\input{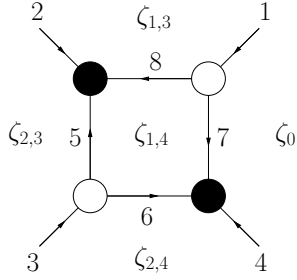}}
\end{center}
\caption{Four-point tree-level.}
\label{Fig:FourTxy}
\end{figure}

Let us start with the form of the three-point deformed amplitudes written explicitly in $(t,x,y)$ variables. 
We will decorate them with arrows indicating now the flow of helicities in the diagram. It is important to note that these orientations are not related to the perfect orientations discussed in previous sections. In the case where all arrows are incoming we have 
\begin{eqnarray}\label{ch4.def3b}
\mathcal{R}_{\bullet}&=&\frac{\delta^{4}(P^{\alpha\dot\alpha})\delta^{8}(Q^{\alpha A})}{t_1^{2+z_1}t_2^{2+z_2}t_3^{2+z_3}(x_2-x_1)^{1+z_3}(x_3-x_2)^{1+z_1}(x_1-x_3)^{1+z_2}}\\ \label{ch4.def3w}
\mathcal{R}_{\circ}&=&\frac{\delta^{4}(P^{\alpha\dot\alpha})\delta^{4}(\tilde Q^{ A})}{(y_2-y_1)^{1-z_3}(y_3-y_2)^{1-z_1}(y_1-y_3)^{1-z_2}}
\end{eqnarray}
with the constraint $z_1+z_2+z_3=0$. By  flipping the direction of an arrow, one changes the sign of the related $z_i$. 
In the above formulas the conservation of momentum is given by \eqref{ch4.cons.mom}, while the super-momentum conservation delta functions are given by
\begin{eqnarray}
\delta^8(Q^{\alpha A})&=&\delta^4\left(\sum_i t_i\, \eta_i\right)\delta^4\left(\sum_i t_i\,x_i\,\eta_i\right)\,, \\
\delta^4(\tilde Q^{A})&=&\delta^4\left ((y_2-y_1)\eta_3+(y_3-y_2)\eta_1+(y_1-y_3)\eta_2\right) \,.
\end{eqnarray}
The on-shell diagram encoding the four-point tree-level amplitude is depicted in figure \ref{Fig:FourTxy}.
The calculation can be broken down to the following precise steps. First, we focus on the 16 bosonic delta functions. 
After singling out 4 delta functions that will produce total momentum conservation, the remaining ones allow us 
to find a solution for the 12 $(t_i,x_i,y_i)$ variables, where $i=5,\ldots,8$ label the internal particles. 
As there is one three-dimensional on-shell integration for each such particle, the left-over 12 bosonic delta functions are saturated\footnote{There are two distinct solutions to the bosonic delta functions. 
They are exactly the two different solutions to the quadruple cut of the one-loop box integral.}. 
In the second step we turn to the 24 fermionic delta functions. 
Since we have 16 fermionic integrations, we are left with 8 unintegrated delta functions which express the conservation of total super-momentum. 
Finally, we collect the Jacobians produced by this procedure. A factor $\left(\prod_{i=5}^8 t_i\right)$ comes from the integration measure  
and the denominators stem from the three-point vertices using formulas \eqref{ch4.def3b} and \eqref{ch4.def3w}. 
Using the solution for the $(t_i,x_i,y_i)$ corresponding to the internal particles, we express the final result as a function of the external data, and obtain
 \begin{eqnarray}\label{ch4.R4}
\mathcal{R}_4&=&\frac{\delta^{4}(P^{\alpha\dot\alpha})\delta^{8}(Q^{\alpha A})}{t_1^2\, t_2^2\, t_3^2\, t_4^2 (x_1-x_2)(x_2-x_3)(x_3-x_4)(x_4-x_1)} \, \mathcal{F}(z_i),
\end{eqnarray}
where the multiplicative spectral-parameter deformation reads
\begin{align}
\mathcal{F}(z_i) &= t_1^{-z_1}t_2^{-z_2}t_3^{-z_3}t_4^{-z_4} (x_1-x_2)^{-z_8-z_2}(x_2-x_3)^{z_8}(x_3-x_4)^{-z_8+z_1}(x_4-x_1)^{z_8-z_1-z_4}(x_1-x_3)^{z_2+z_4} \nn\\
&= \left( \frac{\vev{23}\vev{41}}{\vev{12}\vev{34}}\right )^{z_{8}}
\, \left( \frac{\vev{34}}{\vev{41}}\right )^{z_{1}}\, \left( \frac{\vev{13}}{\vev{12}}\right )^{z_{2}}\, \left( \frac{\vev{13}}{\vev{41}}\right )^{z_{4}}\, 
\,.
\end{align}
Using (\ref{AngleSquare}) we reproduce the R-matrix \eqref{spectralA} after setting 
$z_8 =z$ and $z_{1}=z_{2}=z_{3}=z_{4}=0$, which corresponds to the deformed four-point amplitude in the case where all external particles carry physical helicities\footnote{Note that we
have been rather nonchalant about overall signs in this computation.}. Furthermore, identifying the $z_i$ with the face spectral parameters of the general deformed amplitudes \eqref{deformed.amplitudes} for the case $n=2$ as
\be
z_{1}= \zeta_{0}-\zeta_{1,3}\, , \qquad z_{2}= \zeta_{1,3}-\zeta_{2,3}\, , \qquad z_{3}= \zeta_{2,3}-\zeta_{2,4}\, , \qquad
z_{4}=\zeta_{2,4}-\zeta_{0}\, \qquad z_{8}= \zeta_{1,4}-\zeta_{1,3}\,,
\ee
compare to figure \ref{Fig:FourTxy}, we recover the deformed amplitude of \eqref{deformed.amplitudes} where the external particles are
deformed as well. Note that the Yangian invariance condition of \eqref{Yangianinvariance} for the $n=2$
case translates into
\be
\zeta_{2,4}+\zeta_{1,3} = \zeta_{2,3}+ \zeta_{0} \quad \Rightarrow \quad z_{2}+z_{4}=0
\quad \text{or} \quad z_{1}+ z_{3}=0\, ,
\ee
compare to \eqref{equals1} and \eqref{equals2} with $z_i \to s_i$ and a change of orientations.  

\begin{figure}[ht]
\begin{center}
\scalebox{0.25}{\input{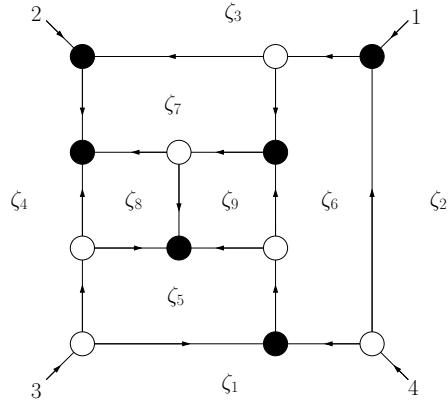}}
\end{center}
\caption{The dressed on-shell diagram for the one-loop four-point amplitude.}
\label{OneLoopPlabic}
\end{figure}

\begin{figure}[ht]
\begin{center}
\scalebox{0.20}{\input{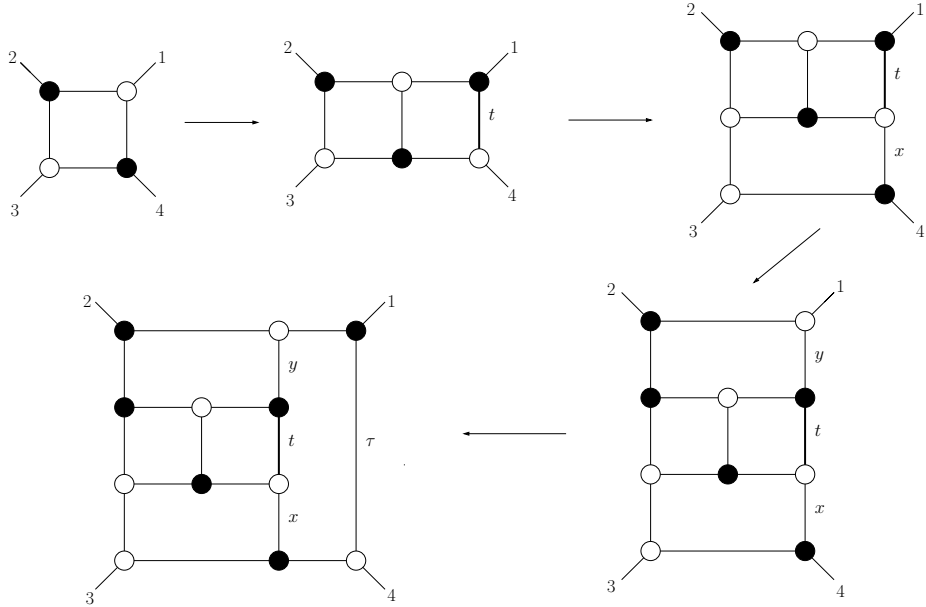}}
\end{center}
\caption{Attaching four bridges.}
\label{Fig:FourBridges}
\end{figure}

Let us proceed to the one-loop correction to the four-point amplitude. 
The proper on-shell diagram can be constructed from the all-loop BCFW recursion relation of \cite{ArkaniHamed:2012nw}. 
Starting from the diagram for the tree-level six-point NMHV amplitude, one has to identify two particles and add a BCFW bridge. 
There is a whole class of equivalent diagrams, which all encode the one-loop result, and are related by moves as was reviewed in section \ref{Sec:ch3_Moves}.
For our calculation we chose the specific representative of this class drawn in figure \ref{OneLoopPlabic}.
One notices that it can be obtained from the tree-level diagram by attaching four BCFW bridges. This process is depicted in figure \ref{Fig:FourBridges}.

\begin{figure}[ht]
\begin{center}
\scalebox{0.25}{\input{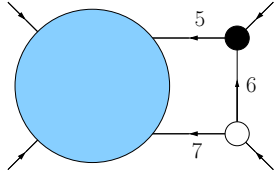}}
\end{center}
\caption{Attaching a single bridge.}
\label{Fig:SingleBridge}
\end{figure}

Again, we can partition the calculation into basic steps, each one consisting in consecutively attaching bridges. 
At each step, the starting point is an expression depending on four momenta, to which we attach two three-point vertices, as shown in figure \ref{Fig:SingleBridge}.

We proceed further along the same lines already described for the tree-level four-point amplitude. 
The only, and important, difference is that now the number of internal variables is larger than the number of delta functions.  
Therefore we cannot fix all internal momenta in terms of external ones. 
By a simple calculation, one can convince oneself that the number of free parameters for every bridge is 1.
After a careful analysis one finds that not all parameters can be left unintegrated. 
There are only three cases allowed: we can keep either $x_5$, $t_6$ or $y_7$ of figure
\ref{Fig:SingleBridge}. 
For the one-loop calculation, which means attaching four bridges, we end up with four unintegrated variables. 
If we want to rewrite the final integral with the use of an off-shell loop momentum, 
we end up with $(t,x,y,\tau)$ according to \eqref{ch4.off.par}. The labels on figure \ref{Fig:FourBridges}  show which parameters are left unintegrated at each step of the bridge-gluing procedure. 
Remembering that for each black vertex all incident $y$ variables are equal and for all white vertices all incident $x$ variables are equal, we see that $(t,x,y)$ yield a parametrization of the 
momentum on the bold line. It is exactly the on-shell momentum of particles we identified with each other while solving the BCFW recursion relation. One observes also that every time we attach a bridge we introduce one new spectral parameter which corresponds to the new face we create. At the end we get a diagram with 9 faces and, correspondingly, 9 face spectral parameters $\zeta_i$. They are marked in figure \ref{OneLoopPlabic}.

The final result is obtained as follows. Let us first suppress the spectral parameter dependence. Then after gluing four bridges  to the tree-level result \eqref{ch4.R4} one indeed obtains the following correction
\begin{equation}
\int t dt\,dx\,dy\, \frac{d\tau}{\tau} \frac{(p_1+p_2)^2 (p_2+p_3)^2}{(q+p_1)^2 (q+p_1+p_2)^2 (q-p_4)^2}
= \int d^4q \frac{s_{12}\,s_{23}}{q^2 (q+p_1)^2 (q+p_1+p_2)^2 (q-p_4)^2}\, ,  
\end{equation}
where we have used \eqref{onshellQ} to go from the parameters $(t,x,y,\tau)$ to the conventional notation for off-shell momenta. The reference momentum 
is  $(\tilde t,\tilde x,\tilde y) = (1, x_4, y_1)$. This expression matches $I^{\mbox{\tiny box}}_4$ of \eqref{Ibox4}.

We can now proceed and bring spectral-parameter dependence into the previous calculation. By using the deformed three-point vertices and following exactly the same steps as before, we finally arrive at
\begin{equation}\label{ch4.1loopint}
\mathcal{A}_4^{\mbox{\tiny 1-loop}}(\zeta_i) =  \mathcal{A}_{4,2}^{\mbox{\tiny
tree}} \int d^4q \, \mathcal{I}^{\mbox{\tiny box}}_4 \, \mathcal{G}(\zeta_i),
\end{equation}
where $\mathcal{I}^{\mbox{\tiny box}}_4$ is the integrand of the underformed scalar box integral $I^{\mbox{\tiny box}}_4$ , which is now modified by the following spectral-parameter dependent factor $\mathcal{G}(\zeta_i)$ 
\begin{eqnarray}
\mathcal{G}(\zeta_i) &=& q^{2(\zeta_3-\zeta_5+\zeta_8-\zeta_7)}  (q+p_1)^{2(\zeta_5-\zeta_8)}   (q+p_1+p_2)^{2(\zeta_8-\zeta_4)}   (q-p_4)^{2(\zeta_7-\zeta_8)}(\langle 34\rangle[12])^{\zeta_5+\zeta_7-\zeta_6-\zeta_8} \nonumber\\
&& \langle 3|q+p_1|2]^{\zeta_4+\zeta_9-\zeta_5-\zeta_7}  \langle p3\rangle^{\zeta_6-\zeta_9-\zeta_1+\zeta_8} \langle p4\rangle^{\zeta_1-\zeta_3} [p2]^{\zeta_6+\zeta_8-\zeta_3-\zeta_9} 
\left(\frac{\tau}{t_4}\right)^{\zeta_2+\zeta_5+\zeta_7-\zeta_3-\zeta_6-\zeta_8}  
\end{eqnarray}
Let us analyze in more detail this important result.
In the first line we see that the four propagators of the scalar box integral are modified by some powers. Furthermore, there is a normalization factor depending only on external particles.
In the second line new terms that depend on the loop momentum appear. We will demand that all of them vanish. One can establish this by fixing all powers appearing in the second line to 0. This gives some relations between the spectral parameters, reducing the parameter space of our problem. One notices that these relations are exactly the same ones described in section \ref{Sec:ch3_ManyPoint}. In other words, we demand that the cluster mutations may be performed for some of the faces in the diagram in figure \ref{OneLoopPlabic}. Importantly, we do not demand all mutations to be possible because it would trivialize our result too much. However, this 
choice also entails the Yangian non-invariance of the construction.
A deeper analysis of the relation between cluster mutations, Yangian invariance and the possibility of regularization is desirable. We postpone this problem to future work.

After elimination of the unwanted terms we are left with the following expression
\begin{eqnarray}\label{ch4.G}
\mathcal{G}(\zeta_i) &=& q^{2(\zeta_3-\zeta_5+\zeta_8-\zeta_7)}  (q+p_1)^{2(\zeta_5-\zeta_8)}   (q+p_1+p_2)^{2(\zeta_8-\zeta_4)}   (q-p_4)^{2(\zeta_7-\zeta_8)}(\langle 34\rangle[12])^{\zeta_4-\zeta_3} \, .
\end{eqnarray}
In order to regularize the integral \eqref{ch4.1loopint} it is sufficient to choose the exponents of the propagators to be all positive. All such choices will give a finite integral. Here we focus on a particular one, where we take all powers on propagators in \eqref{ch4.G} to be equal to $\epsilon$. It fixes the face spectral parameters from the figure \ref{OneLoopPlabic} to be
\begin{equation}
\zeta_1=0\,, \,\,\, \zeta_2=4\epsilon\,, \,\,\,\zeta_3=0\,, \,\,\,\zeta_4=-4\epsilon\,, \,\,\,\zeta_5=-2\epsilon\,, \,\,\,\zeta_6=3\epsilon\,, \,\,\,\zeta_7=-2\epsilon\,, \,\,\,\zeta_8=-3\epsilon\,, \,\,\,\zeta_9=0\,.
\end{equation}
We then arrive at the following modification
\begin{equation}
\mathcal{G}(\epsilon) = \frac{(\langle 34\rangle[12])^{-4\epsilon}}{q^{-2\epsilon}(q+p_1)^{-2\epsilon} (q+p_1+p_2)^{-2\epsilon}(q-p_4)^{-2\epsilon} } \,.
\end{equation}
This is the key result of this section. By attaching consecutive deformed bridges, we were able to derive a spectral parameter modification of the one-loop amplitude.
With our choice of spectral parameters we are left with a result that depends only on one parameter.
Significantly, it is reminiscent of  analytic regularization \cite{AnalyticReg1, AnalyticReg2}.
The regulator scale is set dynamically by the kinematics of the process.
However, in our approach the regularization of the propagators is not chosen ad-hoc, but is derived   in a very natural and meaningful way.
After integration we finally obtain
\begin{equation}
\label{oneloopfinal}
\mathcal{A}_4^{\mbox{\tiny 1-loop}}(\epsilon) =  
\mathcal{A}_{4,2}^{\mbox{\tiny tree}} \, \left(\frac{[34]}{[12]}\right)^{4\epsilon} 
\left[\frac{1}{\epsilon^2}  \left(\frac{s_{12}}{s_{23}}\right)^{-2\epsilon} - \frac{1}{2} \log^2\left(\frac{s_{12}}{s_{23}}\right) - \frac{7}{6} \pi^2 + \mathcal{O}(\epsilon)\right]\,,
\end{equation}
which is to be compared with the dimensionally regulated result \eqref{deformed.amplitudes}.
Hence, the divergences again generate poles in the spectral parameter $\epsilon$, but there is no scale-breaking parameter $\mu$.
Let us stress that in order to regulate the integral it is crucial to assume non-vanishing central charges also for the external particles,   
for which we have chosen $(c_1,c_2,c_3,c_4)=(-4\epsilon,-4\epsilon,4\epsilon,4\epsilon)$.
It should be stressed that the result \eqref{oneloopfinal} is invariant under 
superconformal transformations. The Yangian invariance is necessarily violated in order to
have a regulated integral.



\section{Conclusions and Outlook} 
\label{Sec:conclusions}

In this paper we investigated the introduction of spectral parameters into the perturbative scattering amplitude problem of $\mathcal{N}=4$ SYM. We managed to consistently deform arbitrary on-shell diagrams. We find it exciting that this is possible, and actually very natural. We argued that it brings the amplitude problem technically and conceptually closer to the integrable spectral problem. However, we did not yet gain a good understanding on how to recombine the deformed diagrams into deformed generic perturbative amplitudes. Clearly this is the crucial next step for rendering our approach useful, and in particular in order to proceed to higher loop levels.

We anticipate that numerous readers might still not be convinced that it is at all necessary to introduce a spectral parameter into the amplitude problem. However, we can easily dispel such a skepticism. Here is the {\it argument:} It is well known that the cusp anomalous dimension appears in both the spectral problem as well as in the (logarithm of) the four-point amplitude. An exact equation for this quantity has been derived using the integrable structure of the spectral problem \cite{Beisert:2006ez}. Looking at the structure of this cusp equation, we claim that it is impossible to eliminate the spectral parameter (there it is encoded into the letter $t$, after a Fourier-Laplace transform from $u=i z$) from the equation, and to write a simpler equation for the cusp as a function of only the coupling constant $g=\sfrac{\sqrt{\lambda}}{4 \pi}$. {\it q.e.d.} It is thus an important open challenge for {\it any} all-loop/non--perturbative approach to the $\mathcal{N}=4$ amplitude problem to {\it derive} the exact cusp dimension from first principles, as opposed to merely using it as an input. And we just argued that this will be impossible without employing a spectral parameter. It would be exciting if a properly deformed version of the BCFW recursion relation could somehow be turned into the cusp equation.

In this context it is interesting to investigate whether Bethe ansatz methods can directly be applied to the scattering problem. Recall that the cusp equation was derived from an asymptotic all-loop Bethe ansatz. For a first step, albeit in a simplified situation, towards a Bethe ansatz for Yangian invariants see \cite{NRYMtoappear}.

It might be easier to first gain an understanding of the spectral deformation at strong coupling, and in particular in the dual string picture. In fact, a spectral parameter for scattering amplitudes was already used in the literature at strong coupling 
in the dual description of scattering amplitudes via minimal surfaces with light-like
polygonal edges, where a certain Y-system was identified \cite{Alday:2010vh}. See also the recent works \cite{Sever:2012qp,Basso:2013vsa,Basso:2013aha}, where the spectral parameter has been put to further good use. It would be interesting to explore the relation of the integrable structures used in these papers to the ones studied in our present work. 

Finally, the deformed on-shell diagram technique used in this article is interesting for the general theory of quantum integrable models. As we pointed out before, the Quantum Inverse Scattering Method usually centers around four-legged R-matrices, while the said on-shell diagram technique starts from the three-legged Hodges vertices. It would be interesting to further explore this feature for general symmetry algebras, and to investigate the precise relation with the mathematical theory of quantum groups and the Yang-Baxter equation.


\section*{Acknowledgments}
We thank Yuri Aisaka, Niklas Beisert, Lance Dixon, Gregory Korchemsky, Vladimir Mitev,  Nicolai Reshetikhin, Emery Sokatchev, Jaroslav Trnka and especially Rouven Frassek, Nils Kanning, and Yumi Ko for very useful discussions. T.~{\L}ukowski is supported by a DFG grant in the framework of the SFB 647 {\it ``Raum - Zeit - Materie. Analytische und Geometrische Strukturen''}. C.~Meneghelli is partially supported by a DFG grant in the framework of thee SFB 676 {\it Particles, Strings, and the Early Universe}. The work of J.~Plefka and L.~Ferro is supported by the Volkswagen-Foundation.


\appendix

\section{Fourier Transform of Super-Twistor Variables}
\label{app.Fourier}

In section \ref{Sec:Deformationsoffourpoint} we used the fact that a function of the super-twistor variables $\mathcal{Z}_i^{\mathcal{A}}=(\tilde \mu^{\alpha}_{i},{\tilde \lambda}^{\dot\alpha}_{i}, \eta^{A}_{i})$
can be written in spinor-helicity space $(\lambda^{\alpha}_{i},{\tilde \lambda}^{\dot\alpha}_{i}, \eta^{A}_{i})$ through a Fourier transform on the $\tilde \mu^{\alpha}_{i}$. In this appendix we want to give some more detail.  In particular, the $\tilde\mu$-dependent part of the Gra\ss mannian integral is a product of delta functions, \emph{i.e.}
\be
\prod_{a=1}^k \delta^{2}\left(\tilde\mu^{\alpha}_a - \sum_{j=k+1}^n c_{aj} \tilde\mu^{\alpha}_{j}\right),
\ee
which can be written through the integral representation as
\begin{eqnarray}
\prod_{a=1}^k \int d^2\rho_a \, e^{i \left(\rho_a \cdot \tilde\mu_a - \sum_{j=k+1}^n c_{aj} \rho_a \cdot \tilde\mu_j\right)} \,.
\end{eqnarray}
Since we consider $k$ ingoing and $n-k$ outgoing particles in the text, we need to be careful with momentum conservation when starting from the Gra\ss mannian integral, see \eqref{deltafour} for the specific case of four points. Hence we should perform a Fourier transform on the $\tilde\mu_a$ and an inverse Fourier transform on the $\tilde\mu_i$
\begin{eqnarray}
\label{fouriertr}
&& \prod_{a=1}^k   \int d^2\rho_a \, d^2\tilde\mu_a \,  e^{i \rho_a \cdot \tilde\mu_a }  \,
e^{-i  \lambda_a \cdot \tilde\mu_a}  \prod_{j=k+1}^n  \int \, d^2\tilde\mu_j \,  e^{-i \sum_{b=1}^k c_{bj} \rho_b \cdot \tilde\mu_j}  \, e^{i  \lambda_j \cdot \tilde\mu_j} \nn\\
&=& \prod_{a=1}^k  \int d^2\rho_a  \delta^{2}\left(\lambda_a -  \rho_a \right)  \, \prod_{j=k+1}^{n} \delta^{2}\left(\lambda_j - \sum_{b=1}^k \, c_{bj} \rho_b \right)  \nn \\
&=&  \prod_{j=k+1}^{n} \delta^{2}\left(\lambda_j - \sum_{b=1}^k c_{bj} \lambda_b \right) \,, 
\end{eqnarray}
where in the last step we have used the delta functions to perform the integration over the variables $\rho_a$.
Here we considered $(2,2)$ signature for simplicity, where all bosonic variables are real. In Minkowski space with signature $(1,3)$ one has to choose suitable contours of integration, and the transform gets more involved.


\section{Gra\ss mannian Formula from Dual Diagrams}
\label{app.quivers}

In section \ref{Sec:ch3_ManyPoint} we described a gluing procedure that allows us to associate a Gra\ss mannian integral to any dressed on-shell diagram. The most compact form of the result is obtained if one writes it in terms of the face variables. For some applications, however, it is better to know its explicit form in the variables $c_{ai}$. As we already pointed out, the variable change from face variables to variables $c_{ai}$ simplifies the form of the matrix appearing in delta functions, but complicates the form of the integration measure in the Gra\ss mannian integral. As it is highly non-trivial to find this measure in the general case, we give here an algorithm allowing to write it down, at least in case of integrals related to the top cells of the positive Gra\ss mannian. Our algorithm is similar to, but different from the one presented in e.g.~\cite{Keller}. The formulas we will derive are particularly useful when checking the Yangian invariance of expressions given by deformed on-shell diagrams. We used them to derive the results of section \ref{Sec:ch4_Symmetries}.

\begin{figure}[ht]
\begin{center}
\scalebox{0.55}{\input{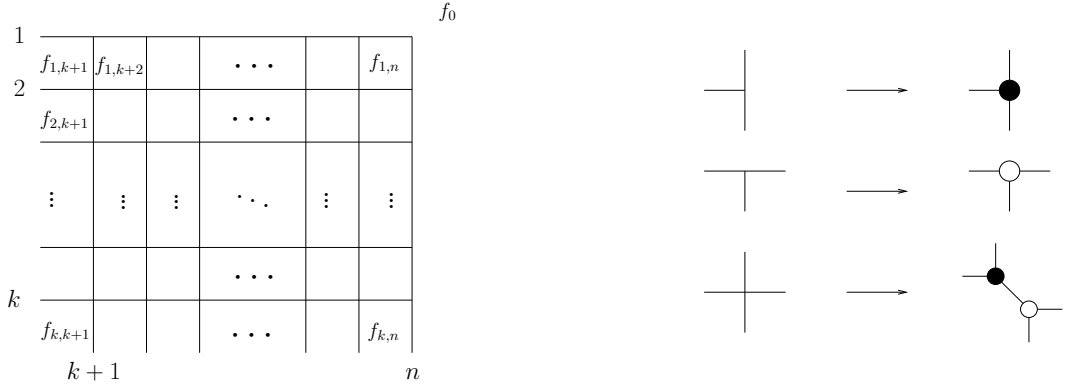}}
\end{center}
\caption{(a) $\Gamma$-diagram relevant for the top cell of Gra\ss mannian $G(n,k)$. (b) Dictionary from $\Gamma$-diagrams to on-shell diagrams.}
\label{Fig:Gamma}
\end{figure}

\begin{figure}[ht]
\begin{center}
\scalebox{0.45}{\input{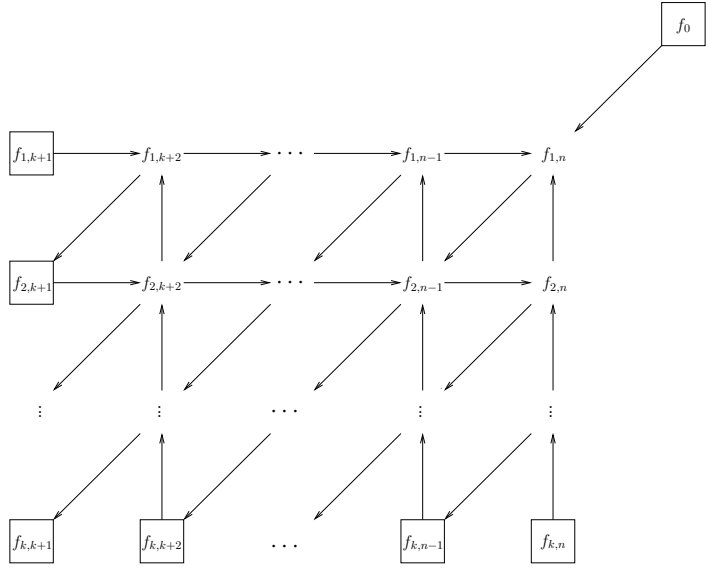}}
\end{center}
\caption{Dual graph for the top cell of $G(n,k)$.}
\label{Fig:Quiver}
\end{figure}

Let us focus on the top cell of the positive Gra\ss mannian $G(n,k)$. In order to find a deformed Gra\ss mannian integral that corresponds to the top cell we need to find a proper dressed on-shell diagram, which for top cells can be easily obtained from the so-called $\Gamma$-diagrams as described in \cite{Postnikov}\footnote{Note that we took a mirror version of $\Gamma$-diagrams compared to \cite{Postnikov} and \cite{Ferro:2012xw}.}. The relevant $\Gamma$-diagrams, together with a dictionary on how to find an on-shell diagram from a $\Gamma$-diagram, were presented already in \cite{Ferro:2012xw} and we depict them in figure \ref{Fig:Gamma}. 
We labeled particles counterclockwise and took particles $i=1,\ldots,k$ to be incoming, while particles $i=k+1,\ldots,n$ to be outgoing. With each face of the $\Gamma$-diagram we associate a face variable $f_{ij}$. Later on we also introduce face spectral parameters $\zeta_{i,j}$ corresponding to the face $(i,j)$. In the case of the top cell, the number of independent face variables is equal to $k(n-k)$, which is exactly the dimension of $G(n,k)$.
As we already pointed out in the main text, the dual diagram of any on-shell diagram is an oriented graph. For the top cell of the Gra\ss mannian $G(n,k)$ this dual graph takes a particularly simple form as presented in figure \ref{Fig:Quiver}.
We marked nodes corresponding to external faces of the on-shell diagram. In the language of cluster algebras, they correspond to the so-called frozen nodes.

Let us take a $k\times n$ matrix $C$ as in \eqref{C-matrix}. For $1\leq i\leq k$ and $1\leq j\leq n$, let $A_{i,j}$ be the largest square submatrix of $C$ whose upper right corner is $(i,j)$. Put
\begin{equation}
a_{i,j}=(-1)^{k-i+1}\det A_{i,j} \,.
\end{equation}
We refer to $a_{i,j}$ as $\mathcal{A}$-variables, see \cite{Golden:2013xva} for a short review. In order to find face variables for the diagram related to the top cell we define the extended dual graph as shown in figure \ref{Fig:ExtQuiver}. 
\begin{figure}[ht]
\begin{center}
\scalebox{0.45}{\input{extquiver.pstex_t}}
\end{center}
\caption{Extended dual graph for the top cell of $G(n,k)$.}
\label{Fig:ExtQuiver}
\end{figure}
The extended dual graph consists of all nodes of the original graph and new nodes, which are marked by hexagons. To each node of the extended dual graph we assign one $\mathcal{A}$-variable $a_{i,j}$ as shown in the figure. Notice that the labeling of nodes in figures \ref{Fig:Quiver} and \ref{Fig:ExtQuiver} is different.

The face variables $f_{i,j}$, which in the language of cluster algebra are usually referred to as $\mathcal{X}$-variables, are then given by
\begin{equation}\label{fofa}
f_{i,j}=\prod_{(m,n)\to (i,j)}a_{m,n} \left(\prod_{(i,j)\to (m,n)}a_{m,n} \right)^{-1}\,,
\end{equation} 
where the first product is over all $\mathcal{A}$-variables for faces of extended dual graph connected to the face $(i,j)$ by an incoming arrow, while the second product is over all $\mathcal{A}$-variables for faces connected by outgoing arrows. Additionally, from the fact that the product over all face variables has to be equal to 1, we find
\begin{equation}
f_{0}=\left(\prod_{i,j} f_{i,j}\right)^{-1}=\frac{a_{2,n-1}}{a_{1,n}}\,.
\end{equation}

\begin{figure}[!ht]
\begin{center}
\scalebox{0.50}{\input{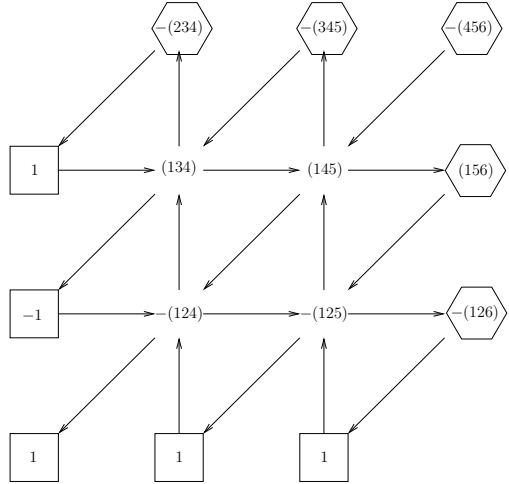}}
\end{center}
\caption{Example of finding face variables for top cells of $G(6,3)$.}
\label{Fig:NMHV6Quiver}
\end{figure}

Let us give an explicit example and show how to find face variables for the top cell of the Gra\ss mannian $G(6,3)$. The extended dual graph for the case $n=6$ and $k=3$, together with its $\mathcal{A}$-variables are given in figure \ref{Fig:NMHV6Quiver},
where by $(i_1 i_2 i_3)$ we denote the determinant of the $3\times 3$ matrix made out of the columns $i_1,i_2,i_3$ of the matrix $C$. Then, according to \eqref{fofa}, we find face variables and using \eqref{deformed.amplitudes} we can write a deformed integral measure associated to the top cell of $G(6,3)$ in the variables $c_{ai}$ as
\begin{align}\nonumber
F_{6,3}(C)=&\frac{1}{(123)(234)\ldots(612)}\left(\frac{(145)}{(456)}\right)^{-\zeta_{0}}\left(-\frac{(234)}{(134)}\right)^{-\zeta_{1,4}} \left(\frac{(345)(124)}{(234)(145)}\right)^{-\zeta_{1,5}}\left(\frac{(456)(125)(134)}{(345)(561)(124)}\right)^{-\zeta_{1,6}}\\
&\hspace{-0.6cm}\left(-\frac{(134)}{(124)}\right)^{-\zeta_{2,4}}\left(\frac{(145)}{(134)(125)}\right)^{-\zeta_{2,5}}
\left(\frac{(561)(124)}{(612)(145)}\right)^{-\zeta_{2,6}}\left(-(124)\right)^{-\zeta_{3,4}}\left(\frac{(125)}{(124)}\right)^{-\zeta_{3,5}}\left(\frac{(612)}{(125)}\right)^{-\zeta_{3,6}}\,.
\end{align}

We present here also a general formula for deformations of MHV amplitudes. They are given as a Gra\ss mannian integral with the measure
\begin{align}\label{MHVFC}
F_{n,2}(C)=&\frac{1}{(12)(23)\ldots(n1)}\left(\frac{(1\, n-1)}{(n-1\, n)}\right)^{-\zeta_{0}}\left(-\frac{(23)}{(13)}\right)^{-\zeta_{1,3}}\left(-\frac{(13)}{(12)}\right)^{-\zeta_{2,3}}\nn\\
&\prod_{i=4}^n \left(\frac{(i-1\, i)(1\, i-2)}{(i-2\, i-1)(1\, i)}\right)^{-\zeta_{1, i}}\left(\frac{(1\, i)}{(1\, i-1)}\right)^{-\zeta_{2, i}} .
\end{align}
We used this formula in section \ref{Sec:ch4_Symmetries}.

\section{Modification of Yangian Generators for Invariance of Top Cells}
\label{app.alphas}

In this appendix we present some examples of the local deformation values $\alpha_i$ in the level-one Yangian generators \eqref{levelonetwistor} found by demanding the Yangian invariance of expressions for top cells \eqref{Rtop}, i.e.~we present values of $\alpha_i$ for which \eqref{eq.YangianF} holds. For simplicity we restrict the presentation to the MHV cases. 
In order to render the results more symmetric we consider all external particles to be incoming. We follow here the notation of figure \ref{Fig:MHVnex}. For any number of particles, while solving relations \eqref{Yangianinvariance}, we can eliminate some face spectral parameters, which leads to two distinct cases. For odd number of external legs all face spectral parameters can be expressed in terms of only external ones. On the contrary, for even $n$ we get, on top of the conservation of central charges, also additional constraints. In that case all $\alpha_i$ can be expressed with use of external face spectral parameters and an internal one, which we combine to define $w=2\zeta_{1,4}-\zeta_0-\zeta_{2,3}$. We express the final results in terms of the central charges of external particles $s_i$, which are differences of external face spectral parameters as in section \ref{Sec:ch3_Deformations}, and $w$. In the following table we collect the results for $\alpha_i$ for $n=4,5,6,7$, together with all constraints on external central charges. In all cases we fix $\alpha_1=0$, see the comment below \eqref{alphas4}.   

\bigskip

\begin{tabular}{ll}
\begin{tabular}{c|c}
$n=4$&\begin{tabular}{l}
$s_1+s_3=0$\\
$s_2+s_4=0$\\
\hline
$\alpha_1=0$\\
$ \alpha_2=w$\\
$ \alpha_3=0$\\
$ \alpha_4=w$
\end{tabular}
\end{tabular}&\hspace{0.4cm}\begin{tabular}{c|c}
$n=5$&\begin{tabular}{l}
$s_1+s_2+s_3+s_4+s_5=0$\\
\hline
$\alpha_1=0$\\
$ \alpha_2=s_1+s_2+2s_3+2s_5$\\
$ \alpha_3=s_1+s_3$\\
$ \alpha_4=s_1+2s_2+2s_3+s_4+2s_5$\\
$ \alpha_5=s_1+2s_3+s_5$
\end{tabular}
\end{tabular}\\
\vspace{0.5cm}\\
\begin{tabular}{c|c}
$n=6$&\begin{tabular}{l}
$s_1+s_3+s_5=0$\\
$s_2+s_4+s_6=0$\\
\hline
$\alpha_1=0$\\
$ \alpha_2=w$\\
$ \alpha_3=s_1+s_3$\\
$ \alpha_4=w+s_2+s_4$\\
$ \alpha_5=s_1+2s_3+s_5$\\
$\alpha_6=w+s_2+2s_4+s_6$
\end{tabular}
\end{tabular}&\hspace{0.4cm}\begin{tabular}{c|c}
$n=7$&\begin{tabular}{l}
$s_1+s_2+s_3+s_4+s_5+s_6+s_7=0$\\
\hline
$\alpha_1=0$\\
$ \alpha_2=s_1+s_2+2s_3+2s_5+2s_7$\\
$ \alpha_3=s_1+s_3$\\
$ \alpha_4=s_1+2s_2+2s_3+s_4+2s_5+2s_7$\\
$ \alpha_5=s_1+2s_3+s_5$\\
$\alpha_6=s_1+2s_2+2s_3+2s_4+2s_5+s_6+2s_7$\\
$\alpha_7=s_1+2s_3+2s_5+s_7$
\end{tabular}
\end{tabular}
\end{tabular}

\bigskip
\noindent One can observe that the following relation holds for all cases we presented above 
\begin{equation}
\alpha_{i+2}-\alpha_{i}=s_i+s_{i+2}\,,
\end{equation}
where we identify $\alpha_{i+n}$ with $\alpha_i$. We claim that this relation holds true for all MHV amplitudes and generalizations of it can be also found for all N${}^k$MHV cases.


\bibliographystyle{nb}
\bibliography{bibliography}

\end{document}